\documentclass[reprint,superscriptaddress,aps,prx,twocolumn,amsmath,amssymb,floatfix]{revtex4-2}

\usepackage{blindtext}
\usepackage{epsfig}
\usepackage{xcolor}
\usepackage{amsmath}
\usepackage{amsfonts}
\usepackage{amssymb}
\usepackage{graphicx}
\usepackage{url}
\usepackage[breaklinks, hypertexnames=false]{hyperref}
\usepackage[english]{babel}
\usepackage{esvect}
\usepackage[normalem]{ulem}
\usepackage[inline]{enumitem}
\usepackage{xfrac}
\usepackage{dsfont}
\usepackage{bm}
\usepackage{bbold}
\usepackage{cleveref}
\usepackage{mathtools}
\usepackage{natbib}
\usepackage{verbatim}
\usepackage[toc, title, titletoc]{appendix}
\usepackage{extarrows}

\hypersetup{colorlinks,citecolor=blue,filecolor=blue,linkcolor=blue,urlcolor=blue}

\def\re{\text{Re}}
\def\im{\text{Im}}
\def\tr{\text{Tr}}

\def\sgn{\text{sgn}}
 \Crefname{equation}{Eq.}{Eqs.}
\Crefname{figure}{Fig.}{Figs.}
\def\tot{\text{tot}}
\def\eff{\text{eff}}
\def\u{\underline}

\bibpunct{[}{]}{,}{n}{}{}

\makeatletter
\renewcommand{\fnum@figure}{FIG. \thefigure}
\makeatother

\numberwithin{equation}{section}
\renewcommand{\theequation}{\arabic{section}.\arabic{equation}}
\addto{\captionsenglish}{}

\begin{document}

\title{Liouvillian skin effect in an exactly solvable model}

\author{Fan Yang}
\affiliation{Department of Physics, Stockholm University, AlbaNova University Center, 106 91 Stockholm, Sweden}
\ 
\author{Qing-Dong Jiang}
\affiliation{Department of Physics, Stockholm University, AlbaNova University Center, 106 91 Stockholm, Sweden}
\affiliation{Tsung-Dao Lee Institute and School of Physics and Astronomy, Shanghai Jiao Tong University, 200240, China}

\author{Emil J. Bergholtz}
\affiliation{Department of Physics, Stockholm University, AlbaNova University Center, 106 91 Stockholm, Sweden}

\date{\today}

\begin{abstract}

The interplay between dissipation, topology and sensitivity to boundary conditions has recently attracted tremendous amounts of attention at the level of effective non-Hermitian descriptions. Here we exactly solve a quantum mechanical Lindblad master equation describing a dissipative topological Su-Schrieffer-Heeger (SSH) chain of fermions for both open boundary condition (OBC) and periodic boundary condition (PBC). We find that the extreme sensitivity on the boundary conditions associated with the non-Hermitian skin effect is directly reflected in the rapidities governing the time evolution of the density matrix giving rise to a {\it Liouvillian skin effect}. This leads to several intriguing phenomena including boundary sensitive damping behavior, steady state currents in finite periodic systems, and diverging relaxation times in the limit of large systems. We illuminate how the role of topology in these systems differs in the effective non-Hermitian Hamiltonian limit and the full master equation framework.  

\end{abstract}

\maketitle

\section{Introduction}

Topological phenomena in the non-Hermitian (NH) realm has attracted ample interest during the past few years  \cite{emil2021,gong2018,fu2018}. Compared to their conventional Hermitan counterparts \cite{HasanKane,QiZhang,Armitage}, NH effective Hamiltonians exhibit an entirely different catalog of gapped and gapless topological phases \cite{emil2021,gong2018,fu2018,kawabataprx,symmetry2,jansym,yoshida2019,kawabataprl,tsuneyaprl,ipsitaprl,marcus22}. The arguably most dramatic effect is caused by a macroscopic piling up of eigenstates at the boundaries of the system \cite{Lee,Alvarez,Xiong}, leading to a spectral sensitivity that grows exponentially with system size when coupling the boundaries \cite{flore2018}. This phenomenology has been dubbed the NH skin effect \cite{yao2018} whereby the celebrated bulk-boundary correspondence in Hermitian systems is replaced by a dichotomy described by either non-Bloch band invariants \cite{yao2018} or in terms of a biorthogonal bulk boundary correspondence \cite{flore2018}. This has led to a blossoming field of research \cite{yao2018,flore2018,Lee,fu2018,flore2019n,Regnault_2019,Okuma_2020,Edvardsson_2019,Longhi,Leykam,LeeThomale,CHLee,secondorder,fleckenstein2022nonhermitian,Rosenow,Borgnia,Murakami,ESpinLiquids,elisabet2020,koch2020,Alvarez,Xiong,CHLee,Schomerus2020,Hyart,NTOS,zhesen,Brandenbourger_2019,Ghatak2020,helbig2020, Neupert_2020,photonicNHBBC, weidemann2020}.

Experiments displaying the NH bulk-boundary correspondence have so far been limited to classical systems including mechanical \cite{Brandenbourger_2019,Ghatak2020}, electrical \cite{helbig2020, Neupert_2020}, and photonic \cite{photonicNHBBC, weidemann2020} platforms. Moreover, the remarkable sensitivity to boundary conditions has recently been suggested to be harnessed in applications such as NH topological sensors \cite{NTOS}, and a quantum input-output theory of such systems has been developed \cite{QNTOS,clerk2020}. A fully consistent quantum mechanical description in terms of Lindbland master equations \cite{lindblad1976} appropriate for Markovian dissipative systems \cite{langen2015,diehl2008, kraus2008, verstraete2009, krauter2011, breuer2007} has earlier been studied and fruitfully employed in the context of preparing or stabilizing Hermitian topological phases \cite{diehl2011, bardyn2013, budich2015, liu2021, he2021}. Recent pioneering work has highlighted that also in such fully quantum mechanically consistent descriptions, the sensitivity to boundary conditions remains \cite{fei2019, wanjura2020,HelicalDamping,mao2021,WanjuraPRL,NENH,ueda2021,zhou2021}, but it is fair to say that a comprehensive understanding is still lacking.

\begin{figure}[b]
        \includegraphics[width=0.95\linewidth]{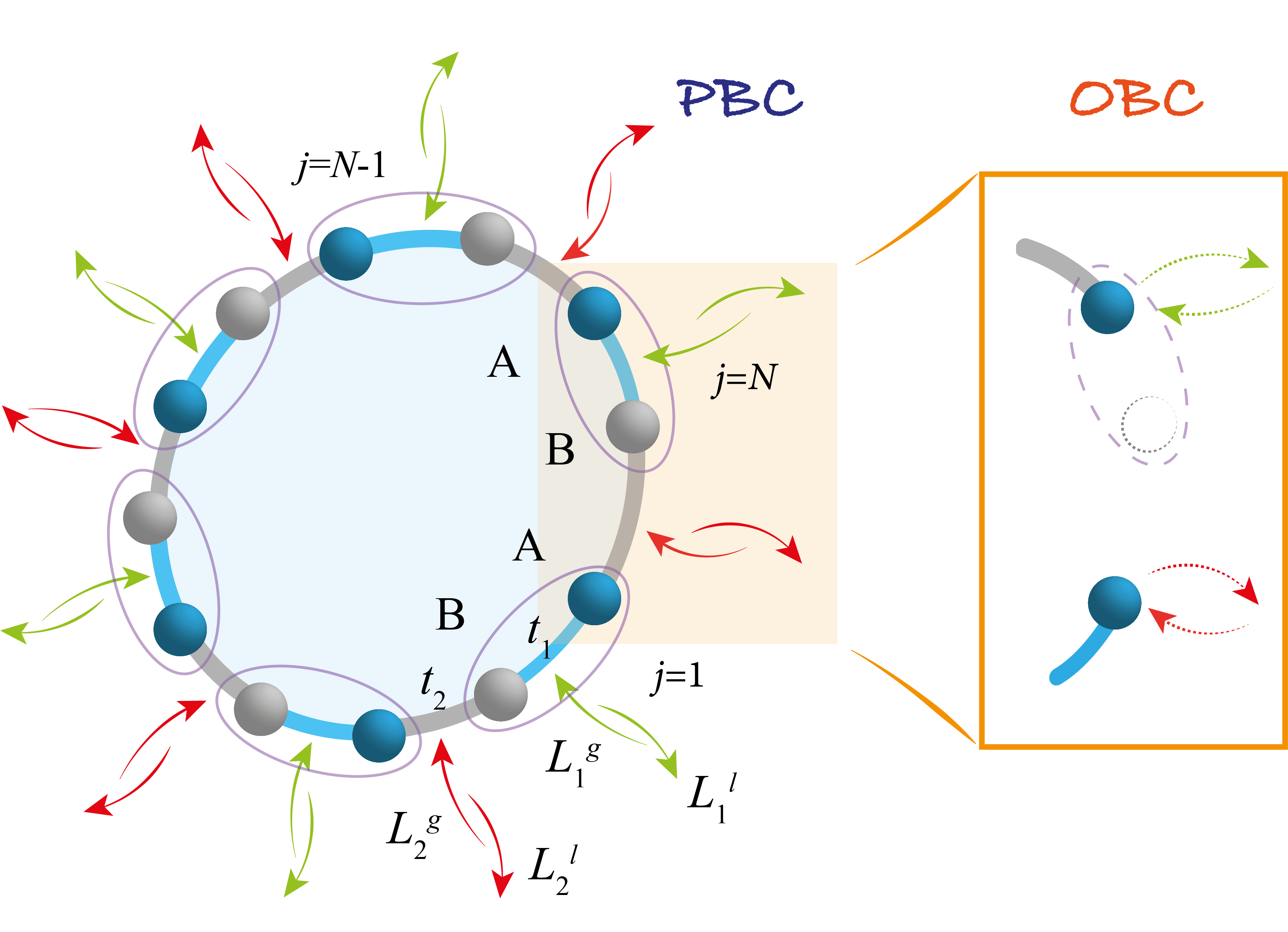} 
            \vskip -0.2cm \protect\caption[]
  {Illustration of a dissipative SSH chain under periodic and open boundary conditions. Lindblad jump operators (solid arrows) are introduced to a total of $2N$ bonds with an explicit form in Eqs.~(\ref{eq:b_dis}), representing the gain and loss from the environment. To switch from PBC to OBC, we remove one $B$ site in the last unit cell. As a consequence, the bond dissipators at the boundary  (dashed arrows)  only act on single $A$ sites marking two ends of the open chain, and share the form of Eqs.~(\ref{eq:e_dis}).}
    \label{fig:ssh_chain} 
    \vskip -0.5cm
\end{figure}

Here we advance the understanding of this problem by providing a complete analytical solution of a class of dissipative fermionic chains for both open and periodic boundary conditions (\Cref{fig:ssh_chain}). Remarkably, we find that there is a NH SSH Hamiltonian $H_\text{S}$ (\ref{eq:h_g}) that fully diagonalizes the Liouvillian: All normal modes can be expressed in terms of the eigenstates $H_\text{S}$, and the rapidities, $\beta_m$, of the Liouvillian are simply related to the {energy eigenvalues} $E_m$ of $H_\text{S}$ as  
\begin{gather}
  \beta_m = \text{const} + iE_m. \label{eq:spec}
\end{gather}
This directly implies that the topological properties and the skin effect of $H_\text{S}$ carry over to the quantum context, including fluctuations and quantum jumps, although the interpretation, as we illustrate, is somewhat altered and there is no skin effect in the steady state reached at sufficiently long times. 

\begin{figure}[t]
        \includegraphics[width=0.95\linewidth]{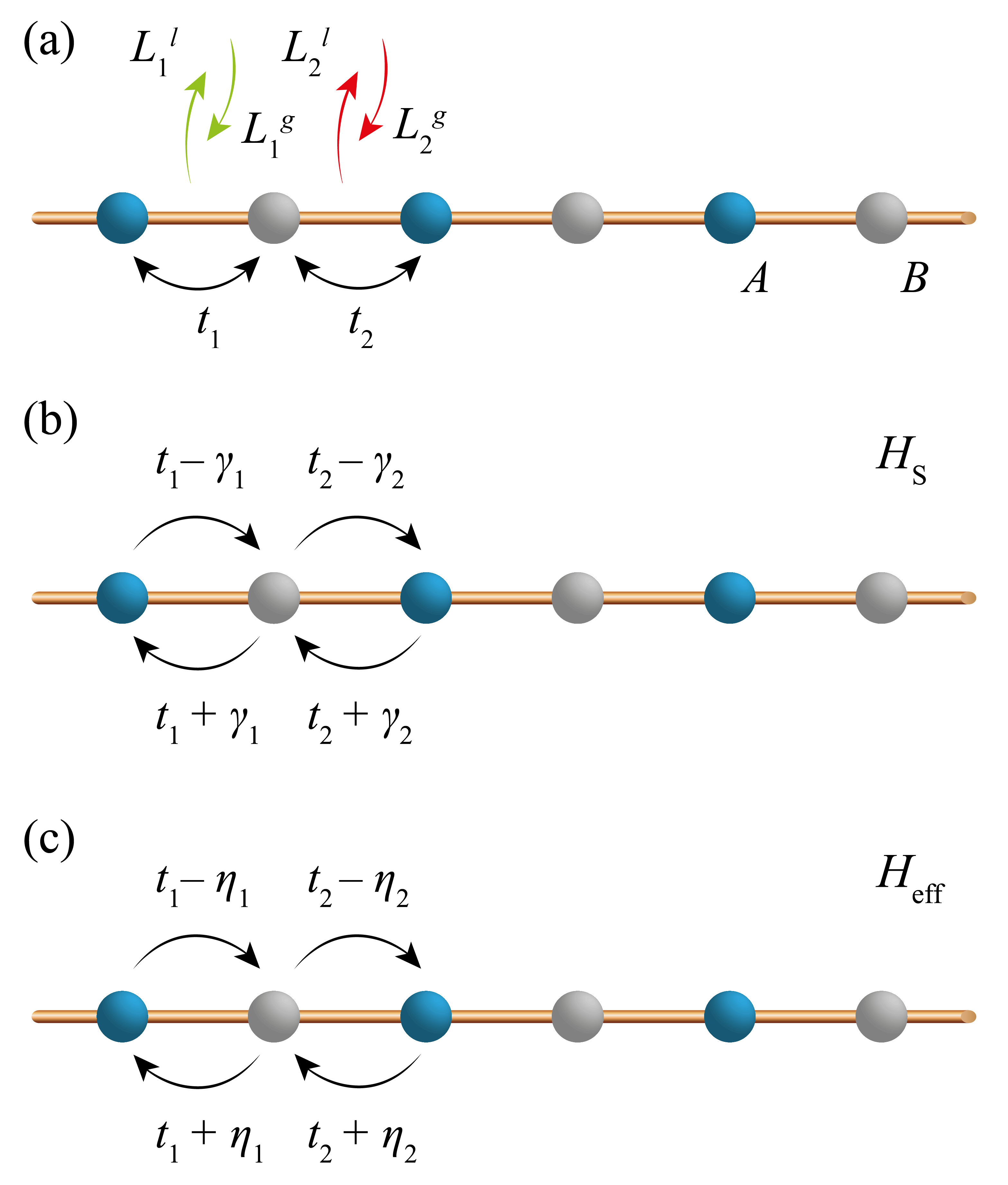} 
            \vskip -0.2cm \protect\caption[]
  {
  Illustration of the full dissipative model in (a). (b) shows the Hamiltonian $H_\text{S}$ in \Cref{eq:h_g} whose eigenstates diagonalizes the full dissipative model (a), and (c) illustrates the effective short-time Hamiltonian $H_\text{eff}$ in \Cref{eq:h_eff} describing the short-time dynamics before any quantum jump has taken place. Intriguingly, both (b) and (c) are NH SSH models although their asymmetric hopping parameters $\gamma_i$ and $\eta_i$ in Eqs.~(\ref{eq:eg}) reflect different aspects of the microscopic dissipative processes in (a). $\gamma_i$ evaluates the total strength of loss and gain dissipations on a given bond, while the imbalance between the two gives rise to $\eta_i$. All models exhibit skin effects as discussed in the text.}
    \label{fig:effectiveH} 
    \vskip -0.5cm
\end{figure}

The effective Hamiltonian, $H_\text{eff}$, of our dissipative model ignoring quantum jumps is also of the form of a NH SSH model in \Cref{eq:h_eff} though the effective parameters reflect different aspects of the underlying dissipation as illustrated in \Cref{fig:effectiveH}. Using recent insights into exactly solving the full set of eigenstates and eigenvalues of NH SSH models \cite{elisabet2020}, this allows us to carry out a detailed study of dynamical phenomena in the Lindblad setting, comparing to the much more studied NH phenomenology and revealing several interesting features, including anomalous damping behavior and diverging relaxation times (in large systems).

The structure of the paper is organized as follows. In \Cref{sec:model}, we introduce the fermionic bond-dissipative SSH model in a Majorana fermion representation. The exact solutions to the quadratic Lindbladian, which can be mapped to a generalized NH SSH chain up to a total rapidity shift, are constructed according to the periodic and open boundary conditions. We point out the differences and links between the full Liouvillian spectrum and the energy spectrum of the truncated effective Hamiltonian that neglects quantum jumps. In \Cref{sec:ness}, we analyze the configurations of the non-equilibrium steady state (NESS). When the boundary is closed, the Liouvillian gap vanishes in an intermediate window of hopping amplitudes and a persistent current flow is identified there. In \Cref{sec:anom}, we discuss the Liouvillian skin effects in an open quantum system. The anomalous quantum dynamics are manifested in the relaxation of the current, the modulated damping behavior, and the lifetime dependence of an edge mode on the system size.

\section{The Model}
\label{sec:model}
Our model is built on a bond-dissipative SSH chain of spinless fermions, shown in \Cref{fig:ssh_chain}.  There are hopping terms  along the chain with alternating amplitudes
$t_1$ and $t_2$: $\mathcal{H} = \sum_j t_1 a_{j,A}^\dagger a_{j,B} + t_2 a_{j+1,A}^\dagger a_{j,B} + {\text{H.c}}$. In the presence of single-particle loss and gain bond dissipations, the set of Lindblad operators takes the general form: $L^l = \sum_{(j,\alpha)} f^l_{(j,\alpha)} a_{j, \alpha}$ and $L^g =  \sum_{(j,\alpha)} f^g_{(j,\alpha)} a^\dagger_{j, \alpha}$. $a_{j, \alpha}^\dagger$ ($a_{j,\alpha}$) creates (annihilates) a fermion in the $j$th unit cell that belongs to the sublattice $\alpha = A$ or  $B$
and they satisfy the fermionic anti-commutation relations: $\{a_{j,\alpha}, a_{j',\alpha'}^\dagger \} = \delta_{j,j'}\delta_{\alpha, \alpha'}$.  Under different boundary conditions, we choose a total of $n  = 2N (2N-1)$ sites for PBC (OBC) with $N$ an integer. Subjected to open boundary, the last unit cell is broken with an empty $B$ site. The odd number of sites helps to stabilize a zero-energy boundary mode and eventually leads to an exact Liouvillian spectrum.

The full dynamics of the generic model is captured by the Lindblad equation \cite{breuer2007, lindblad1976},
   \begin{gather}
     \frac{d \rho}{dt} = \hat{\mathcal{L}} \rho \coloneqq  -i [ \mathcal{H}, \rho ] + \sum_{\mu} ( L_\mu \rho L_\mu^\dagger
     - \frac{1}{2}\{ L_\mu^\dagger L_\mu, \rho \} ),
      \label{eq:lin}
   \end{gather}
 where $\mu$ denotes the summation over all types of Lindblad dissipators. 
 
It is important to note that with linear dissipators, the master \Cref{eq:lin} generates quadratic Lindbladians, 
from which the relaxation process of any observable can be studied by solving the equation of motion either in the Majorana fermion representation \cite{prosen2008, prosen2010ex, prosen2010sp, prosen2011, diehl2011, bardyn2013, caspel2019, cooper2020, zhou2021}
 or in the original complex fermion basis \cite{fei2019, mao2021}.
While the two approaches are equivalent, we employ the first method due to its advantage of obtaining the exact Liouvillian spectrum instantaneously.

\subsection{Liouvillians of non-interacting fermions}
Before addressing the specifics of our model, we give a brief review of this general {\it third quantization} approach
through which the quadratic Lindbladians are diagonalized in the canonical basis of Majorana fermions \cite{prosen2008, prosen2010ex, prosen2010sp}.

Starting with $n$ complex fermions, the reduced density matrix of the system $\rho = \sum_{jk} \rho_{jk} |a_{j}\rangle \langle a_k |$ lives in a Hilbert space of dimension $2^{n} \times 2^n$.
One can always construct a pair of Majorana fermions out of one complex fermion, for instance, $\omega_j = a_l^\dagger + a_l$ and $\omega_k = -i(a_l^\dagger - a_l)$ such that they become their own anti-particles $\omega_j^\dagger = \omega_j$ and satisfy anti-commutation relations $\{ \omega_j, \omega_k \} = 2\delta_{j,k}$.
The Hamiltonian and the dissipators take the matrix form depending on the details of the mapping:
     $\mathcal{H} = \sum_{j,k} w_j H_{j,k} w_k$,
     $L^{\nu} = \sum_j l_{j}^\nu w_j$.

The Hilbert space is now represented in the $2^{2n}$-dimensional Liouville space $\mathcal{K}$
  expanded by the new set of Majorana operators: $P_{\underline{\alpha}} = w_1^{\alpha_1} w_2^{\alpha_2} \cdots w_{2n}^{\alpha_{2n}}$ with $\alpha_j \in \{ 0, 1\}$.
  We apply the notation $\u{x} = (x_1, x_2, \dots, x_k)^T$ to represent a vector of scalars or operators.
   Over the space $\mathcal{K}$,
it is convenient to define the adjoint creation and annihilation linear maps through $\varphi$ fermions: $\varphi_j |P_{\underline{\alpha}} \rangle = \delta_{\alpha_j, 1} | w_j P_{\underline{\alpha}} \rangle$, $\varphi_j^\dagger |P_{\underline{\alpha}} \rangle = \delta_{\alpha_j, 0} | w_j P_{\underline{\alpha}} \rangle$, which obey $\{\varphi_j, \varphi_k^\dagger \} = \delta_{j,k}$. One verifies that
    $|w_j P_{\u{\alpha}} \rangle =  (\varphi_j^\dagger + \varphi_j)  | P_{\u{\alpha}} \rangle$,
    $ |P_{\u{\alpha}} w_j \rangle = \mathcal{P}_F (\varphi_j^\dagger - \varphi_j)  | P_{\u{\alpha}} \rangle$.
The operator $\mathcal{P}_F = (-1)^{\hat{\mathcal{N}}}$  denotes the Fermi parity associated with the fermion number $\hat{\mathcal{N}} = \sum_{j=1}^{2n} \varphi_j^\dagger \varphi_j = \u{\varphi}^\dagger \cdot \u{\varphi}$.
Since $[ \hat{\mathcal{L}}, \mathcal{P}_F ] = 0$ and $ (\mathcal{P}_F)^2 = 1$, the parity is conserved and takes the value $\pm 1$. 

In the canonical basis $| P_{\underline{\alpha}} \rangle$, the Liouvillian in \Cref{eq:lin} contains the even ($+$) and odd ($-$) parity sectors: 
   \begin{align}
       \hat{\mathcal{L}}= &-4i \u{\varphi}^\dagger \cdot H \u{\varphi} \notag \\
      & + \frac{(1+ \mathcal{P}_F)}{2} \left[ - \u{\varphi}^\dagger \cdot (M + M^T) \u{\varphi} + \u{\varphi}^\dagger \cdot (M - M^T) \u{\varphi}^\dagger \right] \notag \\
     &  + \frac{(1- \mathcal{P}_F)}{2} \left[ -\u{\varphi} \cdot (M + M^T) \u{\varphi}^\dagger + \u{\varphi} \cdot (M - M^T) \u{\varphi} \right].
   \end{align}
Here,  the matrix $M = \sum_\nu M^\nu$ arises from the loss and gain dissipators: $M^\nu = (l^\nu)^T (l^\nu)^*$ and $\nu = g, l$. 
   
Since a physical observable must contain an even number of fermionic operators, one may focus on only a definite parity sector, e.g., the even parity sector ($\mathcal{P}_F=1$) in this paper. 
Correspondingly, the dynamics of the system is governed by
   \begin{gather}
     \hat{\mathcal{L}}_+ = \frac{1}{2} 
       \begin{pmatrix}
           \u{\varphi}^\dagger \cdot & \u{\varphi} \cdot
       \end{pmatrix} 
       \begin{pmatrix}
         -X^\dagger & iY \\
         0 & X
       \end{pmatrix}
         \begin{pmatrix}
           \u{\varphi} \\
          \u{\varphi}^\dagger
       \end{pmatrix}
       - A_0, \label{eq:lplus}
    \end{gather}
with $X = -4i H + M + M^T$, $Y = -2i(M - M^T)$, and $A_0 = \frac{1}{2}\tr[X]$. 
After proper diagonalization, 
one is able to express the Liouvillian in terms of  {\it rapidities}  $\beta_m$ and {\it normal master modes} (NMMs) $b'_m$, $b_m$:
  \begin{gather}
      \hat{\mathcal{L}}_+ = - \sum_{m=1}^{2n} \beta_m b'_m b_m, \label{eq:diag}
   \end{gather}
with the band index $m$ and NMMs satisfying the anti-commutation relations $\{ b'_{m}, b_{l} \} = \delta_{m,l}$.

Notably, the complex rapidity spectrum contains rich physics. An initial state $\rho_0 $ of positive fermion parity approaches the NESS after a long-time evolution: $\rho_\text{ss} = \left. e^{\hat{\mathcal{L}}_+ t} \rho_0 \right|_{t \to \infty}$.
While the imaginary part of the rapidity $\beta_m$ encodes the phase oscillation frequency,  the real part of $\beta_m$ reveals the relaxation speed of the system to the steady state and $\re ({\beta_m}) \ge 0$ is required naturally. To describe the asymptotic decay rate quantitively, a spectral Liouvillian gap \cite{minganti2018, ueda2021} can be defined as
  \begin{gather}
     \Delta = 2\min \{ \re[\beta_m]\}.
   \end{gather}

Extra insight comes from the upper triangular structure of $\hat{\mathcal{L}}_+$. It infers that the rapidity spectrum must coincide with the eigenvalues of the matrix $X$ in the diagonal block, which is NH and also called {\it the damping matrix}.  The off-diagonal block $Y$, on the other hand, shapes the configurations of NMMs and NESS \cite{prosen2011, caspel2019}.

 \subsection{Exactly solvable models}
Next, we  apply the third quantization approach to the bond-dissipative SSH chain in \Cref{fig:ssh_chain} and derive exact PBC and OBC Lindblad spectra. Compared with previous studies, our model is more general and include the special cases of Refs.~\cite{cooper2020, fei2019}.
We start from a generic set of linear bond dissipators acting on both $t_1$ and $t_2$ bonds, 
  \begin{align}
   &\begin{cases}
      L^l_{1,j} &= \sqrt{\gamma^l_{1}} (a_{j,A} - i  a_{j,B})  \\
      L^g_{1,j} &= \sqrt{\gamma^g_{1}} (a_{j,A}^\dagger + i a_{j,B}^\dagger),
   \end{cases} \notag \\
  &\begin{cases}
     L^l_{2,j} &= \sqrt{\gamma^l_{2}} (a_{j,B} - ia_{j+1,A})  \\
     L^g_{2,j} &= \sqrt{\gamma^g_{2}} (a_{j,B}^\dagger + ia_{j+1,A}^\dagger).
  \end{cases}  \label{eq:b_dis}
  \end{align} 
Under PBC, the index $j$ runs over $N$ unit cells. 
When the boundary opens up with the last $B$ site taken away, the associated dissipators are curtailed simultaneously: 
    \begin{gather}
    \begin{cases}
      L^l_{1,N} &= \sqrt{\gamma^l_{1}} a_{N,A}  \\
      L^g_{1,N} &= \sqrt{\gamma^g_{1}} a_{N,A}^\dagger,
   \end{cases} \quad
   \begin{cases}
     L^l_{2,N} &= (-i)\sqrt{\gamma^l_{2}} a_{1,A}  \\
     L^g_{2,N} &= i\sqrt{\gamma^g_{2}}  a_{1,A}^\dagger.
  \end{cases} \label{eq:e_dis}
  \end{gather} 
It turns out that the damping matrix $X$ in the Majorana representation can be transformed to the NH SSH Hamiltonian in \Cref{eq:x}:
     \begin{align}
       \mathcal{H}_{\text{S}} &= \sum_j (t_1 + \gamma_1) \varphi_{j,A}^\dagger \varphi_{j,B} + (t_1 - \gamma_1) \varphi_{j,B}^\dagger \varphi_{j,A} \notag \\
                           &\phantom{=} + (t_2 + \gamma_2) \varphi_{j,B}^\dagger \varphi_{j+1,A} + (t_2 - \gamma_2) \varphi_{j+1,A}^\dagger \varphi_{j,B}.
                           \label{eq:h_g}
    \end{align}
The strengths of  asymmetric hopping terms  $\gamma_1$, $\gamma_2$ take the value  $2\gamma_{i} =  |\gamma_i^l| + |\gamma^g_i|$. 
In the Appendix, we construct the exact spectrum of $\mathcal{H}_\text{S}$ as a  direct generalization of Ref.~\cite{flore2018, elisabet2020} to the new limit $\gamma_2 \ne 0$. 
  It enables us to build an exact solution to $\hat{\mathcal{L}}_+$ 
 under both PBC and OBC (with a total number of sites $n = 2N$ and $n = 2N-1$, respectively), using the eigenvectors of  $\mathcal{H}_\text{S}$.
 
 \subsubsection{Majorana representation}
 As a first step, let us define the Liouville space $\mathcal{K}$ by a mapping from $n$ spinless fermions to $2n$ Majorana particles:
   \begin{gather} 
       \begin{cases}
         a_{j,A} &= \frac{1}{2} (c_{j,A} -  id_{j,A}), \\
         a_{j,B} &= \frac{1}{2} (d_{j,B} + ic_{j,B}).
       \end{cases} \label{eq:m_rep}
   \end{gather}
For later convenience, we regroup $c$ and $d$ Majorana fermions  into a whole set $\{w\}$ under the vector notation:
  $\underline{w} = (w_1, w_2, \dots, w_{2n})^T = (c_1,  \dots, c_n, d_1, \dots, d_n)^T$. Accordingly,   in the Lindblad \Cref{eq:lin}, the operators  $\mathcal{H} = \sum_{j,k} w_j H_{j,k} w_k$ and 
     $M_{ij} = \sum_{\nu = g, l} (l^\nu_{i,\mu})^T  (l^\nu_{\mu, j})^*$ with $L_\mu^{\nu} = \sum_j l_{\mu, j}^\nu w_j$ take the following matrix forms:
  \begin{align}
    \u{w}^T H \u{w} &= 
    \begin{pmatrix}
      \u{c}^T & \u{d}^T
    \end{pmatrix}               
    \begin{pmatrix}
              H_0 & 0 \\
              0 & H_0
          \end{pmatrix}     
          \begin{pmatrix}
          \u{c} \\ \u{d}
    \end{pmatrix}, \notag \\
     \u{w}^T M \u{w} &=
    \begin{pmatrix}
      \u{c}^T & \u{d}^T
    \end{pmatrix}               
    \begin{pmatrix}
             M_1 & iM_2 \\
               -iM_2 & M_1
          \end{pmatrix}     
          \begin{pmatrix}
          \u{c} \\ \u{d}
    \end{pmatrix}. 
       \end{align}
$H_0$ and $M_{1,2}$ are $n \times n$ matrices holding entries, for instance, under OBC:
      \begin{align}
      H_0 &= \frac{i}{4}
          \begin{pmatrix}
           0 & t_1 & & & & & & & \\
           -t_1 & 0 & -t_2 & & & & & & \\
            & t_2 & 0 & t_1 & & & & & \\
             &  & -t_1 & 0 &  & & & & \\
                   &   & & & & \ddots &  & & \\
           & &  &  & & & 0 & t_1 &   \\        
        & &   & & & & -t_1 & 0 & -t_2 \\
       & &  & & & & &  t_2 & 0   
         \end{pmatrix}, \notag \\
               M_1 &= \frac{\gamma}{2}  \cdot \mathbb{1}_{n \times n} + \frac{1}{2} \notag \\
         &\times \begin{pmatrix}
           0 & \gamma_1 & & & & & & & \\
           \gamma_1 & 0 & -\gamma_2 & & & & & & \\
            &  -\gamma_2 & 0 & \gamma_1 & & & & & \\
            &  & \gamma_1 & 0 & & & & & \\          
            & & & & & \ddots &  & & \\
       &   &    & & & & 0 & \gamma_1 &   \\
     &  &    & & & & \gamma_1 & 0 & -\gamma_2 \\
     &   &  & & & & & -\gamma_2  & 0   
         \end{pmatrix},  \notag \\
       M_2 &= M_1(\gamma_i \rightarrow \eta_i).
         \label{eq:hb}
         \end{align}
$\gamma_i$'s and $\eta_i$'s stand for the sum
    and the imbalance of loss and gain dissipations:
    \begin{align}
         &\gamma = \gamma_1 + \gamma_2, \quad 2\gamma_{i} =  |\gamma_i^l| + |\gamma^g_i|, \notag \\
         &\eta = \eta_1 + \eta_2, \quad 2\eta_{i} =  |\gamma_i^l| - |\gamma^g_i|. \label{eq:eg}
    \end{align}
    
It can be seen immediately that the Majorana representation in \Cref{eq:m_rep} is better adapted to diagonalize the Liouvillian in \Cref{eq:lplus}:
in the adjoint fermion basis $\u{\varphi}^T = (\u\varphi_c^T, \u\varphi_d^T)$, it incorporates the matrix blocks
  \begin{gather}
     X = \begin{pmatrix}
             X_c & 0 \\
             0 & X_d
            \end{pmatrix}, \quad
     Y = 4 \begin{pmatrix}
              0 &  M_2 \\
               - M_2 &  0 
          \end{pmatrix}, \label{eq:xyo}
       \end{gather}
with $X_c = X_d = -4iH_0 + 2M_1$. Notice that the damping matrix $X$ is diagonal and depends only on the total strength of dissipations $\gamma_i$. By contrast, the matrix $Y$ is off-diagonal (thus couples $\varphi_c$ and $\varphi_d$-fermions) and depends only on the imbalance of gain and loss $\eta_i$. Taking into account the identical structure shared by $X_c$ and $X_d$,
one concludes that the rapidity spectrum {determined by the full damping matrix} is at least doubly degenerate,
    \begin{gather}
      \hat{\mathcal{L}}_+ = - \sum_{m=1}^{n} \beta_m (b'_{c,m} b_{c,m} + b'_{d,m} b_{d,m} ), \label{eq:nmm}
   \end{gather}
where $\beta_m = \beta_{c,m} = \beta_{d,m}$.

It is not difficult to find a unitary transformation $U_{n \times n} =  \text{diag} \{1, i, 1, i, \dots, 1, i, 1\}$ under which $X_{c(d)}$ is mapped to the generalized NH SSH Hamiltonian in \Cref{eq:h_g}:
  \begin{gather}
      X_c = X_d = \gamma \cdot \mathbb{1} + i U H_{\text{S}} U^{-1}. \label{eq:x}
  \end{gather}
  The matrix form $H_\text{S}$ is given in \Cref{eq:ssh}.
We thus reveal one important relation aforementioned in \Cref{eq:spec} for the rapidity spectrum of our model:
      \begin{gather}
        \beta_m = \gamma + iE_m, \label{eq:beta}
      \end{gather}
where $E_m$ represents the eigenvalues of $H_{\text{S}}$.

It can be checked directly that the equality in \Cref{eq:beta} holds true for the PBC spectrum as well. Going to the momentum space, we set the lattice spacing to unity and adopt the Fourier transform $\varphi_{c,{(j,\alpha)}} = \frac{1}{\sqrt{N}} \sum_q e^{iqj} \varphi_{c,\alpha}(q)$ with $q = 2\pi m'/N$, $m' = -N/2, -N/2+1, \dots, 0, \dots, N/2-1$, such that the anti-commutation relations are satisfied: $\{ \varphi_{c,\alpha} (q),\varphi_{c,\alpha'}^\dagger (q') \} = \delta_{q,q'} \delta_{\alpha, \alpha'}$. The same transform is applied to $\varphi_d$-fermions. In the basis of $\u{\varphi} (q) = (\varphi_{c,A}(q), \varphi_{c,B}(q), \varphi_{d,A} (q), \varphi_{d,B} (q))^T$,
the Liouvillian in \Cref{eq:lplus} turns into
     \begin{align}
     \hat{\mathcal{L}}^{\text{PBC}}_+ = &\frac{1}{2} \sum_{q}
       \begin{pmatrix}
           \u{{\varphi}}^\dagger (q) \cdot & \u{{\varphi}} (-q) \cdot
       \end{pmatrix}
       \notag \\
       & \times \begin{pmatrix}
         -X^\dagger (q) & iY(q) \\
         0 & X(q)
       \end{pmatrix}  \begin{pmatrix}
           \u{{\varphi}} (q) \\
          \u{{\varphi}}^\dagger(-q)
       \end{pmatrix} \notag \\
         &- A_0.
    \end{align}
$X(q)$ and $Y(q)$ become $4 \times 4$ matrices inheriting the same structures as before in \Cref{eq:xyo} with ingredients expressed in terms of Pauli matrices:
\begin{align}
   H_0(q)  &=  (-it_2\sin q) \cdot \sigma^x +  i(t_1+t_2\cos q) \cdot \sigma^y, \notag \\
   M_1(q) &=  \frac{1}{2}[ \gamma \cdot \mathbb{1} + (\gamma_1 - \gamma_2 \cos q ) \cdot \sigma^x -\gamma_2 \sin q  \cdot \sigma^y], \notag \\
   M_2(q) &= \left. M_1(q)\right|_{\gamma_i \to \eta_i}.
 \end{align}
The mapping in \Cref{eq:x} from the damping matrix to the NH SSH Hamiltonian
follows naturally under the unitary transformation $U = \text{diag} \{1, i\}$ with
 $H_\text{S} (q)$ shown in Eqs.~(\ref{eq:sshq}). Therefore, we confirm the validity of the equality relation in \Cref{eq:beta} for the PBC rapidity spectrum.

 \subsubsection{Changing boundaries from PBC to OBC}
We proceed to construct the complete set of eigenvectors of the damping matrix based on the mapping in \Cref{eq:x}.
 Let us write the generic eigenvalue equations of 
the NH SSH Hamiltonian:
  \begin{gather}
    H_{\text{S}} \  \tilde{\u{\psi}}_{Rm} = E_m \tilde{\u{\psi}}_{Rm}, \quad
    H_{\text{S}}^\dagger \  \tilde{\u{\psi}}_{Lm}  = E_m^* \tilde{\u{\psi}}_{Lm},
    \label{eq:ssh_es}
  \end{gather}
of which the exact solutions under different boundary conditions are derived in the Appendix.
It renders that the pair of eigenvectors of $X_{c(d)}$ can be constructed as 
  \begin{gather}
     \u{\psi}_{Rm}  = U  \ \tilde{\u{\psi}}_{Rm}, \quad   \u{\psi}_{Lm}  = U  \ \tilde{\u{\psi}}_{Lm}, \label{eq:cb}
    \end{gather}
 with corresponding eigenvalues $\beta_m$ in consistency with the relation (\ref{eq:beta}):
  \begin{gather}
     X_{c(d)} \  {\u{\psi}}_{Rm} = \beta_{m} {\u{\psi}}_{Rm}, \quad
     X_{c(d)}^\dagger \  {\u{\psi}}_{Lm}  = \beta_{m}^* {\u{\psi}}_{Lm}. \label{eq:eox}
       \end{gather}
Under this construction, the biorthogonal normalization of the left and right eigenstates \cite{brody2013, flore2018, elisabet2020} are respected:
  \begin{gather}
     {\u{\psi}}_{Lm}^* \cdot {\u{\psi}}_{Rl} = \tilde{\u{\psi}}_{Lm}^* \cdot \tilde{\u{\psi}}_{Rl}  = \delta_{m,l}. \label{eq:bio}
  \end{gather}

When the boundary is switched from PBC to OBC as depicted in \Cref{fig:ssh_chain}, 
we can extract the rapidity spectrum directly from $E_m$ solved in \Cref{eq:ek,eq:e0,eq:eobc}:
  \begin{align}
    \beta_{\pm}^{\text{PBC}} (q) &= \gamma \pm i [ t_1^2 + t_2^2 - (\gamma_1^2 + \gamma_2^2)   \notag \\
       &\phantom{=} + 2(t_1t_2 + \gamma_1 \gamma_2) \cos q + 2i(t_1 \gamma_2 + t_2 \gamma_1) \sin q ]^{1/2}, \notag \\
   \beta_{\pm}^{\text{OBC}} (q)  &= \gamma \pm i [t_1^2 + t_2^2 - (\gamma_1^2 + \gamma_2^2)    \notag \\
                           &\phantom{=}  + 2 \sqrt{(t_1^2 - \gamma_1^2)(t_2^2 - \gamma_2^2)}\cos q]^{1/2}, \notag \\
     \beta_{m=0}^{\text{OBC}}   &= \gamma.                      \label{eq:rap}
      \end{align}
In the OBC spectrum, the band index $m \in \{(\pm, q), 0\}$ is assigned to $n = 2N-1$ bands with discrete modes $q = \pi m'/N$, $m' = 1, 2, \dots, N-1$. Given an odd number of sites, there emerge two right and left boundary modes at zero energy $E_0 = 0$ with exponential localization factors: $r_R = -(t_1 - \gamma_1)/(t_2 + \gamma_2)$, $r_L = -(t_1 + \gamma_1)/(t_2 - \gamma_2)$ [see also \Cref{eq:e0}]. As for the bulk spectrum,
  analogous to the simplified NH SSH model with only one asymmetric hopping term $\gamma_1$  \cite{yao2018, flore2019n, elisabet2020}, one finds up to a shift in $q$ a general relation of rapidities between two boundary conditions:
  \begin{gather}
     \beta_{\pm}^{\text{OBC}} (q) =  \beta_{\pm}^{\text{PBC}} (q-i\ln (r)), \label{eq:shift}
  \end{gather}
where $r = \sqrt{(t_1-\gamma_1)(t_2 - \gamma_2)/[(t_1 + \gamma_1)(t_2 + \gamma_2)]}$.

It is important to note that under OBC, in the region $|r_L^* r_R| < 1$ or, equivalently,
  \begin{gather}
    \left| t_1^2 - \gamma_1^2 \right| < \left| t_2^2 - \gamma_2^2 \right|, \label{eq:gbb}
  \end{gather}
\textcolor{black}{the complete set of eigenvectors exhibits a non-zero biorthogonal polarization \cite{flore2018} and holds a non-trivial non-Bloch topological invariant \cite{yao2018}.}
It further indicates at $|r_L^* r_R|  = 1$,  the gap of $E_m$ closes [an alternative argument is given above \Cref{eq:gc}]. As a result, when $\mathcal{H}_\text{S}$ in \Cref{eq:h_g} reduces to the Hatano-Nelson model at $t_1 = t_2$, $\gamma_1 = \gamma_2$ \cite{hatano96, hatano97, hatano98, gong2018},  a collapse of the exact bulk states is expected. One is nevertheless able to study the behavior of the system around these gap closing points via approximate variational states \cite{elisabet2020}. Meanwhile,  in a
dissipative quantum system composed of tight-binding bosons and non-linear (quadratic) Lindblad operators, 
a Liouvillian can be constructed in such a way that its diagonal subspace resembles the Hatano-Nelson model, thus giving rise to a similar Liouvillian skin effect \cite{ueda2021}.

 \subsubsection{Exactly solvable NMMs}
We are now prepared to get an analytical set of NMMs for the Liouvillian in \Cref{eq:lplus}, in particular, under the open boundary condition.
 The essence is to remove in the upper triangular structure the off-diagonal block $Y$ that entangles adjoint fermions $\varphi_c$ and $\varphi_d$.
More precisely, 
  \begin{gather}
     \begin{pmatrix}
          -X^\dagger & iY \\
          0 & X
        \end{pmatrix} 
         = W  \begin{pmatrix}
               -X^\dagger & 0 \\
               0 & X
           \end{pmatrix}
           W^{-1}.    \label{eq:trans_Y}
          \end{gather}
 We find a solution for the transformation above
   \begin{gather}
     W = \begin{pmatrix}
                \mathbb{1}_{2n \times 2n} & C \\
               0 & \mathbb{1}_{2n \times 2n} 
              \end{pmatrix}, 
       \end{gather}
where  the covariance matrix $C$ satisfies
  \begin{gather}
     X^\dagger C + CX = iY. \label{eq:cv}
  \end{gather}
It is easy to discern that if ${\eta}/\gamma = {\eta_1}/{\gamma_1}$ or ${\eta_2}/{\gamma_2}$,
the covariance matrix holds a simple structure
  \begin{gather}
     C = \frac{i\eta}{\gamma} \begin{pmatrix}
    0 &  \mathbb{1}_{n \times n}  \\
    - \mathbb{1}_{n \times n} & 0
     \end{pmatrix}.
    \end{gather}
  By definition in  Eqs.~(\ref{eq:eg}), the solvable limit encompasses the following possibilities: 
  \begin{gather}
    \gamma_1\gamma_2 = 0, \quad \frac{\gamma_1}{\eta_1} = \frac{\gamma_2}{\eta_2} \ne 0 \quad \text{and} \quad \eta_1 = \eta_2 = 0. \label{eq:sl}
  \end{gather}

A deeper understanding of the covariance matrix comes from the pairing function of Majorana fermions:
${C}_{jk} (t) = - \tr [ w_j w_k \rho (t)] + \delta_{jk}$. In its time evolution, by applying the Lindblad \Cref{eq:lin}
in combination with the anti-commutation relations of Majorana fermions, one arrives at 
  \begin{gather}
    \partial_t {C}(t) = -{C}(t)X - X^\dagger {C}(t) + iY. \label{eq:eom1}
  \end{gather}
For a steady state, $\partial_t {C}_s = 0$. 
Therefore, 
  \begin{gather}
     C_s \equiv C.
   \end{gather}
It can be interpreted that in the solvable limit of the Lindbladian in Eqs.~(\ref{eq:sl}), the covariance matrix encodes a stationary pairing pattern favoured by Majorana fermions at large times:
  \begin{align}
    \langle c_{j,\alpha} d_{j',\alpha'} \rangle_{\text{ss}} &= -\frac{i\eta}{\gamma} \delta_{j,j'} \delta_{\alpha, \alpha'}, \notag \\
    \langle c_{j,\alpha} c_{j',\alpha'} \rangle_{\text{ss}} &= \langle d_{j,\alpha} d_{j',\alpha'} \rangle_{\text{ss}} = \delta_{j,j'} \delta_{\alpha, \alpha'}. 
  \end{align}
A non-vanishing covariance matrix plays a role in counteracting the effect of the imbalanced gain and loss dissipations ($\eta$).  Back to the original Hilbert space expanded by spinless fermions $a$,
a more physical picture can be drawn as follows. Since $n_{j,\alpha} = a_{j,\alpha}^\dagger a_{j,\alpha} = (1-ic_{j,\alpha}d_{j,\alpha})/2$,  the dissipative fermionic SSH chain has a tendency to evolve towards a uniformly distributed 
configuration with an occupation number:
  \begin{gather}
    n_{\text{ss}, (j,\alpha)} = \frac{\gamma - \eta}{2\gamma} \in [0, 1]. \label{eq:nss}
  \end{gather}
There is no correlation between $a$ fermions. In general,  
one can express the steady state in the form of a product state,
    \begin{gather}
      \left| \Psi_{\text{ss}} \right> = \bigotimes_{j = 1}^{n_{\text{tot}}} \left( \sqrt{\frac{\gamma + \eta}{2\gamma}} \left| 0 \right>_j + e^{i\theta_j} \sqrt{\frac{\gamma - \eta}{2\gamma}} \left| 1 \right>_j \right), \label{eq:ss}
    \end{gather}
    {where $\theta_j$ denotes an arbitrary phase difference.} 

In the end, with a covariance matrix fulfilling the transformation in \Cref{eq:trans_Y}, we can relate the NMMs in \Cref{eq:nmm} to the left and right eigenvectors of the damping matrix in Eqs.~(\ref{eq:eox}):
   \begin{gather}
     \begin{cases}
       \ b'_{c,m} = \u{\psi}_{Lm}^* \cdot \u{\varphi}_c^\dagger \\
       \ b_{c,m} = \u{\psi}_{Rm} \cdot (\u{\varphi}_c - \frac{i\eta}{\gamma} \u{\varphi}_d^\dagger),
     \end{cases} \notag \\
        \begin{cases}
       \ b'_{d,m} = \u{\psi}_{Lm}^* \cdot \u{\varphi}_d^\dagger \\
       \ b_{d,m} = \u{\psi}_{Rm} \cdot (\u{\varphi}_d + \frac{i\eta}{\gamma} \u{\varphi}_c^\dagger).
     \end{cases}
   \end{gather}
The anticommutation relations are guaranteed by the biorthonormality in Eqs.~(\ref{eq:bio}): $\{ b'_{c, m}, b_{c, l} \} = \{ b'_{d, m}, b_{d, l} \} = \delta_{m,l}$ and all others zero.
   
In the same manner, under PBC, the covariance matrix takes a simple analytical form in the solvable limit in Eqs.~(\ref{eq:sl}),
  \begin{gather}
     C(q) = \frac{i\eta}{\gamma} \begin{pmatrix}
    0 &  \mathbb{1}_{2 \times 2}  \\
    - \mathbb{1}_{2 \times 2} & 0
     \end{pmatrix},
    \end{gather}
and the set of NMMs associated with rapidities $\beta_{\nu=\pm}^{\text{PBC}} (q)$ is given by
   \begin{align}
    & \begin{cases}
       \ b'_{c,\nu}(q) = \u{\psi}^*_{L, \nu}(q)  \cdot \u{\varphi}_c^\dagger(-q) \\
       \ b_{c,\nu} (q) = \u{\psi}_{R,\nu}(q) \cdot \left[ \u{\varphi}_c (-q) - \frac{i\eta}{\gamma} \u{\varphi}_d^\dagger(q) \right],
     \end{cases} \notag \\
   &    \begin{cases}
       \ b'_{d,\nu}(q) = \u{\psi}^*_{L, \nu}(q)  \cdot \u{\varphi}_d^\dagger(-q) \\
       \ b_{d,\nu} (q) = \u{\psi}_{R,\nu}(q) \cdot \left[ \u{\varphi}_d (-q) + \frac{i\eta}{\gamma} \u{\varphi}_c^\dagger(q) \right].
     \end{cases}
   \end{align}

\subsection{Spectrum and topology}
\label{sec:spec}
In this section, we reveal the topology of the Liouvillian starting from the spectral winding of the NH damping matrix. 
Known for a NH system, the conventional bulk-boundary correspondence is broken. Yet the prevalence of the exceptional points (EPs) in the OBC rapidity spectrum
implies the Liouvillian skin effect, of which more physical consequences will be discussed in \Cref{sec:anom}.
We further analyze the formation of the Liouvillian gap that is found to be highly sensitive to the boundary conditions. 

\subsubsection{Spectral winding number and exceptional topology}
Based on different classification schemes \cite{cooper2020, altland2021}, open fermion matter falls into one of the ten NH Bernard-LeClair symmetry classes.
Given a quadratic Liouvillian \cite{cooper2020}, the classification can be defined through the damping matrix $X$, or, in our case, $X_{c(d)}$:
  \begin{align}
    Z &= -iX_{c(d)}, \notag  \\
  &= -i(\gamma_1 + \gamma_2) \cdot \mathbb{1} - [i\gamma_1+t_2\sin q - i\gamma_2 \cos q] \cdot \sigma^x \notag \\
  &\phantom{=}  + [(t_1+t_2\cos q) + i \gamma_2 \sin q] \cdot \sigma^y.
  \end{align}
The matrix $Z$ resembling the Hamiltonian in the closed limit preserves the time-reversal, particle-hole, and pseudo-anti-Hermiticity (PAH, or generalized chiral) symmetries:
$Z = \sigma^x Z^T \sigma^x$, $Z = - Z^*$, and $Z = -\sigma^x Z^\dagger \sigma^x$.
Hence, each subspace of the damping matrix belongs to class BDI with a $\mathbb{Z}$ classification in 1D.
It should be noted that at the edge,  two real Majorana fermions $c$ and $d$ recombine into one complex fermion $a$ as indicated by \Cref{eq:m_rep}. When the dissipations are turned on, this  edge mode shares a finite lifetime {with a contribution coming from the effective Liouvillian gap (non-vanishing as indicated by the purely imaginary total energy shift in $Z$) and another from the non-Hermiticity of the damping matrix [see \Cref{eq:gap_eff}].}

\begin{figure}[t]
            \includegraphics[width=0.72\linewidth]{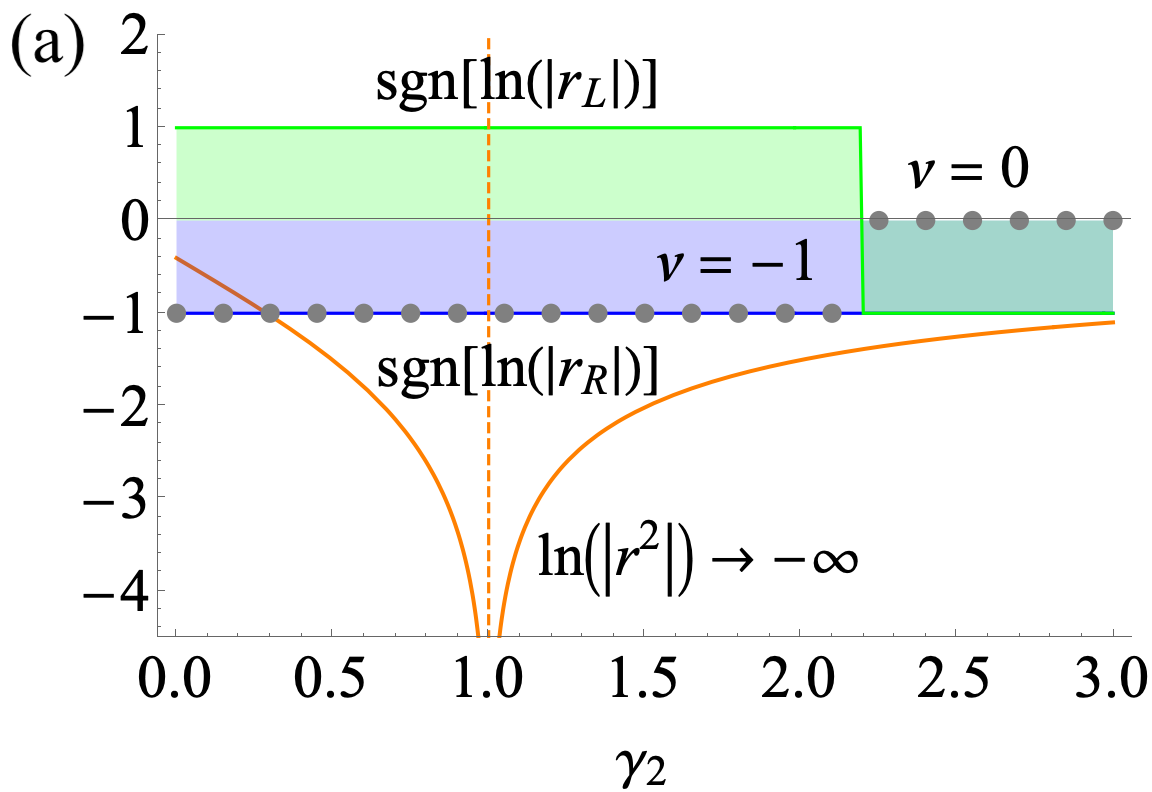} \\
            \includegraphics[width=0.72\linewidth]{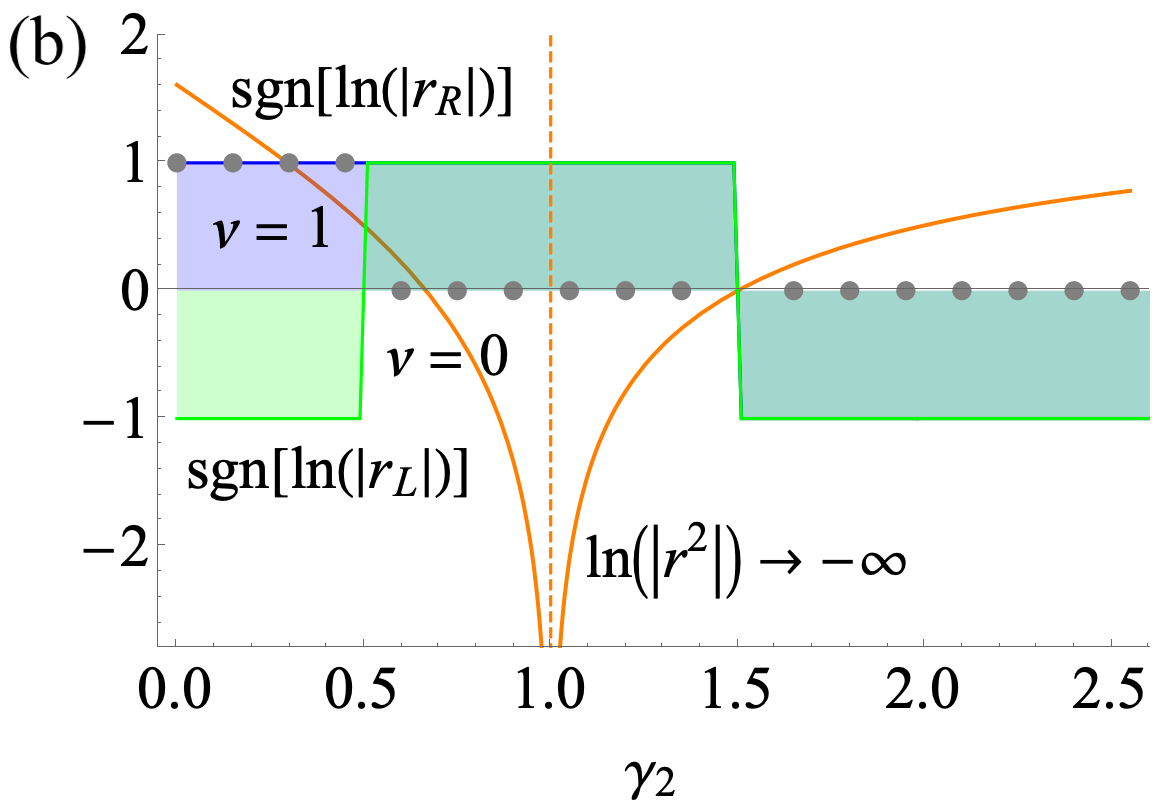}
                                  \vskip -0.2cm \protect\caption[]
  {Spectral winding number (gray dots) and localization determinants for  the left and right boundary modes (green and blue lines) and for the bulk modes (orange curve) as a function of the bond dissipation strength $\gamma_2$ at (a) $t_1 = 1, t_2 = 1, \gamma_1 = 0.2$; (b) $t_1 = -1, t_2 = 1, \gamma_1 = 1.5$. For the evaluation of the spectral winding number $\nu$, we turn the integral in \Cref{eq:nu} into a discrete summation over $N = 500$ unit cells.}
    \label{fig:topo} 
    \vskip -0.5cm
  \end{figure}

In the presence of the PAH symmetry,  a $\mathbb{Z}$ classification is captured by the topological invariant, spectral winding number. By shifting the reference point to $(0, -i\gamma)$ in the complex plane, it is
equivalent to evaluate the winding of the Bloch Hamiltonian $H_{\text{S}}(q)$ \cite{gong2018, fu2018}:  
  \begin{gather}
    \nu = \frac{1}{2\pi i} \int_{-\pi}^{\pi} dq \  \partial_q \ln \{ \det[H_{\text{S}} (q)] \}. \label{eq:nu}
  \end{gather}
Remarkably, when the spectral winding number becomes non-trivial, the left and right boundary modes in the OBC spectrum  in \Cref{eq:e0} start to localize at different ends of the chain:
  \begin{gather}
   |\nu| = 1, \quad  \sgn [\ln (|r_L|)] \ne \sgn [\ln (|r_R|)]. \label{eq:ntri}
  \end{gather}
For $\gamma_1 \gamma_2 \ge 0$, the above topological regime resides in   
  \begin{align}
     & t_1 t_2 > 0, \quad 
       \begin{cases}
           \left| |t_1| - |t_2| \right| < |\gamma_1| + |\gamma_2| \\
           ||\gamma_1| - |\gamma_2|| <  |t_1| + |t_2|,
       \end{cases} \notag \\
       &  t_1 t_2 < 0, \quad 
       \begin{cases}
           \left| |t_1| - |t_2| \right| < ||\gamma_1| - |\gamma_2|| \\
            |\gamma_1| + |\gamma_2|  <  |t_1| + |t_2|,
       \end{cases} \notag \\
       &  \phantom{t_1 t_2 < 0 \ } \text{or} \ 
       \begin{cases}
           \left| |t_1| - |t_2| \right| > ||\gamma_1| - |\gamma_2|| \\
            |\gamma_1| + |\gamma_2|  >  |t_1| + |t_2|.
       \end{cases} \label{eq:topo}
  \end{align}
   In the framework of our model, the bond dissipators entail $\gamma_1 \ge 0, \gamma_2 \ge 0$.
   Figure~\ref{fig:topo} shows the dependence of the spectral winding number on one of the dissipation strengths with the signs of two symmetric hopping terms being either the same ($t_1 = t_2$) or the opposite
    ($t_1 = - t_2$). As soon as $|\nu| \ne 0$, the left and right boundary modes exponentially pile up at different ends. This unique feature of $\mathcal{H}_\text{S}$ can be applied to the design of the NH topological sensors exhibiting anomalous sensitivity that grows exponentially with the system size \cite{NTOS}.
  
Inside the bulk spectrum, however, it is clear to see that the topological invariant $\nu$ obtained from the Bloch Hamiltonian  fails to locate the boundary zero modes at $E_{m \ne 0}^{\text{OBC}} = 0$  [if compared with \Cref{eq:gc}]. 
In a closed NH system, the concept of conventional 
bulk-boundary correspondence has thus been generalized to allow the reconstruction of topological quantities in the biorthogonal basis \cite{flore2018, yao2018} 
such that  the occurrence of the boundary modes are accurately predicted [see also \Cref{eq:gbb}]. What happens to an open quantum system? Similarly, its relaxation dynamics have \textsl{the Liouvillian skin effect} once the NMMs of the Liouvillian  pile up exponentially close to the boundary \cite{fei2019, ueda2021}. It is then crucial to look at the behavior around
 the EPs, arising naturally from the Lindblad master equation \cite{hatano2019}:
  \begin{gather}
       t_i = \pm \gamma_i,  \qquad i = 1, 2. \label{eq:ep}
  \end{gather}
At EPs, the geometric multiplicity of the Liouvillian is smaller than the algebraic multiplicity which holds an order that scales with the system size  \cite{Alvarez, flore2019n, emil2021}.
We check that for $t_i = \pm \gamma_i$, the rapidity spectrum $\beta_{m}^{\text{OBC}}$ in Eqs.~(\ref{eq:rap}) indeed has one or three eigenvalues. Approaching one of the EPs, the set of NMMs merges into one or three linearly independent eigenstates. 
Moreover, after the mapping of the damping matrix to $\mathcal{H}_{\text{S}}$ in \Cref{eq:h_g}, the adjoint fermions are only permitted to hop in one direction when $t_i = \pm \gamma_i$, so all NMMs become exactly localized at that one end. Consequently, we envision the most drastic Liouvillian skin effect in close proximity to EPs,  which, in terms of the momentum shift parameter $r$ defined for linking two rapidities in \Cref{eq:shift}, is manifested as 
\begin{gather}
       t_i \to \pm \gamma_i  \quad \Leftrightarrow \quad \left|\ln |r^2|\right| \gg 0. \label{eq:sk}
  \end{gather}
 As expected, with a sufficiently large system size, the Liouvillian skin effect is fully determined by the localization behavior of the bulk modes [see also Eqs.~(\ref{eq:bl0})$-$(\ref{eq:le_obc})] and the influence of the boundary modes is negligible. 
By varying the dissipation strength, alongside the spectral winding number, \Cref{fig:topo} compares the responses in different localization determinants: $r_R^j (r_L^j)$, $r^j (r^{-j})$ for the piling  of the right (left) boundary and bulk modes at unit cell $j$.
To conclude, while not characterized by the Bloch topological invariant, the Liouvillian skin effect is embodied in the exceptional topology of the Liouvillian, or, more precisely, the damping matrix.

\subsubsection{Liouvillian gap}
\begin{figure}[t]
          \includegraphics[width=0.72\linewidth]{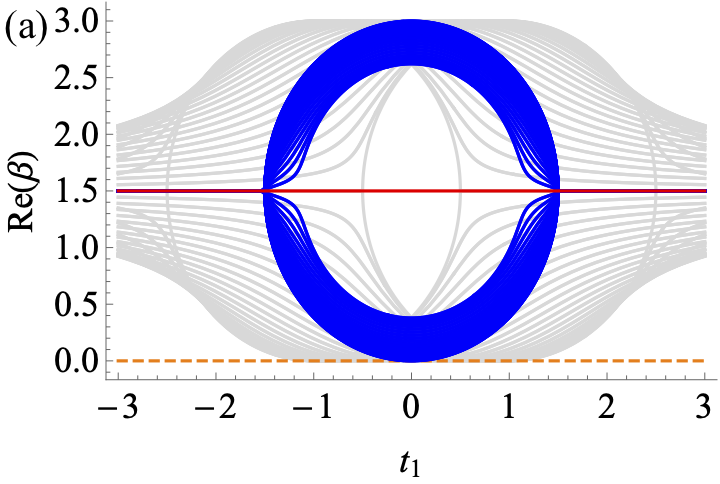}  \\
          \includegraphics[width=0.72\linewidth]{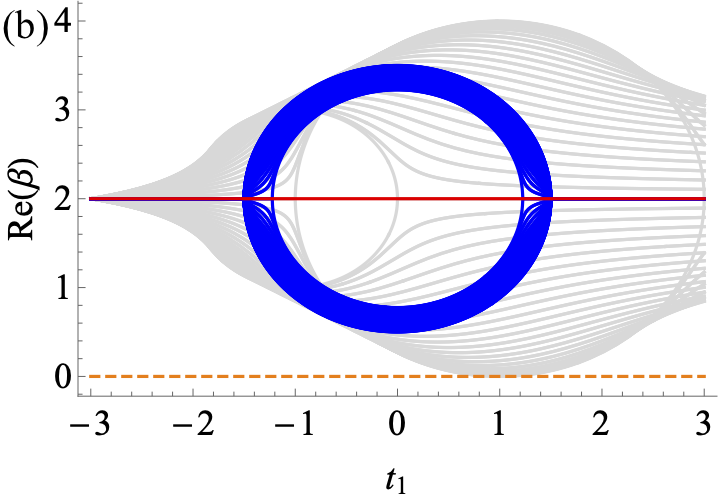}
                      \vskip -0.2cm \protect\caption[] 
  {Real part of the rapidity spectra for $\gamma_1 = 1.5,  t_2 = 1$ and $N = 46$. The grey lines show the structure of the periodic system while the blue and red ones indicate the gap formed by the bulk and edge modes under OBC. We vary the hopping amplitude $t_1$ and set different bond dissipation strength between unit cells: (a) $\gamma_2 = 0$, (b) $\gamma_2 = 0.5$.}
    \label{fig:gap} 
    \vskip -0.5cm
\end{figure}

We go on to study the development of the Liouvillian gap in relation to various hopping amplitudes and dissipation strengths together with its response to
different boundary conditions.
Apparently, the imaginary part of complex energy $E$ in Eqs.~(\ref{eq:rap}) is bounded by $\pm (\gamma_1 + \gamma_2)$. Hence,
$\Delta = 2  \min \{ \re[\beta_m] \} \ge 0 $.
Figure~\ref{fig:gap} shows the real part of the rapidity spectrum as a function of $t_1$ under PBC and OBC. 

Subjected to a periodic boundary,
with finite bond dissipation inside unit cells $\gamma_1 \ne 0$, 
one observes drastically different gap closing behaviors with and without the second inter-unit-cell dissipation:
  \begin{align}
    \gamma_2 &= 0, \quad \Delta^{\text{PBC}} = 0 \text{ for } |t_1| \le |t_2|; \notag \\
    \gamma_2 &\ne0, \quad \Delta^{\text{PBC}} = 0 \text{ at } t_1 = t_2.
  \end{align}
We find that after the introduction of the second bond dissipation, the gap closing line of the Liouvillian discovered in Ref.~\cite{fei2019} 
becomes unstable and shrinks to a point in the phase diagram. It would lead to a collapse of the non-trivial (quasi-)NESS and a termination of the steady-state current in a wide parameter range
(see more details in \Cref{sec:ness_pbc}). 

Meanwhile, the gap-closing points vanish completely when the boundary opens up. In this circumstance, we resolve 
the Liouvillian gap from the exact spectrum in Eqs.~(\ref{eq:rap}):
    \begin{gather}
    \Delta^{\text{OBC}} = 2 \gamma - \sum_{i = 1,2} 2\sqrt{\gamma_i^2 - t_i^2} \cdot \theta(\gamma_i - |t_i|). \label{eq:gapo}
      \end{gather}
Here, $\theta(x)$ denotes the Heaviside step function which is defined as: $\theta (x) = 0$ if $x < 0$; $\theta (x) = 1$ if $x > 0$.
Meeting the EPs at $t_i = \pm \gamma_i$, a discontinuity appears in the derivatives of the gap with respect to the symmetric hopping term $\partial_{t_i} \Delta^{\text{OBC}}$
(see \Cref{fig:gap}) as well as the asymmetric hopping term $\partial_{\gamma_i} \Delta^{\text{OBC}}$ (see \Cref{fig:beta_gamma}).

Lastly, we mention briefly the effects of single-site loss and gain on the Liouvillian. Given a general set of loss and gain dissipators acting 
 on individual sites,
\begin{gather} 
      L^l_{0,(j, \alpha)} = \sqrt{\gamma^l_{0}} a_{j,\alpha},  \quad
      L^g_{0,(j, \alpha)} = \sqrt{\gamma^g_{0}} a_{j,\alpha}^\dagger,
\end{gather}
 the only change to the Liouvillian becomes a modified constant term in the damping matrix that ultimately lifts the minimum of the Liouvillian gap,
\begin{gather}
   \gamma = \gamma_0 + \gamma_1 + \gamma_2, \quad  \Delta \ge 2\gamma_0,
\end{gather}
with $2\gamma_0 = |\gamma_0^l| + |\gamma_0^g|$.
The gap now can no longer be removed by closing the boundary. Fortunately, the on-site dissipations will not add any NH term to the damping matrix, thus
not altering the Liouvillian skin effect. Therefore, we implicitly assume $\gamma_0^l = \gamma_0^g = 0$. 

\subsection{Correlation function and comparison with the effective Hamiltonian}
In this section, we derive a closed form of the single-particle correlation function  from the exact eigenmodes of the damping matrix. The rapidity spectrum
determines the time-dependent part of the two-point correlator. 
Any observable consisting of even number fermionic operators can then be constructed by Wick's theorem. From the perspective of the correlation function,
we compare the physics of the effective Hamiltonian that neglects Lindblad quantum jump operators with the picture of the full Liouvillian. Notably, when the bonds are
subjected to purely loss dissipations, the two mechanisms become identical.

\subsubsection{Single-particle correlator}
Let us start by resolving the time-dependent pairing function for Majorana fermions. In the equation of motion in \Cref{eq:eom1},
the constant matrix $Y$ can be replaced by the covariance (or steady state) matrix from \Cref{eq:cv}. Through a change of variable
$\tilde{C}(t) = C(t) - C_{\text{ss}}$, one reaches
  \begin{gather}
    \partial_t \tilde{C}(t) = -\tilde{C}(t)X - X^\dagger \tilde{C}(t).
  \end{gather}
Starting from an arbitrary initial configuration that is not trivial $\tilde{C}(0) \ne 0$, we can integrate the above equation and implement the diagonalized damping matrix in the exponential:
  \begin{gather}
   X = \sum_m \sum_{\mu = c,d} \beta_{m} |\Theta^\mu_{Rm}\rangle \langle \Theta^\mu_{Lm}|,  \label{eq:decom}\\
      |\Theta^c_{R(L)m}\rangle = \begin{pmatrix}
                                             \u{\psi}_{R(L)m} \\
                                             0
                                           \end{pmatrix}, \quad
             |\Theta^d_{R(L)m}\rangle = \begin{pmatrix}
                                              0 \\
                                              \u{\psi}_{R(L)m} 
                                           \end{pmatrix}.        \notag                                                           
  \end{gather}
Taking into account the biorthogonality of the basis and the fact
that the damping matrix is real, $X^* = X$,
we arrive at
 \begin{gather}
    \tilde{C}(t) = \sum_{m,m'} \sum_{\mu, \mu'} e^{-(\beta_m + \beta_{m'})t} |\Theta^{\mu' *}_{Lm'} \rangle \langle \Theta^{\mu' *}_{Rm'} | \tilde{C}(0) |\Theta^\mu_{Rm} \rangle \langle \Theta^\mu_{Lm} |. \label{eq:pmf}
 \end{gather}
At $t=0$, without loss of generality, throughout the text we choose the system to be in a static configuration with each site completely filled: $| \Psi_0\rangle = \bigotimes_{j = 1}^{n_{\text{tot}}} | 1 \rangle_j$, which corresponds to
  \begin{gather}
    \tilde{C}(0) = -\frac{i(\gamma + \eta)}{\gamma} 
     \begin{pmatrix}
    0 &  \mathbb{1}_{n \times n}  \\
    - \mathbb{1}_{n \times n} & 0
     \end{pmatrix}.  \label{eq:ini}
  \end{gather}
Apparently, $\tilde{C}(0)$ selects $\mu \ne \mu'$. With no pairing between Majorana fermions of the same species in the initial state,
 $\langle c_j c_k \rangle_t =  \langle d_j d_k \rangle_t = 0$. 

Next, we go back to the physical space and define the single-particle correlation function in the spinless fermion language:
$Q_{jk} (t) =  \tr [a_j^\dagger a_k \rho(t) ]$. After the mapping to Majorana fermions in \Cref{eq:m_rep},
one rewrites it in terms of the pairing function,
  \begin{gather}
   Q_{jk}(t) =
          \frac{i}{4} \sigma(j,k) \left[ C_{j,k+n} (t) + C_{k,j+n} (t) \right],
  \end{gather}
with $n$ the total number of sites.  The phase factor depends on whether the correlation resides on the same sublattice or not:
  \begin{gather}
    \sigma(j,k) = \begin{cases}
                 1, & j+k = \text{even} \\
                 (-1)^{j}\cdot(-i), & j+k = \text{odd}.
               \end{cases}
  \end{gather}
Combined with \Cref{eq:pmf,eq:ini}, the single-particle correlator takes the explicit form in terms of the exact solutions of the damping matrix in Eqs.~(\ref{eq:eox}):
 \begin{align}
    &Q_{jk} (t)  = \left( \frac{\gamma + \eta}{2\gamma} \right) \sigma(j,k) \sum_{m,m'} \sum_{l =1}^{n_{\text{tot}}} e^{-(\beta_m + \beta_{m'})t} \notag \\
    &\phantom{===} \psi_{Lm}^* (j) \psi_{Lm'}^* (k) \cdot \psi_{Rm} (l) \psi_{Rm'} (l). \label{eq:spc}
  \end{align}

\subsubsection{Effective Hamiltonian in the absence of quantum jumps}
For non-Gaussian Lindbladians, on the other hand, it is useful to study the short-time dynamics by ignoring quantum jumps in the Lindblad equation \cite{diehl2011}: $\sum_{\mu} L_{\mu} \rho  L_\mu^\dagger$.
Here, we compare the effective Hamiltonian description with the full Lindblad master equation framework.

Without quantum jumps,  the time evolution of the density can be described by
  \begin{gather}
    \partial_t \rho = -i (\mathcal{H}_{\eff} \rho - \rho \mathcal{H}_\eff^\dagger), \label{eq:eom2}
  \end{gather}
where $H_\eff  = H - \frac{i}{2} \sum_\mu L_\mu^\dagger L_\mu$. 
 It turns out that the structure of the effective Hamiltonian becomes drastically different from the damping matrix \cite{fei2019}. 
We verify that  rather than the total strength $\gamma$, the non-Hermiticity of $\mathcal{H}_{\text{eff}}$ is related to the imbalance $\eta$ between loss and gain dissipations:
  \begin{align}
    \mathcal{H}_{\eff} =& \sum_{j=1}^{N-1} (t_1 - \eta_1) a_{j,A}^\dagger a_{j, B} + (t_1 + \eta_1) a_{j,B}^\dagger a_{j,A} \notag  \\
                 &+ (t_2 - \eta_2) a_{j,B}^\dagger a_{j+1, A} + (t_2 + \eta_2)a_{j+1,A}^\dagger a_{j,B} \notag  \\
                 & -i\eta \sum_{(j, \alpha)} a_{(j,\alpha)}^\dagger  a_{(j,\alpha)}  -i s_0. \label{eq:h_eff}
  \end{align}
The purely imaginary energy shift scales with the size of the system: $s_0= [(\gamma - \eta)/2] \cdot n_\tot = \sum_{i=1,2} (|\gamma_i^g|/2) \cdot n_{\text{tot}}$.
It is convenient to resolve  the single-particle correlator directly from $\mathcal{H}_\eff$ according to the equation of motion in \Cref{eq:eom2},
    \begin{align}
    &Q_{jk, \eff} (t)  = \sigma(j,k) \sum_{m,m'} \sum_{l = 1}^{n_{\text{tot}}} e^{-(\beta_{m, \eff} + \beta_{m', \eff})t} \notag \\
    &\phantom{===} \psi_{Lm}^{\eta *} (j) \psi_{Lm'}^{\eta *} (k) \cdot \psi_{Rm}^{\eta} (l) \psi_{Rm'}^{\eta} (l),
  \end{align}
with an effective rapidity spectrum:
    \begin{gather}
       \beta_{m,\eff}= \eta  + iE_m(\eta) + \left( \frac{\gamma - \eta}{2} \right) \cdot n_{\tot}. \label{eq:beff}
    \end{gather}
To work with the same basis as the damping matrix, we have applied the transformation in Eqs.~(\ref{eq:cb}). In the rapidity spectrum, $s_0$ prevents the Liouvillian gap from turning negative when $\eta  = \sum_{i=1,2} (|\gamma_i^l| - |\gamma_i^g|)/2 < 0$.
 
 By neglecting the quantum jumps in the Lindblad equation, we find that the dynamics of the dissipative system are not properly captured by the effective Hamiltonian at all times. For the short-time interval, regardless of the total dissipation strength $\gamma$, the imbalance $\eta$ determines the Lindblad spectrum and makes the Liouvillian gap increase with the system size rather than stay a constant value as suggested by \Cref{eq:beta}.
In the long-time limit, the EPs of the exact solutions move to $t_i = \pm \eta_i$, leaving the Liouvillian skin effect unpredictable compared with \Cref{eq:sk}. Ultimately, the system always decays to an empty chain as there is no residual matrix $Y$ in the equation of motion  that can add up to a finite stationary occupation of fermions according to Eqs.~(\ref{eq:eom1})$-$(\ref{eq:ss}).
 
 In spite of all the discrepancies, however, once gain dissipators are suppressed on the bonds $\gamma_1^g = \gamma_2^g = 0$, we reach one special point 
where
  \begin{gather}
    \gamma = \eta, \quad Q_{jk, \eff} (t)  = Q_{jk} (t).
  \end{gather}
It infers that for the SSH chain with bond dissipations that only lead to losses, though truncated, the effective Hamiltonian encapsulates the full dynamics at arbitrary times.

\section{Non-equilibrium steady states}
\label{sec:ness}
In \Cref{sec:model}, we have already revealed the configuration of the trivial steady state in \Cref{eq:ss}.
It is a unique NESS under OBC and persists in PBC as long as the Liouvillian gap is not vanishing.
With a gapless rapidity spectrum, NESS becomes degenerate and carries a stationary current that is independent of the dissipation strengths 
when the gain and loss contributions are in balance. 
  \begin{figure}[t]
                \includegraphics[width=0.72\linewidth]{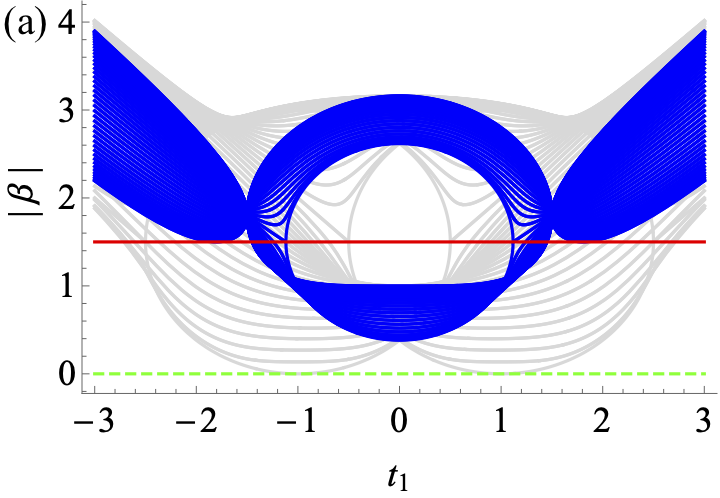}  \\
          \includegraphics[width=0.72\linewidth]{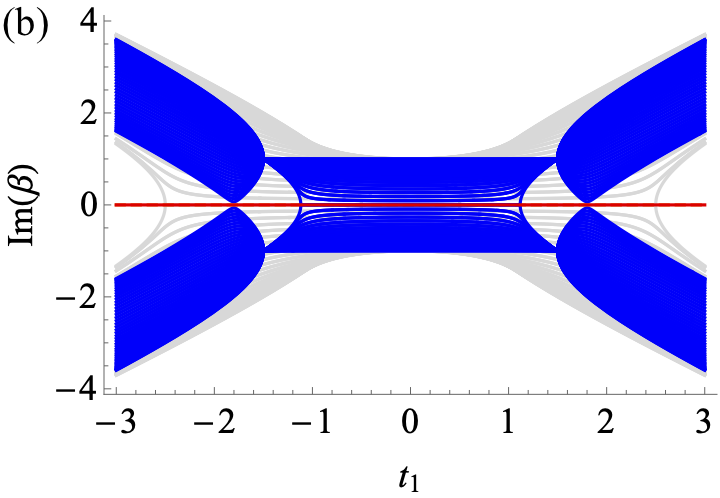}       
            \vskip -0.2cm \protect\caption[]
  {(a) Absolute value and (b) the imaginary part of the rapidity spectra  as a function of $t_1$  for $\gamma_1 = 1.5, \gamma_2 = 0, t_2 = 1$, and $N = 46$. Comparisons are made between the open (blue, red) and periodic  (gray) boundary conditions. For the periodic system, the green dashed line indicates the emergence of a degenerate NESS at $|t_1| = |t_2|$ while the intermediate region $|t_1| < |t_2|$ supports a quasi-NESS mode with a closed Liouvillian gap [see \Cref{fig:gap}\textcolor{blue}{(a)}] and a non-zero imaginary rapidity.}
    \label{fig:beta_s} 
    \vskip -0.5cm
\end{figure}

 \subsection{Open boundary: Uniqueness of NESS}
First, we look at the steady state from the OBC rapidity spectrum. As shown in \Cref{fig:beta_s}\textcolor{blue}{(a)}, every NMM
has a non-zero rapidity:
$\beta_m \ne 0$,  $\forall m$. It implies that the density matrix of the steady state $\rho_\text{ss} = \left. e^{\hat{\mathcal{L}}_+ t} \rho_0 \right|_{t \to \infty}$ is {\it uniquely} determined by the left and right vacua of the Liouvillian in \Cref{eq:nmm}:
$\langle \text{NESS}' | b'_{\mu,m} = 0$, $b_{\mu, m} |\text{NESS} \rangle = 0$ for $\forall m$ and $\mu \in \{ c, d\}$. A proper normalization can be chosen as  $\langle \text{NESS}' | \text{NESS} \rangle  = \tr \rho_{\text{ss}} = 1$.
Moreover, taking into account $\re({\beta_m})>0$ (see \Cref{fig:gap}), any initial state decays to the trivial steady state in \Cref{eq:ss} with a relaxation rate proportional to the strictly positive Liouvillian gap in \Cref{eq:gapo}. 
In particular, at $\gamma_2 = 0$,
  \begin{gather}
    \Delta^{\text{OBC}} =
     \begin{cases}
        \ 2\gamma_1 - 2\sqrt{\gamma_1^2 - t_1^2}, & |t_1| < \gamma_1 \\
        \ 2\gamma_1, & |t_1| \ge \gamma_1. 
    \end{cases} \label{eq:gapo_1}
      \end{gather}
When $|t_1| \ge \gamma_1$, all the bulk and edge modes stabilized by the open boundary share the same Liouvillian gap $2\gamma_1$. Whereas for $|t_1| < \gamma_1$, the Liouvillian gap decreases but stays positive as long as $t_1 \ne 0$. In this regime,  the modes with the slowest decay rate  appear at $m = (\nu, q) \in \{ (+, 0), (-, \pi)\}$.

\subsection{Periodic boundary: Degeneracy, quasi-NESS, and stationary current}
\label{sec:ness_pbc}
From \Cref{sec:spec}, the Liouvillian gap can be closed by switching the boundary condition to PBC, thus lifting the degeneracy of NESS.
For non-zero $\gamma_1$ and $\gamma_2$, \Cref{fig:gap} shows the real part of the PBC rapidity spectrum holds a gapless point at $t_1 = t_2$. 
It is easy to check that the rapidity of  the bulk mode $m^* = (+, -\pi)$ vanishes completely: $\beta_{m^*} = 0$.
 Therefore, 
the right set of the steady states becomes {\it three-fold} degenerate: $s_{R,0} =  |\text{NESS} \rangle$, $s_{R,c} = b'_{c,m^*}|\text{NESS} \rangle$, $s_{R,d} = b'_{d,m^*}|\text{NESS} \rangle $ (accordingly, the left set 
is expanded by $ \langle \text{NESS}'  |, \langle \text{NESS}' | b_{c,m^*}$ and $ \langle \text{NESS}' | b_{d,m^*}$). 
In contrast to OBC,  after a long-time evolution, the final state of the periodic system now depends on the initial configuration and may appear as a superposition among different NESSs: $\rho_\text{ss} = \rho_{\text{ss}} (\rho_0)$.

When $\gamma_2 = 0$,  from \Cref{fig:beta_s}\textcolor{blue}{(a)}, the parameter regime allowing the three-fold degenerate NESS can be extended
 to $|t_1| = |t_2|$ where the zero-rapidity bulk mode appears at $m^* = (+, - \frac{\pi}{2} [\sgn(t_1t_2) + 1])$. Meanwhile, with $\gamma_2$ suppressed, one observes 
interesting features in the relaxation behavior reflected by the Liouvillian gap [see also \Cref{fig:gap}\textcolor{blue}{(a)}]:
  \begin{gather}
    \Delta^{\text{PBC}} =
     \begin{cases}
     \   0, & |t_1| \le  |t_2| \\
     \   2\gamma_1 -  2\sqrt{\gamma_1^2 - (|t_1| - |t_2|)^2}, & |t_2| < |t_1| \le t_c \\
     \   2\gamma_1 - \frac{2\gamma_1 |t_2|}{\sqrt{t_1^2 - \gamma_1^2}}, & |t_1| > t_c,
    \end{cases} \label{eq:gap_p}
  \end{gather}
where the critical value is identified as $t_c =  (|t_2| + \sqrt{t_2^2 + 4 \gamma_1^2})/2$.
In the region $|t_1| < |t_2|$, the Liouvillian gap closes at the bulk modes $m^* = (\pm, \pm \arccos[-t_1/t_2])$. We call them {\it quasi-steady} states, which are stationary states characteristic of a vanishing decay rate and a finite
phase oscillation frequency as shown in \Cref{fig:beta_s}\textcolor{blue}{(b)}: $|\re (\beta_{m^*})| = 0, |\im (\beta_{m^*}) | = \sqrt{t_2^2 - t_1^2}$. Similar to the degenerate NESS, the final state can also select the quasi-steady states without any decay in the probability density.
Once $|t_1| > |t_2|$, the Liouvillian gap of the PBC
spectrum opens up but remains smaller than OBC. The bulk mode that dominates the relaxation process should be the one with a minimal decay rate. 
It changes from $(+, - \frac{\pi}{2} [\sgn(t_1t_2) + 1])$ to $(\pm, \pm \arccos [t_1t_2/(\gamma_1^2-t_1^2)] )$ when the growing amplitude  $|t_1|$ goes past $t_c$. It should be noted that the gap solution in \Cref{eq:gap_p} is valid for both the strong ($\gamma_1 > |t_2|$) and weak  ($\gamma_1 \le |t_2|$) dissipations. 

Compared to the trivial steady state in \Cref{eq:ss}, the degenerate and quasi-NESSs can be viewed as  plane waves of fermions with  fixed momenta $m^*$ on top of a static uniform occupation.  
It is rather important to distinguish the degenerate and quasi-NESSs from the trivial one, especially on account of the former two being a direct consequence of the closing of the Liouvillian gap. 
We find the current flow \cite{caspel2019, benatti2021} is such an ideal observable. Defined as $ j_c (t) = (\frac{i}{n_\text{tot}}) \sum_j [ \langle a_j^\dagger a_{j+1} \rangle_t  -  \langle a_{j+1}^\dagger a_{j} \rangle_t ]$, 
 the time-dependent current flow can be conveniently obtained from the single-particle correlator in \Cref{eq:spc}:
\begin{align}
  j_c (t) 
  &= \frac{1}{2N} \left( \frac{\gamma + \eta}{\gamma} \right) \sum_{j=1}^{N} \sum_{l=1}^{2N} \sum_{m,m' \in \text{all bands}}  e^{-(\beta_m + \beta_{m'})t} \notag \\
  &\times \left[ -\psi_{Lm}^* (2j) \psi_{Lm'}^* (2j-1) + \psi_{Lm}^* (2j) \psi_{Lm'}^* (2j+1) \right]  \notag \\
  &\times  \psi_{Rm} (l) \psi_{Rm'} (l). \label{eq:jt}
\end{align}

Our focal point is to study the behavior of the current in different NESSs. 
At larger times,  only the Liouvillian gapless modes $m^*$ satisfying  $\re (\beta_{m^*}) = 0$ survive. It leads to
\begin{align}
  j_{\text{ss}} &\stackrel{t \to \infty}{=} \frac{1}{2N} \left( \frac{\gamma + \eta}{\gamma} \right) \sum_{j=1}^{N} \sum_{l=1}^{2N} \sum_{m,m' \in m^*}  e^{-i\im ({\beta_m} + \beta_{m'})t} \notag \\
  &\times \left[ -\psi_{Lm}^* (2j) \psi_{Lm'}^* (2j-1) + \psi_{Lm}^* (2j) \psi_{Lm'}^* (2j+1) \right]  \notag \\
  &\times  \psi_{Rm} (l) \psi_{Rm'} (l).
\end{align} 

Let us begin with the special limit $\gamma_2 = 0$ where all three types of NESSs coexist and assume $t_2 > 0$.
For the trivial steady state $m^* = \varnothing$, thus the current vanishes in the end
  \begin{gather}
    \left. j_{\text{ss}}\right|_{\gamma_2 = 0} = 0, \quad |t_1| > t_2. \label{eq:jss_0}
  \end{gather}
At phase transition points,  the degenerate NESS generated by two bulk modes $m^* = (+, - \frac{\pi}{2} [\sgn(t_1t_2) + 1])$ supports a current flow with an amplitude
 \begin{gather}
    \left. j_{\text{ss}}\right|_{\gamma_2 = 0} = \begin{cases}
   			\   0, & t_1 = - t_2; \\
			\   \frac{1}{2N} \left( \frac{\gamma + \eta}{\gamma} \right), & t_1 = t_2.
   			\end{cases} \label{eq:jss_1}
 \end{gather}
 In the intermediate gapless region, there emerges quasi-NESS from the modes $m^* = (\pm, \pm \arccos[-t_1/t_2])$. Analytically, the stationary current can be expanded in the orders of $1/N$:
  \begin{align}
     & \left. j_{\text{ss}}\right|_{\gamma_2 = 0}   \stackrel{t \to \infty}{=} \left( \frac{\gamma + \eta}{\gamma} \right) \left\{  \frac{1}{2N}  \left( \frac{t_1+t_2}{t_2} \right) \right. \notag \\
         &+ \frac{1}{2N^2} \left. \left[ \left( \frac{t_1+t_2}{t_2} \right) \cos(2t_2 \alpha  t) + \alpha \sin(2t_2 \alpha  t)\right]\right\} + \cdots, \label{eq:ord}
   \end{align}
 with $\alpha = \sqrt{1-(t_1/t_2)^2}$. The time dependence in the steady steady current arises from the finite phase oscillation frequency, a unique feature possessed by quasi-NESS.
Given sufficiently large system size, to the leading order $O(N^{-1})$, we get
   \begin{gather}
      \left. j_{\text{ss}}\right|_{\gamma_2 = 0} = \frac{1}{2N} \left( \frac{\gamma + \eta}{\gamma} \right)  \left( \frac{t_1+t_2}{t_2} \right), \quad |t_1| < t_2. \label{eq:jss_2}
   \end{gather}
   
For non-zero $\gamma_1$ and $\gamma_2$, on the other hand, the steady-state current is carried by the gapless bulk mode $m^* = (+, -\pi)$ at $t_1 =  t_2$, while it vanishes elsewhere. Therefore, 
     \begin{gather}
     \left. j_{\text{ss}}\right|_{\gamma_1 \cdot \gamma_2 \ne 0} =  \frac{1}{2N} \left( \frac{\gamma + \eta}{\gamma} \right) \delta_{t_1, t_2}. \label{eq:sc}
    \end{gather} 

\begin{figure}[b]
        \includegraphics[width=0.8\linewidth]{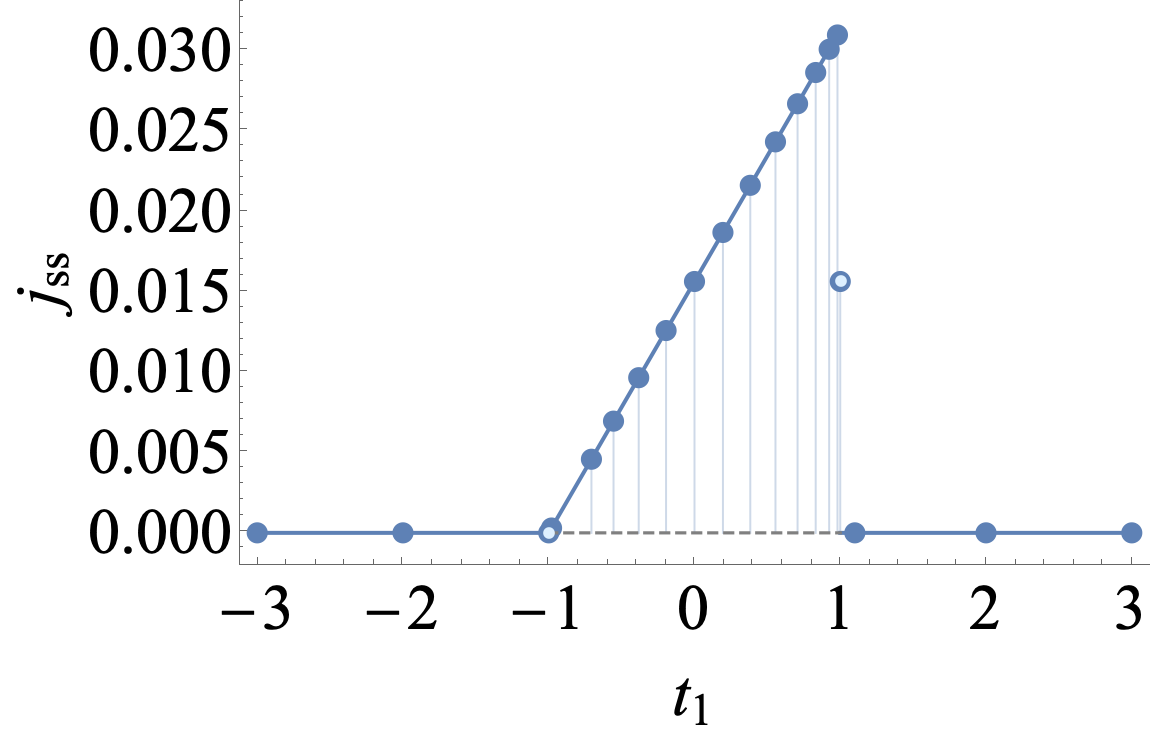}
                       \vskip -0.2cm \protect\caption[]
  {Steady-state current in the periodic chain with balanced gain and loss ($\eta = 0$) as a function of $t_1$ for $\gamma_1 > 0, \gamma_2 = 0, t_2 = 1$, and $N = 32$. The dark dots correspond to the stationary current flow $j_{\text{ss}} = j_c(t)$ measured at the time $\gamma_1 t = 10^5$. At $|t_1| = t_2$, the numerical current matches well with the analytical prediction [light dots, given by \Cref{eq:jss_1}]. For $|t_1| < t_2$, the solid line indicates the leading order approximation in \Cref{eq:ord}.  Outside  this region, the current vanishes due to the opening of the Liouvillian gap. }
    \label{fig:jss} 
    \vskip -0.5cm
\end{figure}

 One immediately notices that when the gain and loss dissipations are in balance, namely $\eta = 0$, the prefactor $(\gamma + \eta)/\gamma \to 1$. 
The steady state current is then independent of the dissipation strengths $\gamma_i$.
 In \Cref{fig:jss}, we confirm the analytical predictions on the steady state current for $\gamma_2 = 0$ in Eqs.~(\ref{eq:jss_0})$-$(\ref{eq:jss_2})  by a measurement of $j_c(t)$ at time $\gamma t = 10^5$.
  With balanced gain and loss bond dissipations, we verify the stationary current remains the same under different values of $\gamma_1$.
When $\gamma_2 \ne 0$, in the same manner, we still find a persistent current at the Liouvillian gap closing point regardless of the choices of $\gamma_{2}$ [see also \Cref{fig:jc}\textcolor{blue}{(a)}]. 

By contrast, in the description of the effective Hamiltonian, once $\gamma \ne \eta$ the real part of the rapidity spectrum for a periodic chain in \Cref{eq:beff} is always gapped and a persistent current will not be observed in
any allowed parameter regime.

\section{Anomalous quantum dynamics}
\label{sec:anom}
In this section, we search for dynamical signatures of the Liouvillian skin effect in dissipative quantum systems, originating from the piling up of the NMMs exponentially close to an open boundary.
Compared with previous studies \cite{fei2019, mao2021, zhou2021}, we show the relaxation behaviors directly obtained from our exact solutions for an odd number of sites $n = 2N-1$ and, at the same time, include 
the impact of the second bond dissipators. 
 Apart from a diverging lifetime without gap closing \cite{ueda2021}, 
 we find that other {\it global} observables such as a tail of a dynamical current flow and a chiral damping wavefront center can also serve as good probes of the Liouvillian skin effect,
 the nature of which will be related to the exceptional topology of the damping matrix. 
 
\subsection{Relaxation of current flow}
  \begin{figure}[t]
        \includegraphics[width=0.47\linewidth]{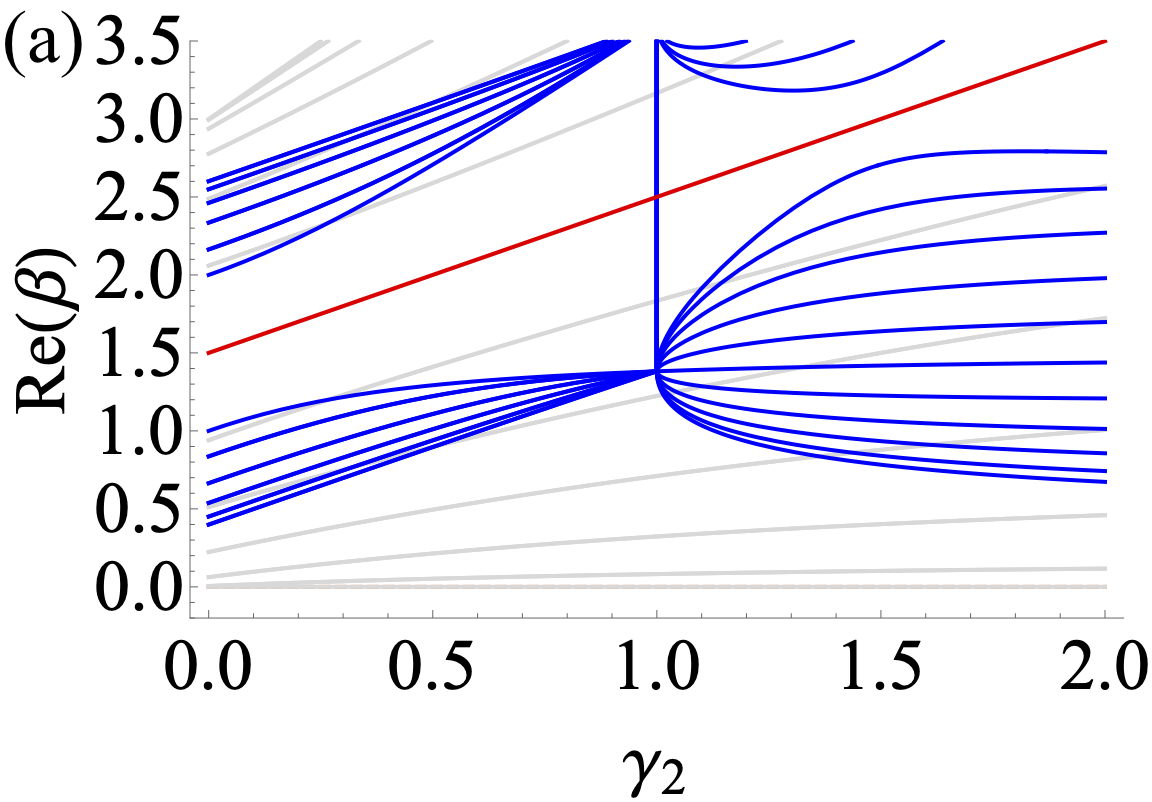}  \ 
          \includegraphics[width=0.47\linewidth]{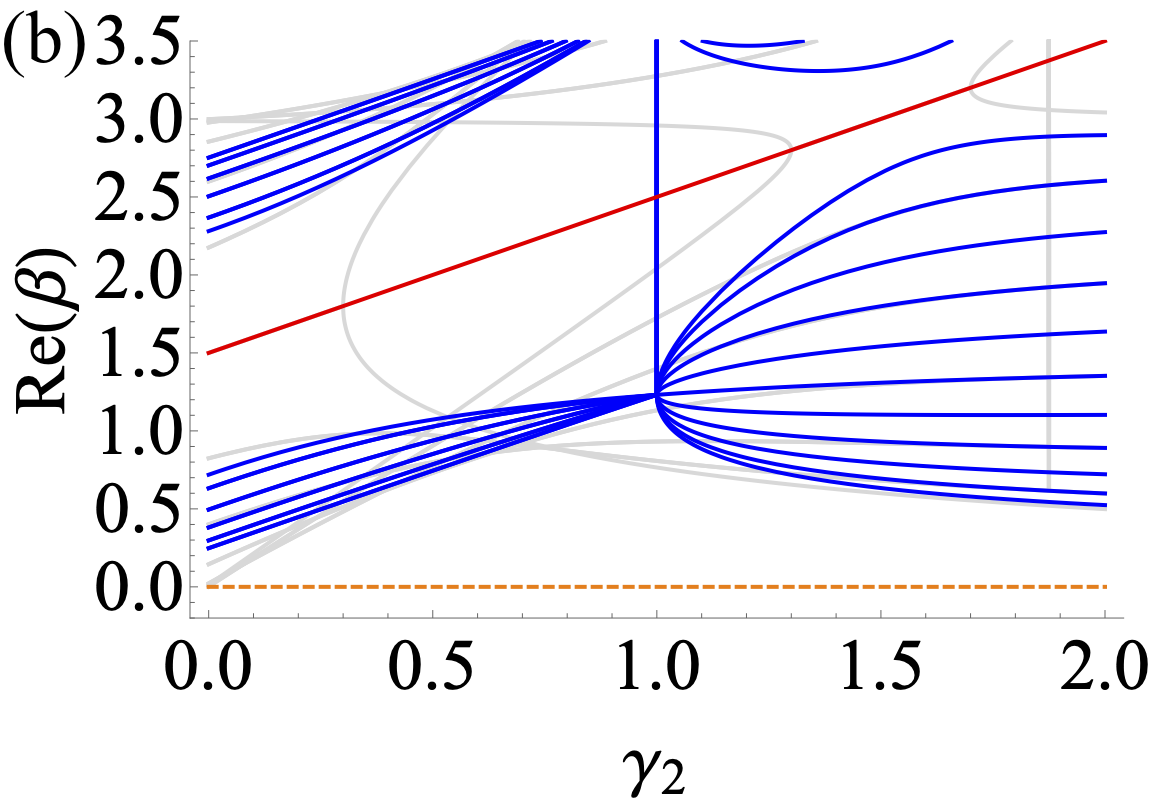}  \\
           \includegraphics[width=0.47\linewidth]{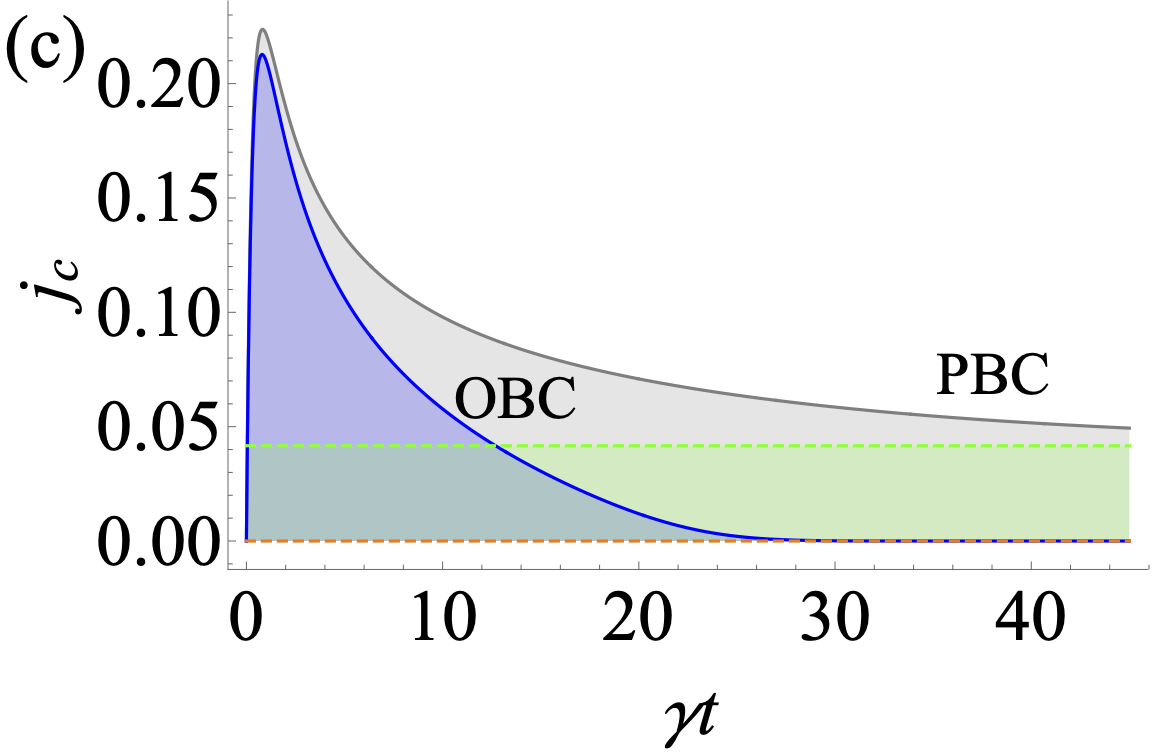}  \ 
          \includegraphics[width=0.47\linewidth]{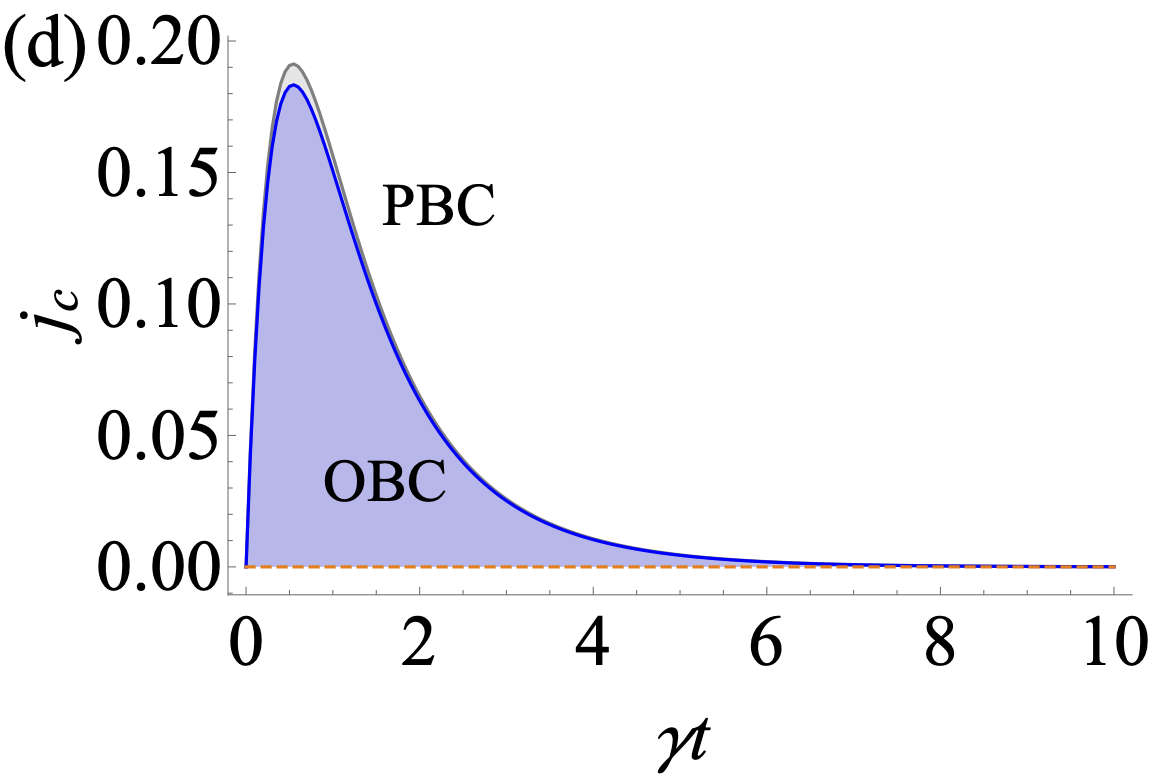}  \\
         \vskip -0.2cm \protect\caption[]
  {The real part of the Lindblad spectrum as a function of the dissipation strength $\gamma_2$  with varied hopping amplitudes: (a) $t_1 = t_2 = 1$, (b) $t_1 = -0.8, t_2 = 1$. We choose $\gamma_1 = 1.5$ and take $N = 12$ unit cells.
Fixing $\eta = 0$ or $|\gamma_1^l| = |\gamma_1^g|$, $|\gamma_2^l| = |\gamma_2^g|$, the lower panel shows the time evolution of the current circulating along the chain with $\gamma_1 = 1.5$, $\gamma_2 = 0.5$, $\gamma = \gamma_1 + \gamma_2 = 2$, and the hopping regimes are chosen in accordance with the upper panel: (c) $t_1 = t_2 = 1$, (d) $t_1 = -0.8, t_2 = 1$. The green line indicates a steady-state current $j_{\text{ss}} = 0.0417 (\simeq 1/24)$ measured at longer times $\gamma t = 10^{4}, 10^{5}$.}
    \label{fig:beta_gamma} 
    \vskip -0.5cm
\end{figure}
As revealed in \Cref{sec:ness_pbc}, for a general set of bond dissipators ($\gamma_1 \ne 0, \gamma_2 \ne 0$),  a non-vanishing current in the steady state in \Cref{eq:sc} helps us to distinguish the degenerate NESS from the trivial one on a periodic chain. Figure~\ref{fig:beta_gamma} depicts the relaxation process of the current before reaching the steady state under different boundary conditions. For PBC, the closing of the Liouvillian gap at $t_1 = t_2$ shown by \Cref{fig:beta_gamma}\textcolor{blue}{(a)} sustains a stationary current in \Cref{fig:beta_gamma}\textcolor{blue}{(c)}
which, after averaging over all sites, scales with the inverse of the system size $\sim 1/N$. Deviating from the gap closing point, the current flow vanishes at relatively short times.  
Changing the boundary condition to OBC, the gapless mode in the rapidity spectrum disappears. All bulk and edge NMMs immediately pile up exponentially at one of the boundaries, thus terminating the current flow in 
\Cref{fig:beta_gamma}\textcolor{blue}{(c)}. If we zoom in to look at the region where the Liouvillian gaps of two spectra are comparable [for instance, $\gamma_2 = 0.5$ in \Cref{fig:beta_gamma}\textcolor{blue}{(b)}], the behaviors of  the current flow seen from \Cref{fig:beta_gamma}\textcolor{blue}{(d)} turn out to be less sensitive to the boundary conditions. 

\begin{figure}[b]
        \includegraphics[width=0.85\linewidth]{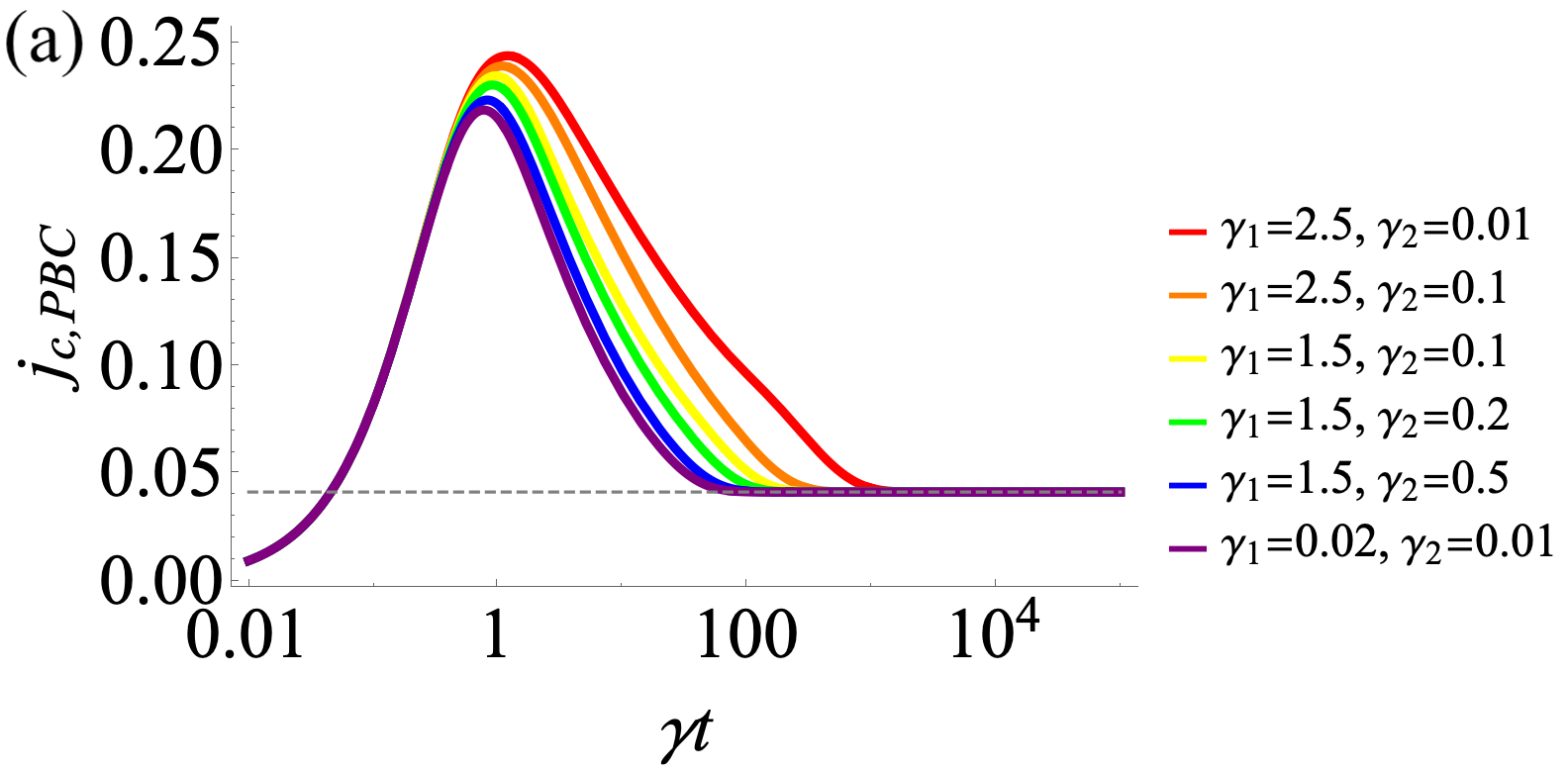} \\
          \includegraphics[width=0.85\linewidth]{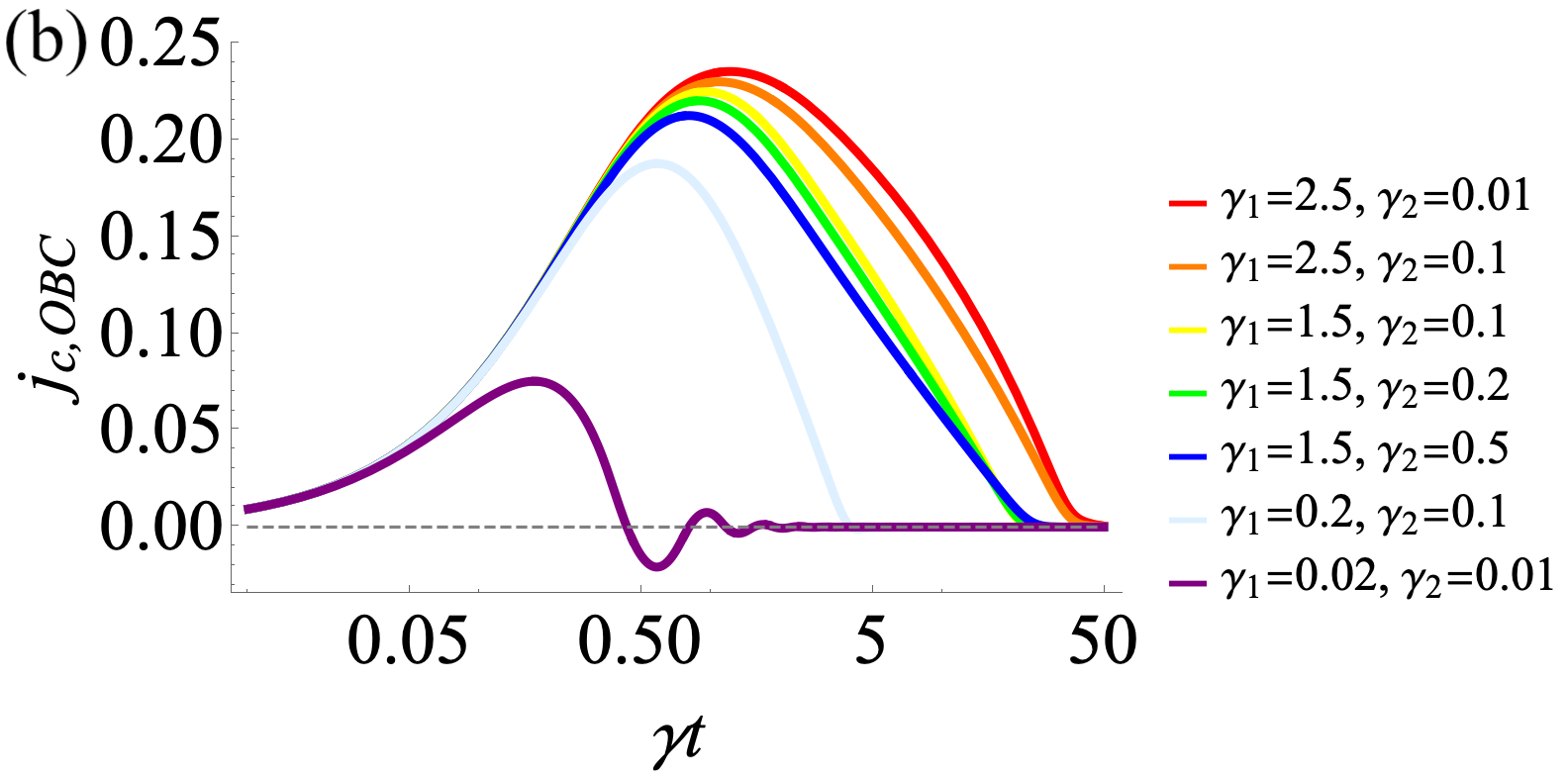}          
            \vskip -0.2cm \protect\caption[]
  {Relaxation of the electronic current under different boundary conditions: (a) PBC, (b) OBC. We vary the bond dissipation amplitudes $\gamma_1, \gamma_2$ while fixing $t_1 = t_2 = 1$ in a system composed of $N = 12$ unit cells.  Given a long-time evolution, the current in a periodic chain saturates to the steady-state value $j_{\text{ss}} = 0.0417 (\simeq 1/24)$ regardless of the strengths of $\gamma_i$. When the open boundary builds up, the current vanishes immediately after reaching the peak. Its relaxation time, equal to the lifetime of the particle at the edge under the Liouvillian skin effect, depends on the system correlation length $\xi (r^2)$ and the effective Liouvillian gap $\Delta_{\text{eff}}$ (\ref{eq:gap_eff}).
  }
    \label{fig:jc} 
    \vskip -0.5cm
\end{figure}

It infers that from the perspective of the current, the Liouvillian skin effect is better captured when the gap of the PBC rapidity spectrum is closed. In \Cref{fig:jc}, we thus fix $t_1 = t_2$ and study the dynamics of the current flow
occurring at the Liouvillian gap-closing point in a wide range of dissipation strengths.  Consistent with our earlier prediction in \Cref{eq:sc} for balanced gain and loss dissipators ($\eta = 0$), \Cref{fig:jc}\textcolor{blue}{(a)} shows that even in the presence of very weak dissipations ($\gamma_1 = 0.02, \gamma_2 = 0.01$), the current flow along a periodic chain saturates to a finite value identical to the limit of strong dissipations. However, once the boundary opens up, driven by weak dissipations, the current flow decays with oscillations and has a much shorter relaxation time as indicated in \Cref{fig:jc}\textcolor{blue}{(b)}. It also represents the behaviors of those points far away from EPs ($t_i = \pm \gamma_i$), bringing about weak Liouvillian skin effect. For stronger dissipations or $|\ln (|r^2|)| \gg 0$, the amplitude of the current is enlarged as the NMMs continue to pile up in the same direction. The current flow also shares a longer relaxation time before the system evolves to the trivial NESS hosting a static uniform distribution of fermions. The relaxation time or the tail of the current flow is determined by two factors, the effective Liouvillian gap $\Delta_\text{eff}$ and the correlation length $\xi(r^2)$ of the system. It is equal to the lifetime of the particle at one boundary in favour of the Liouvillian skin effect, of which more detailed analysis can be found in \Cref{sec:life}. 
\subsection{Density evolution under damping}
Next, we focus on the chiral damping phenomena in the particle-number distribution of the dissipative chain \cite{fei2019}.  We find the damping itself displays the Liouvillian skin effect beyond the gapless point $t_1 = t_2$.
By turning on the second bond dissipators $\gamma_2$,  we show the center of the chiral damping wavefront can be tuned via the localization parameter $r^2$ of the bulk NMMs. It enables us to associate 
the Liouvillian skin effect with the existence of the EPs in the damping matrix at which $|r^2| \to 0$ or $\infty$. We further clarify the sensitivity of dynamical observables to the boundary conditions at longer times.  Our results are consistent with one of the earlier studies \cite{mao2021}, demonstrating that probes in the short-time domain are not sufficient to distinguish the relaxation processes of open quantum matter under different boundary conditions.

\subsubsection{Link with exceptional topology}
First, we associate the chiral damping of the particle-number operator with the exceptional topology of the damping matrix. 
In the full Lindblad master equation framework, the evolution of the particle number to NESS,  $\tilde{n}_j(t) = n_j (t) - n_{j, \text{ss}}$ can be built on our exact 
single particle correlator $\langle a_j^\dagger a_j \rangle_t$ from \Cref{eq:spc}: 
\begin{align}
  \tilde{n}_j(t) 
    &= \left( \frac{\gamma + \eta}{2\gamma} \right) \sum_{m,m'} \sum_{l =1}^{n_{\text{tot}}} e^{-(\beta_m + \beta_{m'})t} \notag \\
    &\phantom{===} \psi_{Lm}^* (j) \psi_{Lm'}^* (j) \cdot \psi_{Rm} (l) \psi_{Rm'} (l). \label{eq:pn}
\end{align}

Under OBC, a large system size enables us to safely neglect the contributions from the boundary eigenmodes in \Cref{eq:e0}.
Given exact solutions of the left and right bulk eigenmodes in \Cref{eq:eox,eq:bl1,eq:le_obc}, one derives an asymptotic scaling of particle number evolution for sites in the odd sublattice $A$: 
  \begin{gather}
  \tilde{n}_{2j-1} (t) \sim \frac{1}{r^{2j}} e^{-\Delta_{\text{eff}} t} \cdot \left( \frac{\gamma + \eta}{2\gamma} \right). \label{eq:asy}
  \end{gather}
A similar expression can be obtained for the even sublattice $B$. Here, the damping factor $r^2$ comes from the localization of the left bulk eigenmodes. When $|r^2| \to 0 (\infty)$, for instance, the left eigenmodes are localized on the right (left) end of the chain, pushing the chiral wavefront towards the same boundary while
the right eigenmodes center on the opposite. In the time-dependent part of the particle number operator, we replace the original gap $\Delta^\text{OBC}$ with an effective Liouvillian gap $\Delta_{\text{eff}}$. The comparison of two quantities is made in \Cref{fig:gap_eff}.
Since we are interested in the full dynamics of the relaxation process, instead of taking the long-time limit and keeping track of the slowest decaying mode, we should include all bulk modes with various decaying rates. 
The summation then leads to an effective Liouvillian gap that is also useful in the evaluation of the particle lifetime in \Cref{eq:gap_eff}. 

The two extremities of the Liouvillian skin effect $|r^2| \to 0$ or $\infty$ occur at EPs: $t_i = \pm \gamma_i$, when the adjoint fermions in the damping matrix  are only allowed to hop in one direction after the mapping to $\mathcal{H}_\text{S}$ in \Cref{eq:h_g}. It is noteworthy that with a large system size, the chiral damping condition can be further relaxed to  
 \begin{gather}
   |r^2| \ll 1 \quad \text{or}  \quad |r^2| \gg 1,
 \end{gather}
 irrespective of whether the PBC rapidity spectrum is gapless or not. 

So far, one discerns that within the framework of the full Lindbladian, the Liouvillian skin effect is closely related to the exceptional topology of the non-Bloch damping matrix, whereas 
it does not depend on
the topology of the Bloch NH Hamiltonian $H_\text{S} (q)$ as illustrated in \Cref{fig:topo}. 
Meanwhile, considering $\gamma_i$'s and $\eta_i$'s belong to two sets of independent parameters in Eqs.~(\ref{eq:eg}), the effective Hamiltonian in \Cref{eq:h_eff} does not play a role here neither. The EPs of $\mathcal{H}_{\text{eff}}$ are shifted to $t_i = \pm \eta_i$ and the topological regime of ${H}_\eff (q)$  for $\eta_1 \eta_2 \ge 0$ lies in Eqs.~(\ref{eq:topo}) with a 
replacement: $\gamma_i \to \eta_i$.

\subsubsection{Modulation of chiral wavefront center}
  \begin{figure}[b]
        \includegraphics[width=0.98\linewidth]{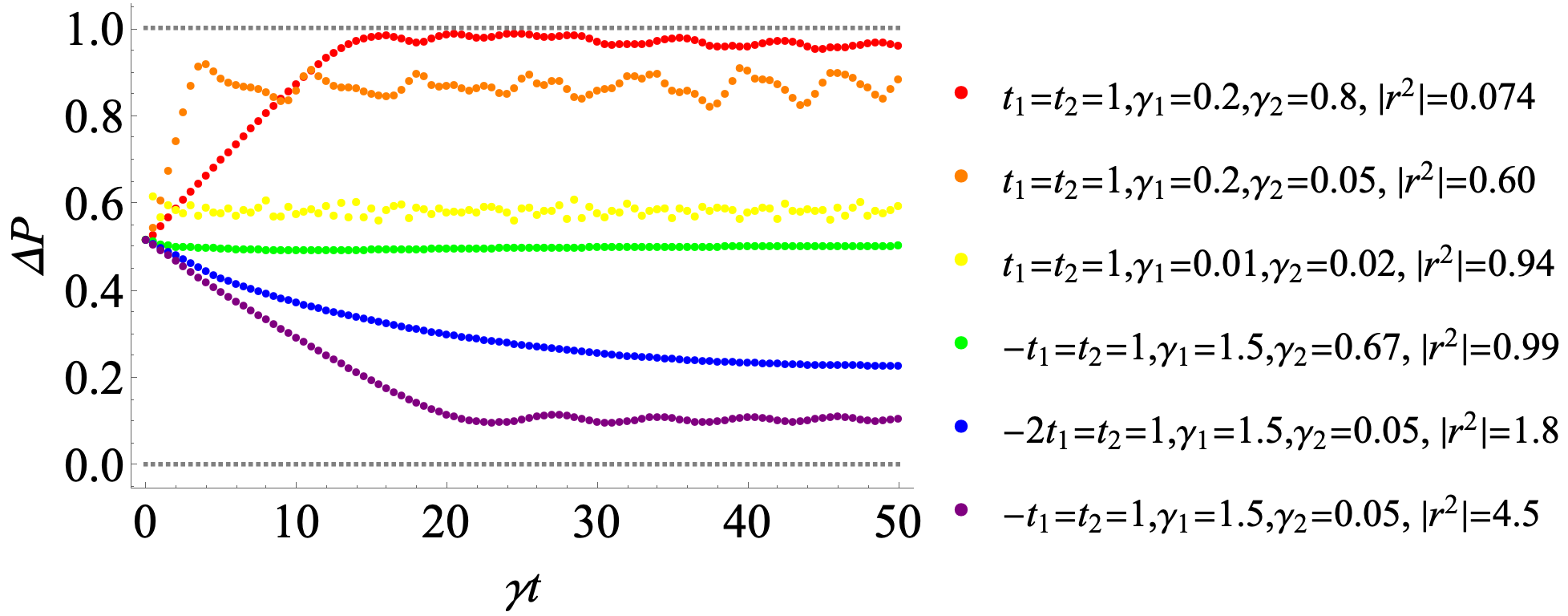}
            \vskip -0.2cm \protect\caption[]
  {Polarization of the chiral damping wavefront under OBC. We choose a system of $N = 14$ unit cells. The initial state of the system is completely filled: $n_j(0) = 1, \forall j$. At $\eta = 0$, the  reference steady state (trivial one)  becomes $n_{j,\text{ss}} =  0.5, \forall j$. When $|r^2| \ll 1$ or $|r^2| \gg 1$, the polarization of $\tilde{n} (t) = n(t) - n_{\text{ss}}$ approaches $1$ or $0$ at longer times. These are cases where the wavefront terminates at the right or left boundary, leaving the chiral damping behavior the most distinguished (see also Fig.~\ref{fig:damping_obc}).}
    \label{fig:pol} 
    \vskip -0.5cm
\end{figure}
\begin{figure}[t]
          \includegraphics[height=3.9cm]{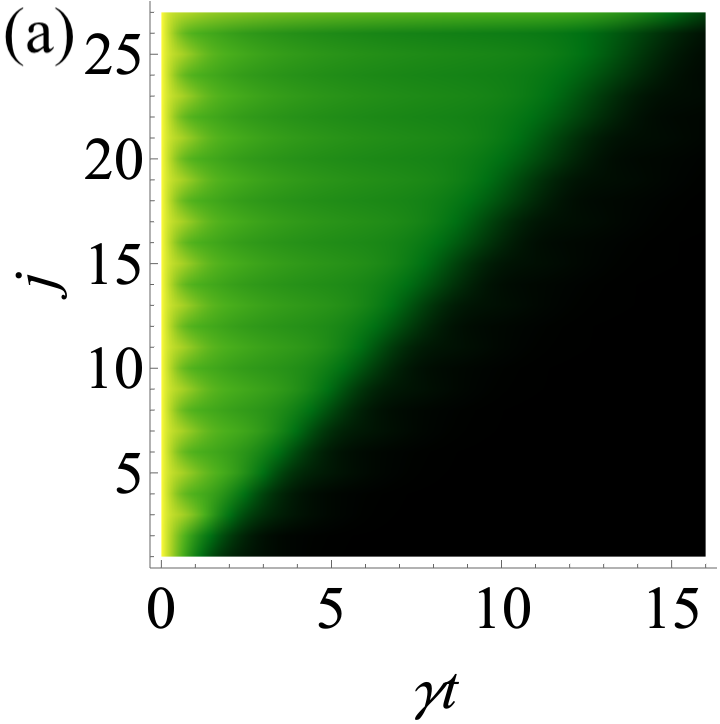} 
          \includegraphics[height=3.9cm]{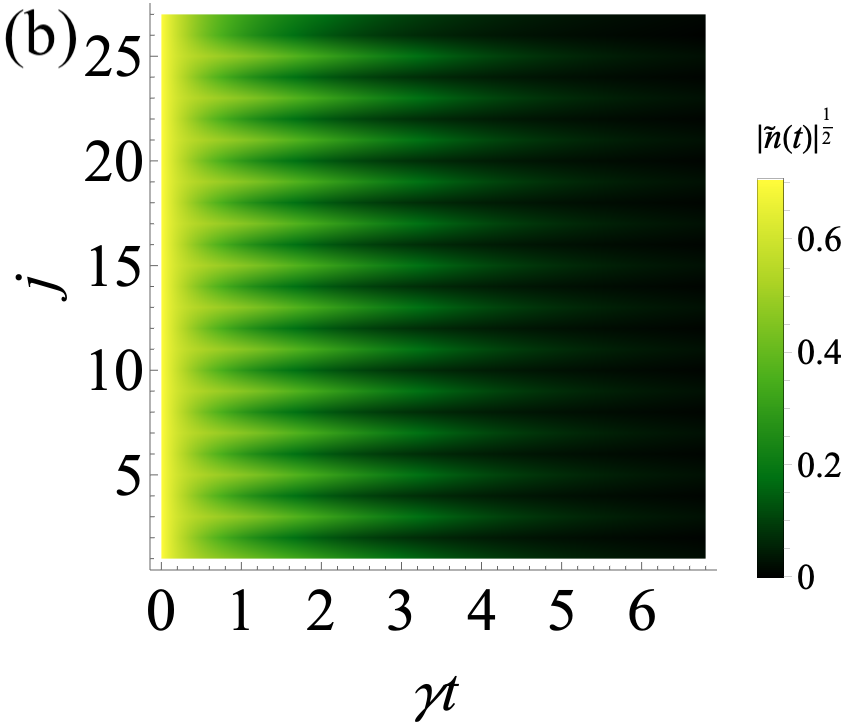}       \\
        \includegraphics[height=3.9cm]{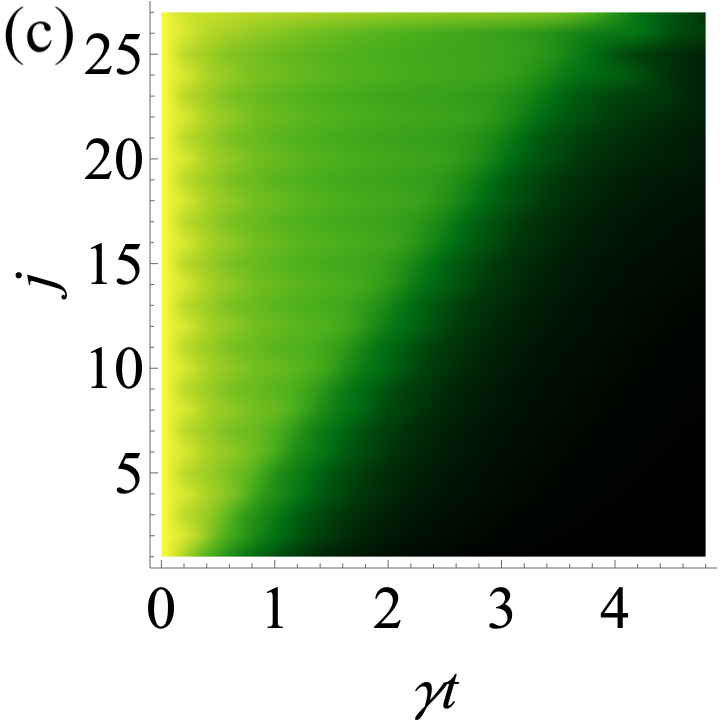} 
          \includegraphics[height=3.9cm]{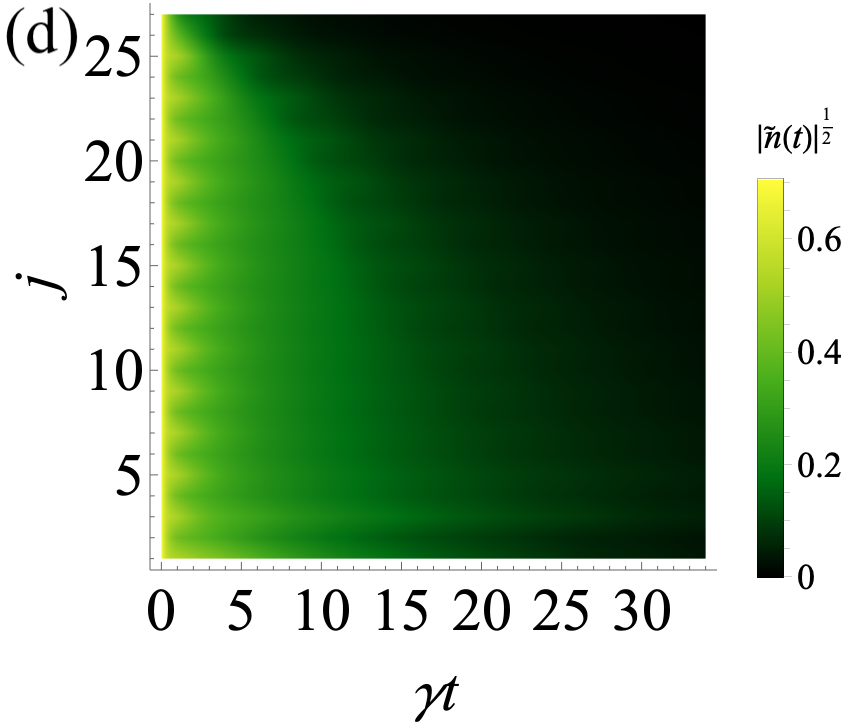}          \\
            \includegraphics[height=3.9cm]{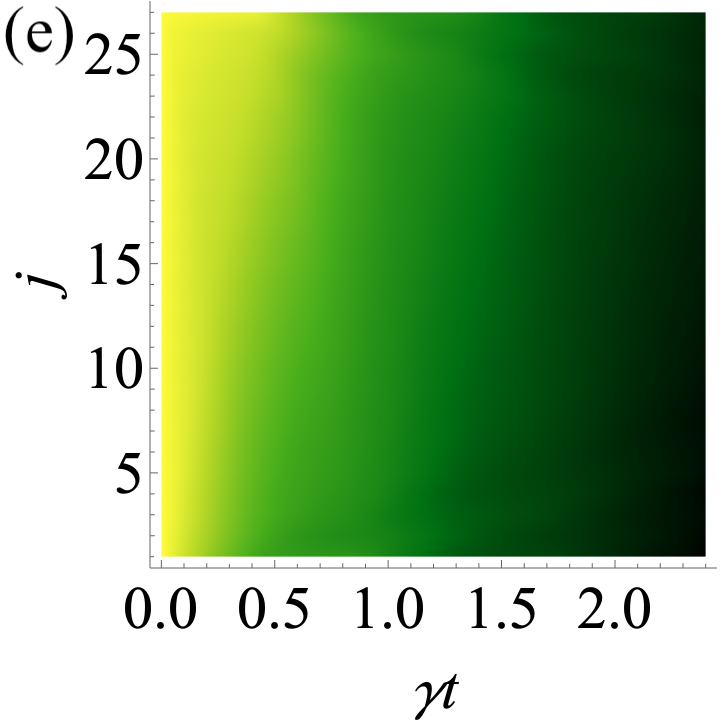} 
          \includegraphics[height=3.9cm]{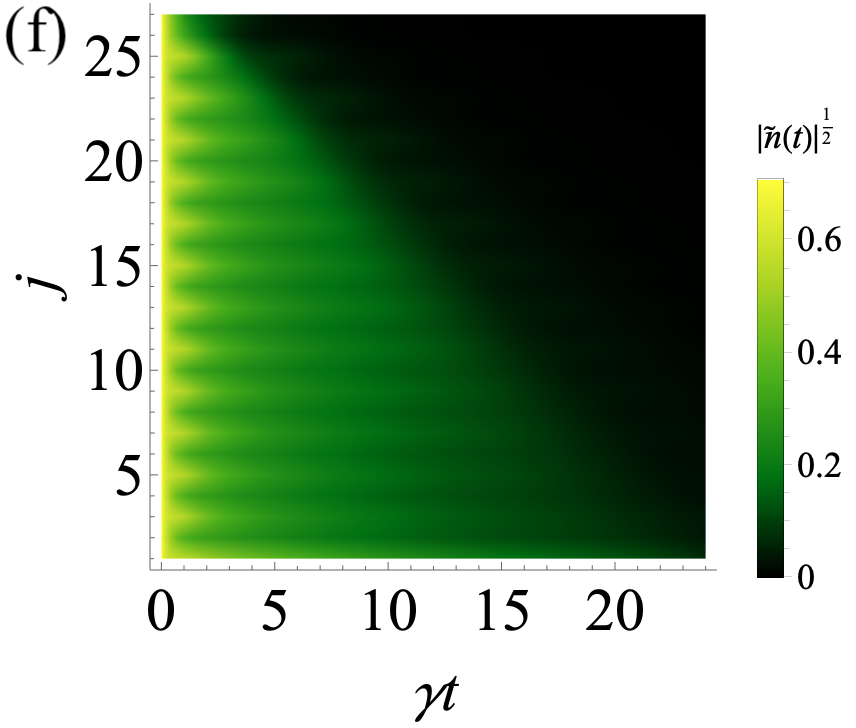}      
            \vskip -0.2cm \protect\caption[]
  {Dynamics of the particle-number operator $\tilde{n}_j(t) = {n}_j(t) - n_{j,\text{ss}}$ under OBC. We keep the same set of Lindblad bond dissipators and the same initial condition as Fig.~\ref{fig:pol}. The left column corresponds to the upper three curves in Fig.~\ref{fig:pol} associated with $|r^2|$: (a) $0.074$, (c) $0.60$, (e) $0.94$, and the right column produces the damping of the lower three curves linked with $|r^2|$: (b) $0.99$, (d) $1.8$, (f) $4.5$. The wavefront is visible in the square root of the relative density.}
    \label{fig:damping_obc} 
    \vskip -0.5cm
\end{figure}

We embark on the characterization of the chiral damping behavior emerging in the dissipative SSH chain when subjected to an open boundary. 
Let us define the polarization of the damping process according to
\begin{gather}
  \Delta P (t) = \frac{\sum_{j=1}^{n_{\text{tot}}} j \cdot \tilde{n}_j(t)}{{n_{\text{tot}}} \cdot \sum_{j=1}^{n_{\text{tot}}} \tilde{n}_j(t)},
  \end{gather}
 where, as depicted in \Cref{fig:ssh_chain}, the length of chain is chosen to be odd ${n_{\text{tot}}} = 2N-1$.
 Without loss of generality,  we restrict ourselves to the case when the gain and loss dissipations are in balance:  $\eta_1 = \eta_2 = 0$. 
Starting from a completely filled chain, the NESS should then be half filling at each site in \Cref{eq:nss}. 
 Plugging the analytical expression for particle number in \Cref{eq:pn}, the tendencies of polarization under different hopping 
 amplitudes and dissipation strengths are shown in \Cref{fig:pol}.  Figure~\ref{fig:damping_obc} further illustrates the motion of the chiral damping wavefront along the chain as time goes by. 

 At the initial time,  $\Delta P = 1/2$, it refers to a uniform particle number distribution
  consistent with our initial condition. Around $|r^2| = 1$, when the Liouvillian skin effect is quite weak, the polarization stays close to $1/2$ and the chiral damping wavefront is absent in Figs.~\ref{fig:damping_obc}\textcolor{blue}{(e)} and \ref{fig:damping_obc}\textcolor{blue}{(b)}. As $|r^2| \ll 1$ ($\gg 1$), the NMMs of the Liouvillian start to pile up towards the right (left) end such that the particles closer to that boundary are granted a longer lifetime,
thus  pushing the polarization to $1$ $(0)$.
From Figs.~\ref{fig:damping_obc}\textcolor{blue}{(a)} and \ref{fig:damping_obc}\textcolor{blue}{(f)}, the chiral wavefront also appears most distinguished in these two limits.
  In the intermediate parameter regime of $|r^2|$, the polarization evolves to a finite value in between $[0, 1/2]$ or $[1/2, 1]$ in accompaniment with a damping wavefront growing obscure, as shown by Figs.~\ref{fig:damping_obc}\textcolor{blue}{(d)} and \ref{fig:damping_obc}\textcolor{blue}{(c)}. Therefore, by tuning the parameter $r^2$, we are able to modulate the polarization or the center of the chiral damping wavefront regardless of the 
topology of the PBC rapidity spectrum before the boundary opens up. The left and right columns of \Cref{fig:damping_obc} correspond to the gap closing ($t_1 = t_2$) and gap opening ($t_1 \ne t_2$) points in the PBC spectrum, respectively.
  
On a side note, in \Cref{fig:pol} even at relatively large times, $\Delta P$ will 
 not converge to $1/2$ for its measurement is targeted on the excess of the particle number over a uniformly half-filled NESS, rather than the real occupation number.

\subsubsection{Boundary sensitivity}
From the chiral damping phenomena, the Liouvillian skin effect takes place upon changing the boundary from PBC to OBC. 
Since the system evolves from a completely filled initial state to a uniform steady state, the damping serves as a global effect.
Nevertheless, we can also resolve the motion of a single particle and study the sensitivity of the open system to boundary conditions on broader grounds. 

It is argued in Ref.~\cite{mao2021} that in the thermodynamical limit, the single-particle Green's function of NH systems is independent of the boundary conditions. 
We show in the following within the Lindblad master equation framework, the argument is indeed true for the evolution of the density matrix at short times. However, as soon as the motion of a particle 
involves the edge, due to the gap closing and the Liouvillian skin effect, there emerge drastic differences in the relaxation process under PBC or OBC. 

Let us start by putting one particle in the middle of the chain: $n_j(0) = \delta_{j, N}$. The initial condition corresponds to a Majorana pairing configuration,
  \begin{gather}
    \tilde{C}(0) =  -\frac{i(\eta - \gamma)}{\gamma} 
     \begin{pmatrix}
    0 &  \mathbb{1}_{n \times n}  \\
    - \mathbb{1}_{n \times n} & 0
     \end{pmatrix} - i   \begin{pmatrix}
    0 &  D  \\
    - D & 0
     \end{pmatrix},
  \end{gather}
where the $n \times n$ matrix $D$ only holds one non-zero element $D_{(N,N)} = 1$. The time evolution of the particle number at the $j$th site
can be described by
\begin{align}
  \tilde{n}_j(t) 
    &= \sum_{m,m'}  e^{-(\beta_m + \beta_{m'})t} \psi_{Lm}^* (j) \psi_{Lm'}^* (j) \notag \\
    &\left[ \psi_{Rm} (N) \psi_{Rm'} (N)+  \left( \frac{ \eta - \gamma}{2\gamma} \right)\sum_{l =1}^{n_{\text{tot}}} \psi_{Rm} (l) \psi_{Rm'} (l) \right]. 
\end{align}
To make a direct comparison with Ref.~\cite{mao2021}, we suppress the gain dissipations on both bonds: $\gamma_1^g = \gamma_2^g = 0$
without violating the solvable constraint in Eqs.~(\ref{eq:sl}): $ {\gamma_1}/{\eta_1} = {\gamma_2}/{\eta_2} = 1$.
Since $\gamma = \eta$, the second term in the particle-number operator vanishes, and the trivial NESS refers to an empty chain indicated by \Cref{eq:ss}.

\begin{figure}[t]
          \includegraphics[height=3.8cm]{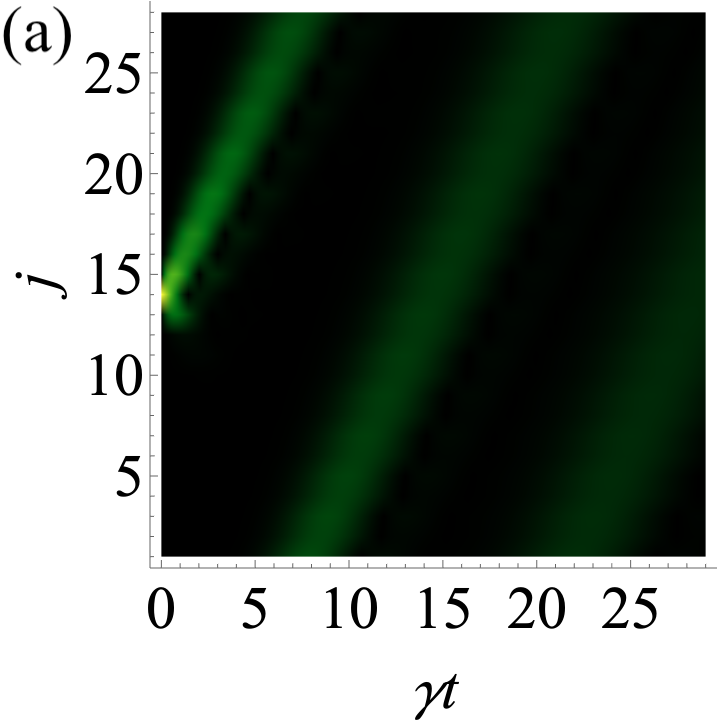} 
          \includegraphics[height=3.8cm]{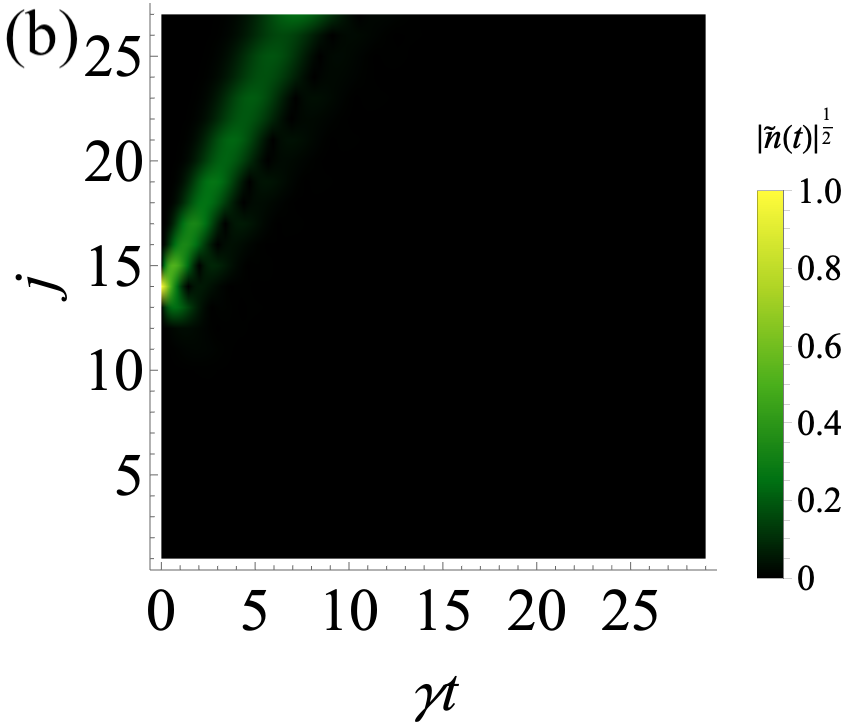}  \\
            \includegraphics[height=3.8cm]{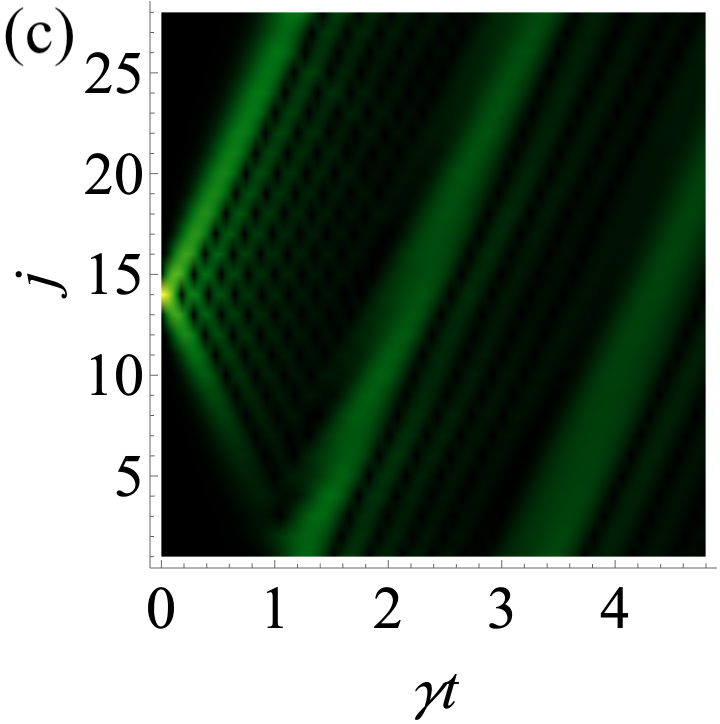} 
          \includegraphics[height=3.8cm]{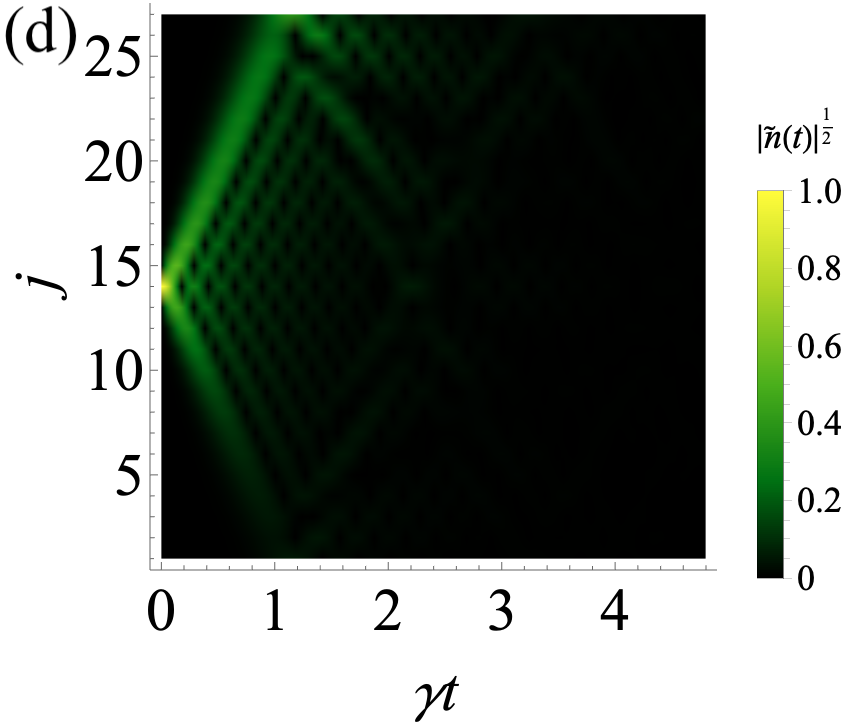}            
            \vskip -0cm \protect\caption[]
  {Boundary effects on the motion of a fermionic particle. At the beginning, we put one particle in the middle of the chain: $n_N(0) = 1$ with $N = 14$ and $n_j(0) = 0$, elsewhere. With gain dissipators suppressed, $\gamma_1^g = \gamma_2^g = 0$, $\gamma  = (|\gamma_1^l| + |\gamma_2^l|)/2 = \eta$. The reference steady state (trivial one) corresponds to an empty chain: $n_{j,\text{ss}} = 0, \forall j$. The top panel shows the case of the strong Liouvillian skin effect at $t_1 = t_2 = 1, \gamma_1 = 0.2, \gamma_2 = 0.8, |r^2| = 0.074$ under boundary conditions: (a) PBC and (b) OBC. The lower panel compares the region where the Liouvillian skin effect becomes weaker: $t_1 = t_2 = 1, \gamma_1 = 0.1, \gamma_2 = 0.05, |r^2| = 0.74$ under boundary conditions: (c) PBC and (d) OBC.}
    \label{fig:boundary} 
    \vskip -0.5cm
\end{figure}

Also, we focus on 
the relaxation process at the same point $t_1 = t_2$ as Ref.~\cite{mao2021}, where the gap of the PBC rapidity spectrum closes for arbitrary dissipation strengths $\gamma_1$ and $\gamma_2$.
The inclusion of $\gamma_2 \ne 0$ offers more tunability on the Liouvillian skin effect parameter $r^2$. 
As shown in \Cref{fig:boundary}, the motion of the particle does not depend on the boundary condition until its trajectory hits the edge. 
Once that long-time evolution is permitted, contrary to Ref.~\cite{mao2021}, the motion differs a lot under PBC or OBC. We observe a persistent current $j_{\text{ss}} = 1/N$ circulating along the periodic chain while 
the current quickly terminates under the influence of an open boundary. 

Meanwhile, we compare the responses of the particle motion to the presence of a strong or weak Liouvillian skin effect. 
When the skin effect dominates [$|r^2| \to 0$ in Figs.~\ref{fig:boundary}\textcolor{blue}{(a)} and \ref{fig:boundary}\textcolor{blue}{(b)}], the wave packet becomes less dispersive and oriented towards the right end under both PBC and OBC. The current also vanishes completely at the open boundary without any reflection. 
Both features are in marked contrast to the case of the weak Liouvillian skin effect [$|r^2| \to 1$ shown by Figs.~\ref{fig:boundary}\textcolor{blue}{(c)} and \ref{fig:boundary}\textcolor{blue}{(d)}].

One concludes that regardless of the 
initial conditions in the Lindblad setup, the Liouvillian skin effect is visible and manifested in the evolution of the density matrix given a probe of open quantum matter at longer times.

\subsection{Lifetime of non-equilibrium particle at the edge}
\label{sec:life}
In the last section, we resolved the relaxation time of dynamical observables under the Liouvillian skin effect.  For a measurement, any local observable of the operator $\hat{Q}$
has reached the steady state if $|\tilde{Q}(t)| = |Q(t) - Q_\text{ss}| \ll ||\hat{Q}|| = (\tr[\hat{Q}^\dagger \hat{Q}])^{1/2}$. Along the lines of Ref.~\cite{ueda2021},
one can thus define the maximal relaxation time $\tau$ of the system as that of the slowest decaying mode: $|c|e^{-\tau \Delta} = e^{-1} \ll 1$ with $\Delta$ the Liouvillian gap.
The presence of the Liouvillian skin effect enhances the amplitude of this mode, showing an exponential localization tendency near the boundary: $|c| \sim e^{\mathcal{O}(L/\xi)}$. Here, $\xi$ denotes 
the correlation length of the system. It leads to
  \begin{align} 
    \tau \sim \frac{1}{\Delta} + \frac{1}{\xi} \cdot \frac{L}{\Delta}. \label{eq:tau_g}
  \end{align}

In our dissipative SSH model, we look at the Liouvillian skin effect region $|r^2| < 1$ and take the dynamical observable as the occupation of a particle residing at the right boundary. 
More importantly, we make fewer assumptions than Ref.~\cite{ueda2021} by defining the lifetime of the non-equilibrium particle according to
  \begin{gather}
     |\tilde{n}_{2N-1} (\tau)| = e^{-l}|\tilde{n}_{2N-1} (0)|, \label{eq:taud}
  \end{gather}
  where $l$ is taken as a positive integer. The precision of the lifetime can be improved by increasing the value of $l$.
 The definition in \Cref{eq:taud} has the advantage of including the contributions from all decaying modes, staying closer to a real measurement. 
 
For simplicity, we start with a completely filled chain and balanced gain and loss dissipators $\eta_1 = \eta_2 = 0$. Then,
 $|\tilde{n}_{2N-1} (0)| = 1 - 1/2 = 1/2$. From the asymptotic scaling of the particle number evolution in \Cref{eq:asy}, we establish that
      \begin{gather}
     \tau_{2N-1} \sim  \frac{1}{\Delta_{\text{eff}}} + \ln(r^{-2}) \cdot \frac{N}{\Delta_{\text{eff}}}. \label{eq:gap_eff}
    \end{gather}
The correlation length of the system in \Cref{eq:tau_g} thus satisfies
    \begin{gather} 
      \xi (r^2) \simeq \frac{2}{|\ln (|r^{2}|)|}, \label{eq:corr}
    \end{gather}  
where the factor of $2$ comes from the identification $L = 2N-1$. Consistent with Ref.~\cite{ueda2021}, we find the lifetime of the particle at the edge grows linearly with the system size
without a closing of the effective Liouvillian gap. 

Next, based on our exact solutions, we verify numerically the above relations by varying the system size and the skin effect parameter $r^2$. It is also important to explore two limits, the weak and strong dissipations where
the formation of the effective Liouvillian gap differs. 
 
\begin{figure}[t]
        \includegraphics[width=0.8\linewidth]{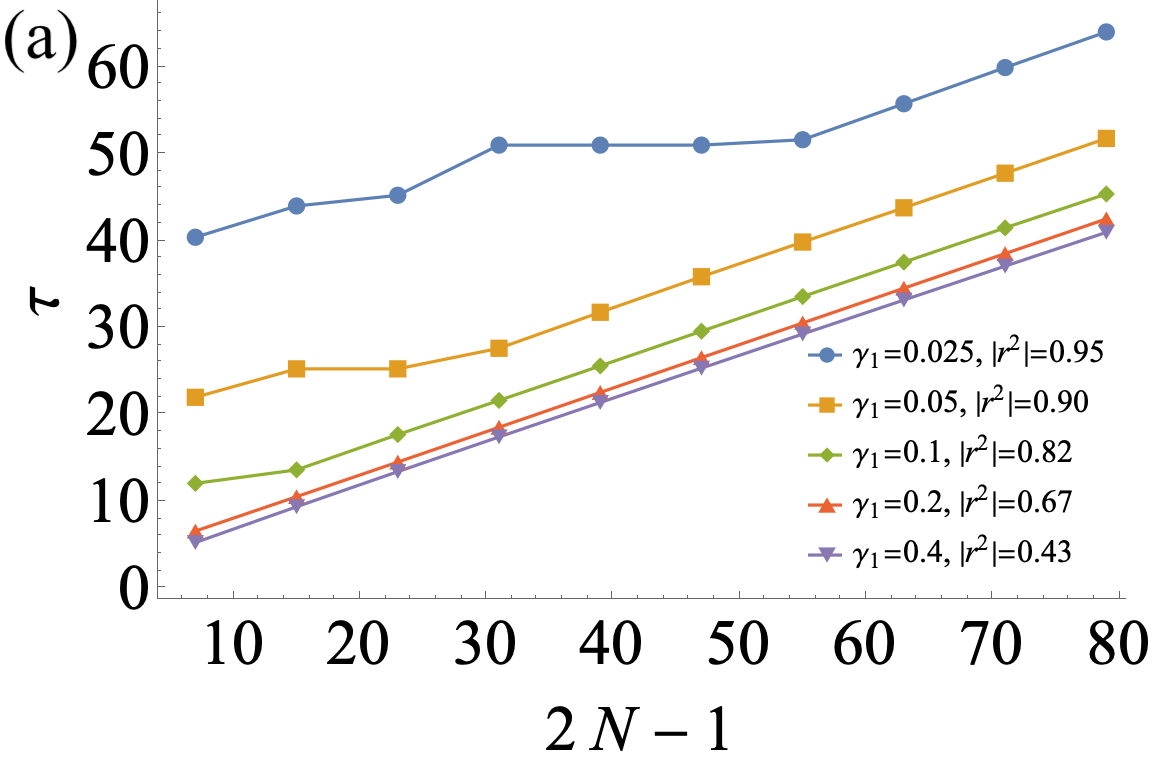} \\
           \includegraphics[width=0.8\linewidth]{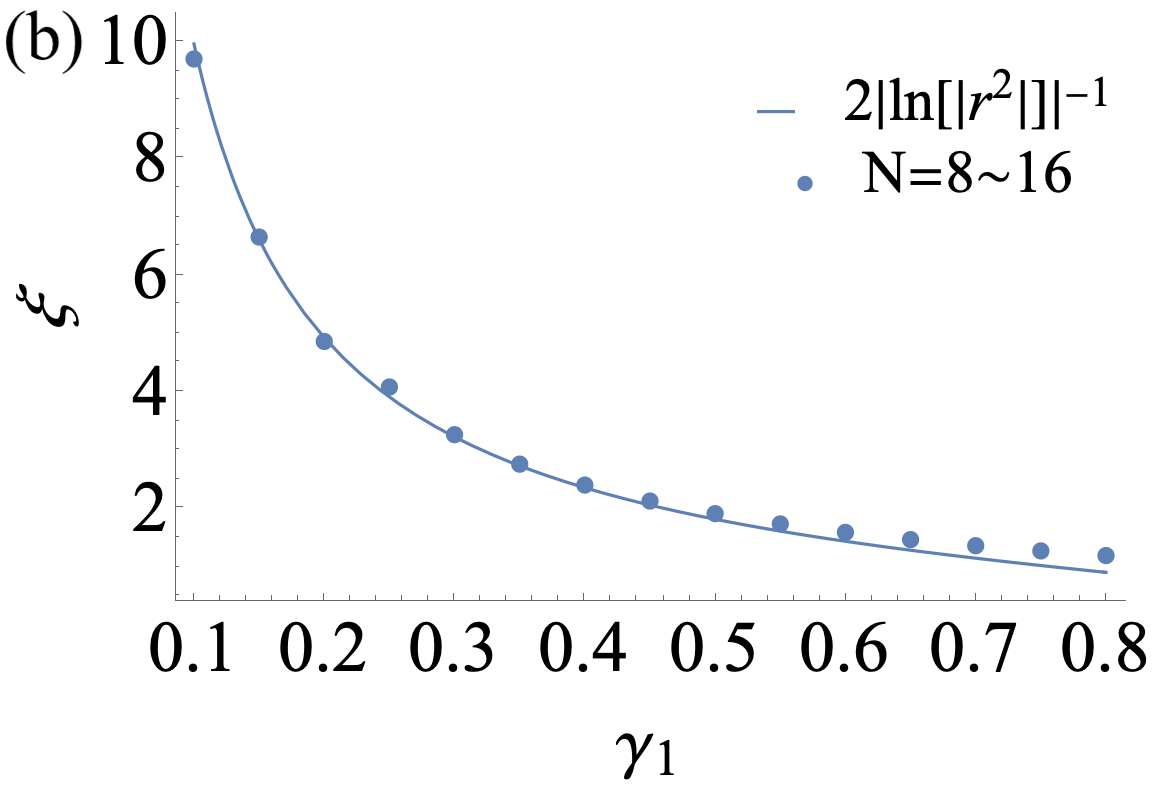}
                                  \vskip -0.2cm \protect\caption[]
  {(a) Lifetime of the out-of-equilibrium mode on the right boundary as a function of the system length $L = 2N-1$. We take $ |\tilde{n}_{2N-1} (\tau)| = e^{-2}|\tilde{n}_{2N-1} (0)| = 0.068$.
  (b) Correlation length as a function of the bond dissipation strength $\gamma_1$.  To extract the correlation length $\xi$ in the linear relation in \Cref{eq:tau_g}, a finite system size in the range of $8$ to $16$ unit cells is employed. Meanwhile, $\tau$ is chosen at the value $|\tilde{n}_{2N-1} (\tau)| = e^{-3}|\tilde{n}_{2N-1} (0)| = 0.025$. For both plots, the initial conditions are kept the same as \Cref{fig:pol}: the filling of each site evolves from $1$ to $1/2$. The remaining parameters are fixed accordingly: $t_1 = t_2 = 1$ and $\gamma_2 = 0$.}
    \label{fig:tau} 
    \vskip -0.5cm
\end{figure}

 At weak dissipations $\gamma_1 < |t_1|$ and $\gamma_2 < |t_2|$, all bulk and boundary modes share the same Liouvillian gap [see Eqs.~(\ref{eq:rap}) and \Cref{fig:gap}]. Hence, $\Delta_\text{eff} = \Delta^{\text{OBC}} = 2 (\gamma_1 + \gamma_2)$. Figure~\ref{fig:tau}\textcolor{blue}{(a)}
plots the numerical scaling of the particle lifetime at the edge $\tau_{2N-1} = \tau$ under different system sizes at $\gamma_2 = 0$. The linear dependence on $N$ in \Cref{eq:gap_eff}
 becomes more visible as the length of the chain increases. With stronger Liouvillian skin effect $|r^2| \ll 1$, the linear scaling holds true for a relatively small system size $N \sim 8 $.
The system correlation length can be extracted from the slope and 
\Cref{fig:tau}\textcolor{blue}{(b)} confirms our analytical prediction in terms of the skin effect parameter in \Cref{eq:corr}. Approaching the EP, $\gamma_1 \to t_1 = 1$,
 the small discrepancy results from the fact that the strong polarization of the bulk eigenmodes challenges the numerical precision of the damping matrix decomposition in Eqs.~(\ref{eq:decom}).

\begin{figure}[b]
               \includegraphics[width=0.8\linewidth]{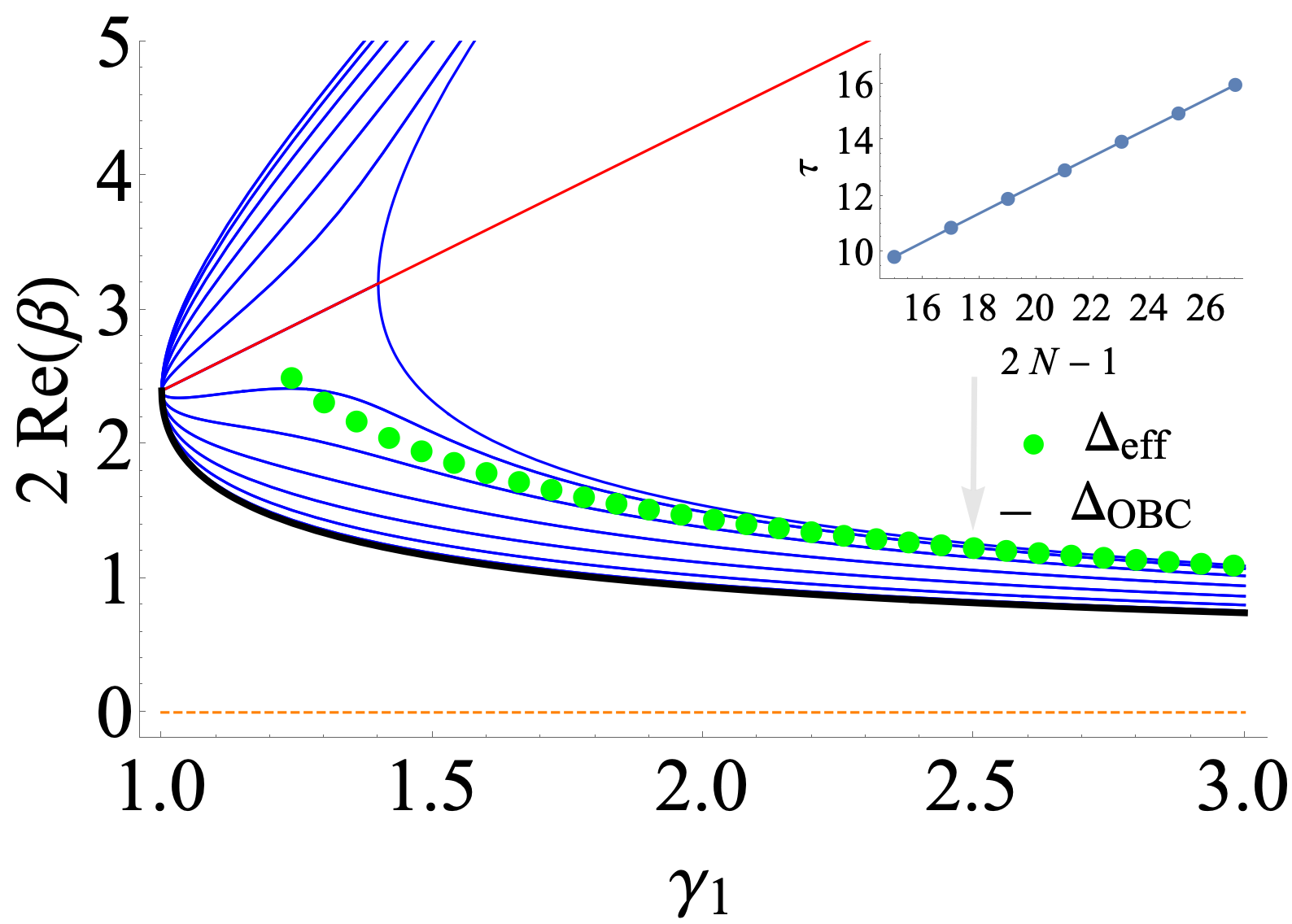}
                       \vskip -0.2cm \protect\caption[]
  {Effective Liouvillian gap (green dots) as a function of $\gamma_1$ beyond weak dissipations: $\gamma_1 > |t_1|$, $\gamma_2 < |t_2|$. We take $t_1 = t_2 = 1, \gamma_2 = 0.2$, and $\tau \to |\tilde{n}_{2N-1} (\tau)| = e^{-5}|\tilde{n}_{2N-1} (0)| = 0.0034$. The fitting of the lifetime $\tau$ in \Cref{eq:gap_eff} is performed with a finite system size $N=8\sim14$. The inset shows its linear dependence at $\gamma_1 = 2.5, |r^2| = 0.286$. The background corresponds to the real part of the rapidity spectrum generated by the bulk modes (blue curves) and boundary modes (red curve) in a system of $N = 14$ unit cells. $\Delta_{\text{OBC}}$ (black curve) provides a lower bound for the effective Liouvillian gap.}
    \label{fig:gap_eff} 
    \vskip -0.5cm
\end{figure}

At strong dissipations, on the other hand, from \Cref{fig:gap} different bulk and boundary modes form distinct real rapidity spectra. 
As a consequence,  the effective Liouvillian gap incorporating effects of all decaying modes differs from $\Delta^{\text{OBC}}$ in \Cref{eq:gapo}. By definition,
  \begin{align}
      \Delta_{\text{eff}} &= \Delta^{\text{OBC}}, \quad \gamma_1 < |t_1|, \ \gamma_2 < |t_2|; \notag \\
      \Delta_{\text{eff}} &> \Delta^{\text{OBC}}, \quad \text{ otherwise }.
  \end{align}
Still, we can estimate its value from the established linear relation for the particle lifetime
 in \Cref{eq:gap_eff} and at the same time, fix the correlation length by the skin effect parameter $r^2$ in \Cref{eq:corr}. In \Cref{fig:gap_eff}, we show the formation of the effective Liouvillian gap beyond weak dissipations: $\gamma_1 > |t_1|$, $\gamma_2 < |t_2|$. Indeed, $\Delta^{\text{OBC}}$ provides a lower bound for the effective Liouvillian gap.

On the contrary, when $\gamma \ne \eta$, the lifetime of a particle at the right boundary under the truncated $\mathcal{H}_\eff$ would scale as
     \begin{gather}
      \left. \tau_{2N-1}\right|_{\mathcal{H}_\eff} \sim \frac{1}{\Delta} \simeq \frac{1}{2(\gamma-\eta)} \cdot \frac{1}{N}
    \end{gather}
   in the large $N$ limit. Without the quantum jumps, the lifetime decreases with the system size. As expected, the effective Hamiltonian becomes problematic in describing the long-time dynamics. 

\section{Discussion} 
We have studied the Liouvillian skin effect in a bond-dissipative SSH model, of which the rapidity spectrum and NMMs are exactly solvable. This has illuminated the relation between the NH skin effect, two different effective Hamiltonians, $H_\text{eff}$, $H_\text{S}$, and the full quantum master equation description. A number of dynamical phenomena, such as diverging relaxation times, inherited in the quantum setup originate from the anomalous boundary sensitivity of the NH Hamiltonians and have been investigated in detail. This paper also provides a solid platform to resolve the entanglement spectrum and identify transitions between different NH topological phases \cite{sayyad2021} whose experimental realization is feasible, e.g., in ultra-cold atoms with a momentum lattice \cite{he2021}.

Dynamical probes of the quantum Fisher information \cite{clerk2020} and the application as NH topological sensors \cite{QNTOS} provide intriguing questions for future investigation. Meanwhile, the current model reveals the Liouvillian skin effect in the relaxation process
while posing the question on the search for steady states that inherit the exceptional topology of the NH damping matrix. It is found that using monitored quantum circuits, periodic measurement allows access to biorthogonal steady state observables \cite{fleckenstein2022nonhermitian}.  Future works could also be centered around the interaction effects on generic Lindbladians. For instance, 
the dynamical mean-field theory can be implemented to uncover the interplay of dissipation and environmental fluctuations in the open quantum matter that consists of interacting fermions \cite{altland2021} or bosons \cite{clerk2021dyn}. 
A mixture of particle species is a good path to the discovery of further variations of the Liouvillian skin effect. 
From a broader theoretical point of view,  effective field theories with bosonization prove powerful in studying dissipative Luttinger liquids \cite{yamamoto2021}. In presence of weak symmetries, one may tackle even strongly interacting Liouvillians \cite{clerk2022}.
The inclusion of disorder and chaos may be possible provided the solvable limit in a disorder-averaging SYK Lindbladian \cite{prosen2021}.

\section*{Acknowledgements}

  This work has benefited from useful discussions with Elisabet Edvardsson, Yuchi He, Lukas K{\"o}nig, Daniel Varjas, and Kang Yang.
  The authors were supported by the Swedish Research Council (VR) and the Wallenberg Academy Fellows 
  program as well as the project Dynamic Quantum Matter of the Knut and Alice Wallenberg Foundation. Q.D.J was also supported by Pujiang Talent Program No.~21PJ1405400. 

\renewcommand{\theequation}{A\arabic{equation}}

\appendix

 \section*{\uppercase{Appendix: Exact solution of the generalized non-Hermitian SSH chain}}
\label{app:ssh}
In the Appendix, we present a detailed derivation on the exact solution of the NH SSH chain in \Cref{eq:h_g}.
The generic results for the simplified model \cite{flore2018, elisabet2020} are extended here to allow two asymmetric hopping terms.

\subsection{Periodic boundary condition}
First, we address the case of a periodic boundary: $\varphi_{N+1, A(B)} = \varphi_{1,A(B)}$. For convenience, we set the lattice spacing to unity. Through the Fourier transform $\varphi_{{j,\alpha}} = \frac{1}{\sqrt{N}} \sum_q e^{iqj} \varphi_{\alpha}(q)$, with $q = 2\pi m'/N$, $m' = -N/2, -N/2+1, \dots, 0, \dots, N/2-1$, the anti-commutation relations for fermionic operators are respected: $\{ \varphi_{\alpha} (q),\varphi_{\alpha'}^\dagger (q') \} = \delta_{q,q'} \delta_{\alpha, \alpha'}$. 
 In the new basis $\u{\varphi}(q) = (\varphi_A(q), \varphi_B(q))^T$, one rewrites the Hamiltonian into 
  \begin{gather}
    \mathcal{H}_{\text{S}} = \sum_q \u{\varphi}^\dagger(q) \cdot H_{\text{S}} (q) \u{\varphi}(q).
  \end{gather}
In terms of Pauli matrices, the matrix elements read
 \begin{align}
        H_\text{S} (q) &=  \vec{h}(q)\cdot \vec{\sigma}, \notag \\
        h_x(q) &= t_1+t_2\cos (q) + i \gamma_2 \sin(q), \notag \\
         h_y(q) &= i\gamma_1 - i\gamma_2 \cos(q) + t_2 \sin(q), \notag \\
         h_z(q) &= 0. 
 \label{eq:sshq}
 \end{align}
From the eigenvalue Eqs.~(\ref{eq:ssh_es}), we solve the energy spectrum, $E_m^{\text{PBC}} = E^{\text{PBC}}_{\nu=\pm} (q)$,
  \begin{align}
      E^{\text{PBC}}_{\pm} (q)
     &= \pm \left[ t_1^2 + t_2^2 - (\gamma_1^2 + \gamma_2^2)+ 2(t_1t_2 + \gamma_1 \gamma_2) \cos q  \right. \notag \\
     &\phantom{=} \left.  + 2i(t_1 \gamma_2 + t_2 \gamma_1) \sin q \right]^{\frac{1}{2}},  \label{eq:ek}
  \end{align}
 together with the right and left eigenvectors:
 \begin{align}
 \tilde{\u{\psi}}_{R, \nu} (q, \gamma_i) &= \frac{1}{\sqrt{2}} 
    \begin{pmatrix}
      (t_1 + \gamma_1 + (t_2-\gamma_2) e^{-iq})/E_{\nu} (q) \\
      1
    \end{pmatrix}, \notag \\
      \tilde{\u{\psi}}_{L, \nu} (q,\gamma_i) &= \tilde{\u{\psi}}_{R, \nu} (q,-\gamma_i).  \label{eq:re_pbc}
 \end{align}
 The second mutual relation can be proved by taking into account $H_{\text{S}}^\dagger(q, \gamma_i) = H_{\text{S}}(q, -\gamma_i)$ and $E_{\nu}^* (q, \gamma_i) = E_{\nu} (q, -\gamma_i) |_\text{PBC}$.
The normalization of eigenvectors is chosen according to the biorthogonality:
  \begin{gather}
     \tilde{\u{\psi}}_{L, \nu}^* (q) \cdot    \tilde{\u{\psi}}_{R, \nu'} (q) = \delta_{\nu, \nu'}.
  \end{gather}
The Hamiltonian now shares a diagonalized structure:
   \begin{gather}
     H_{\text{S}} = \sum_{(\nu, q)} \u{\varphi}^\dagger(q) \cdot \left[E_\nu(q) \tilde{\u{\psi}}_{R, \nu}(q) \cdot \tilde{\u{\psi}}^*_{L, \nu}(q) \right] \u{\varphi}(q).
   \end{gather}

For the evaluation of the single-particle correlation acting on the real space in \Cref{eq:spc}, the components $\tilde{\psi}_{R,(\nu,q)} (j,\alpha)$, $\tilde{\psi}^*_{L,(\nu,q)} (j,\alpha)$ are obtained by the Fourier transform:
     \begin{align}
       \begin{pmatrix}
         \tilde{\psi}_{R, (\nu, q)} (j,A) \\
           \tilde{\psi}_{R, (\nu, q)} (j,B) 
        \end{pmatrix}
         &= \frac{e^{iqj}}{\sqrt{N}}   \ 
               \tilde{\u{\psi}}_{R, \nu} (q) , \notag \\
        \begin{pmatrix}
         \tilde{\psi}^*_{L, (\nu, q)} (j,A) \\
           \tilde{\psi}^*_{L, (\nu, q)} (j,B) 
        \end{pmatrix}
          &= \frac{e^{-iqj} }{\sqrt{N}} \ 
               \tilde{\u{\psi}}^*_{L, \nu} (q),
     \end{align}
  where we have used $\langle 0 | \varphi_{\alpha'}(q')\varphi^\dagger_{\alpha}(q) |0\rangle = \delta_{\alpha, \alpha'}\delta_{q,q'}$.

\subsection{Open boundary condition}
Under OBC, we break the $N$th unit cell by taking away the last $B$ site. 
The NH SSH Hamiltonian is then expressed in an explicit $n \times n$ matrix form: $\mathcal{H}_{\text{S}} = \u{\varphi}^\dagger \cdot H_{\text{S}}  \u{\varphi}$ with $\varphi = (\varphi_{(1,A)}, \varphi_{(1,B)}, \dots,
\varphi_{(N,A)})^T$
and
 \begin{align}
     & \ \ H_{\text{S}} = \notag  \\
      &    \begin{pmatrix}
           0 & t_1 + \gamma_1 & & & & &  \\
           t_1 - \gamma_1 & 0 & t_2 + \gamma_2 & & & & & & \\
            & t_2 - \gamma_2 & 0 & & & & & \\
                    & & & & \ddots & & \\     
           & & & &  & 0 & t_2 + \gamma_2 \\
         & & & & &  t_2 - \gamma_2 & 0   
         \end{pmatrix}.        \label{eq:ssh}
     \end{align}
For  $n = 2N-1$ an odd number of sites,  $H_\text{S}$ is shown to be 
exactly solvable at $\gamma_1 \ne 0$, $\gamma_2 = 0$ \cite{flore2018, elisabet2020}. Below, we generalize the exact OBC solution to the region: $\gamma_1 \ne 0$, $\gamma_2 \ne 0$.
The spectrum recovers the known solution by setting $\gamma_2 \to 0$ and retains the main features in the new limit.

On one hand, there exist two zero-energy boundary states that are fully suppressed on the sublattice $B$,
    \begin{align}
     E_0 = 0: \quad \tilde{\u{\psi}}_{R0}  = \mathcal{N}_R \begin{pmatrix} r_R \\ 0 \\ r_R^2 \\ 0 \\ \dots \\ 0 \\ r_R^N \end{pmatrix}, \ 
      \tilde{\u{\psi}}_{L0}  = \mathcal{N}_L \begin{pmatrix} r_L \\ 0 \\ r_L^2 \\ 0 \\ \dots \\ 0 \\ r_L^N \end{pmatrix}, \label{eq:e0}
    \end{align}
with associated parameters  $r_R = -(t_1 - \gamma_1)/(t_2 + \gamma_2)$, $r_L = -(t_1 + \gamma_1)/(t_2 - \gamma_2)$. The two eigenstates are biorthogonal to each other,
          \begin{gather}
            \tilde{\u{\psi}}_{L0}^* \cdot \tilde{\u{\psi}}_{R0}  = 1,
          \end{gather} 
  which brings about a constraint on the normalization factors:
     $\mathcal{N}_L^* \mathcal{N}_R = {(1-r_L^*r_R)}/[{r_L^* r_R(1-(r_L^*r_R)^N)]}$.
It is easy to discern that, depending on the absolute values of $r_L$ and $r_R$, the left and right boundary modes can be localized at different ends: $\sgn [\ln (|r_L|)] \ne \sgn [\ln (|r_R|)]$. 
We find the parameters satisfying this condition all stay in the topological regime in Eqs.~(\ref{eq:topo}) where the spectral winding number is non-trivial. 

On the other hand, to derive the bulk spectrum, we present an intuitive approach by analogy to Refs.~\cite{yao2018, elisabet2020}. Our first step is the identification of the gap-closing points when the boundary is opened.
One way is to find a transformation matrix $R$ such that the NH SSH Hamiltonian is mapped to its Hermitian counterpart $\bar{H}_{\text{S}}$: 
  \begin{gather}
    \bar{H}_{\text{S}} = R^{-1} H_{\text{S}} R. \label{eq:hssh}
  \end{gather}
A proper construction leads to $R = R_1 R_2$, where $R_1 = \text{diag} \{1, r_1, r_1, r_1^2, r_1^2, \dots, r_1^{N-1}, r_1^{N-1}\}$, $R_2 = \text{diag} \{1, 1, r_2, r_2, r_2^2, r_2^2, \dots,r_2^{N-2}, r_2^{N-1}\}$,
and $r_{1} = \sqrt{(t_1 - \gamma_1)/(t_1 + \gamma_1)}$, $r_{2} = \sqrt{(t_2 - \gamma_2)/(t_2 + \gamma_2)}$.
The  Hermitian SSH chain is embedded with the anisotropic hopping strengths
  \begin{gather}
    \bar{t}_1 = \sqrt{t_1^2 - \gamma_1^2}, \quad   \bar{t}_2 = \sqrt{t_2^2 - \gamma_2^2},
  \end{gather}
with the gapless phase transition occurring at $|\bar{t}_1| = |\bar{t}_2|$. $H_\text{S}$ inherits from the transformation these gap closing points:
  \begin{gather}
    |t_1^2 - \gamma_1^2| = |t_2^2 - \gamma_2^2|. \label{eq:gc}
  \end{gather}
An alternative method of reproducing the gapless points in the OBC spectrum is to study the biorthogonal polarization of the NH system \cite{flore2018, elisabet2020}, which changes its integer value at $|r_L^*r_R| = 1$,
in consistency with our result in \Cref{eq:gc}.

Under the transformation in \Cref{eq:hssh}, one can also link the bulk eigenstates of $H_\text{S}$ to the ones of $\bar{H}_\text{S}$ by $\tilde{\u{\psi}}_R = R \bar{\u{\psi}}_R$ and  $\tilde{\u{\psi}}_L =  \bar{\u{\psi}}_L R^{-1}$, which indicates the piling up of the right (left) states at one end with an exponential localization factor $r^j$ ($r^{-j}$), where $r = r_1r_2 = \sqrt{r_R/r_L}$. It is equivalent to a change of momentum in the Bloch phase factor $e^{iqj}$ from $q$ to $q - i\ln (r)$, thus allowing us to build the OBC bulk spectrum from the PBC one
\cite{yao2018, flore2019n}.
More precisely, at a fixed momentum $q$, like the PBC spectrum the bulk energies always come in pairs ($\nu = \pm$):  $E_{m \ne 0} = E_{\nu}^{\text{OBC}} (q)$.
The two spectra are related by 
  \begin{gather}
     E_{\pm}^{\text{OBC}} (q) = E_{\pm}^{\text{PBC}} (q - i\ln (r))  \notag \\
     = \pm \sqrt{t_1^2 + t_2^2 - (\gamma_1^2 + \gamma_2^2)  +2 \sqrt{(t_1^2 - \gamma_1^2)(t_2^2 - \gamma_2^2)}\cos(q)}.  \label{eq:eobc}
  \end{gather}
  Given an even number of sites $n = 2N$,  the OBC spectrum obtained in this way applies only to the large $N$ limit \cite{yao2018}.
  For an odd number of sites $n = 2N-1$, with additional mirror symmetry \cite{flore2019e, elisabet2020}, it can be shown that the bulk spectrum in \Cref{eq:eobc} becomes exact for any finite $N$.
  
  Let us verify by building the exact left and right eigenvectors. We choose $q = \pi m'/N$, $m' = 1, \dots, N-1$, so that
in total the index $m \in \{0, (\pm, q)\}$ reproduces a complete set of  $2N-1$ bands. A trial function for the right eigenstates can be written as
    \begin{align}
     \tilde{\u{\psi}}_{R\nu} (q, \gamma_i)  
               = \frac{1}{\sqrt{2N}} 
                   \begin{pmatrix}
                      \tilde{\u{\psi}}_{R\nu} (q,1)   \\
                       \tilde{\u{\psi}}_{R\nu} (q,2)  \\
                          \cdots \\
                       \tilde{\u{\psi}}_{R\nu} (q,N)  
                  \end{pmatrix}, \label{eq:re_obc}
     \end{align}
where the component in the $j$th unit cell should be linked with the PBC eigenvector in \Cref{eq:re_pbc} by the same momentum shift:
  \begin{gather}
     \tilde{\u{\psi}}_{R\nu} (q,j)  \sim  r^j e^{iqj} \tilde{\u{\psi}}_{R\nu}^{\text{PBC}} (q-i\ln(r)). \label{eq:bl0}
  \end{gather}
On top of that, mirror symmetry in the spectrum $E_{\pm}^{\text{OBC}} (q) = E_{\pm}^{\text{OBC}} (-q)$ enforces a superposition of two wave functions with opposite momenta.
The relative amplitude of this superposition is determined by the boundary condition of a broken last unit cell: $| \tilde{\u{\psi}}_{R\nu} (q, 0)\rangle_B = | \tilde{\u{\psi}}_{R\nu} (q, N)\rangle_B = 0$. 
One arrives at
     \begin{align}
          \tilde{\u{\psi}}_{R\nu} (q,j)  &=  r^j  \left( e^{iqj} \u{u}_{R\nu} (q)  - e^{-iqj} \u{u}_{R\nu} (-q)  \right),  \label{eq:bl1}\\
          \u{u}_{R\nu} (q)  &= \frac{i}{\sqrt{2}} \begin{pmatrix} (t_1 + \gamma_1 +(t_2-\gamma_2)r^{-1} e^{-iq})/E_{\nu} (q) \\ 1 \end{pmatrix}. \notag
         \end{align}
Taking into account $H_{\text{S}}^\dagger (\gamma) = H_{\text{S}}^T (\gamma) = H_{\text{S}}(-\gamma)$ and $E_{\nu} (q, \gamma_i) = E_{\nu} (q, -\gamma_i) |_\text{OBC}$,  
 the left and right eigenvectors are related  in a way slightly different from PBC:
    \begin{gather}
     \tilde{\u{\psi}}_{L\nu} (q, \gamma_i)  =  \tilde{\u{\psi}}_{R\nu}^* (q, -\gamma_i). \label{eq:le_obc}
    \end{gather}
It is straightforward to check that the energies in \Cref{eq:eobc} and eigenstates in Eqs.~(\ref{eq:re_obc})$-$(\ref{eq:le_obc}) satisfy the eigenvalue Eqs.~(\ref{eq:ssh_es}) for any finite system size $N$. 
The biorthogonal normalization conditions among  the bulk and the boundary states are respected as well:
    \begin{gather}
       \tilde{\u{\psi}}^*_{L\nu} (q) \cdot \tilde{\u{\psi}}_{R\nu'} (q') = \delta_{\nu, \nu'} \delta_{q, q'}, \notag \\
         \tilde{\u{\psi}}^*_{L\nu} (q) \cdot \tilde{\u{\psi}}_{R0} = \tilde{\u{\psi}}^*_{L0} \cdot \tilde{\u{\psi}}_{R\nu} (q) = 0.
    \end{gather}
   

\bibliography{sample}

\begin{thebibliography}{87}%
\makeatletter
\providecommand \@ifxundefined [1]{%
 \@ifx{#1\undefined}
}%
\providecommand \@ifnum [1]{%
 \ifnum #1\expandafter \@firstoftwo
 \else \expandafter \@secondoftwo
 \fi
}%
\providecommand \@ifx [1]{%
 \ifx #1\expandafter \@firstoftwo
 \else \expandafter \@secondoftwo
 \fi
}%
\providecommand \natexlab [1]{#1}%
\providecommand \enquote  [1]{``#1''}%
\providecommand \bibnamefont  [1]{#1}%
\providecommand \bibfnamefont [1]{#1}%
\providecommand \citenamefont [1]{#1}%
\providecommand \href@noop [0]{\@secondoftwo}%
\providecommand \href [0]{\begingroup \@sanitize@url \@href}%
\providecommand \@href[1]{\@@startlink{#1}\@@href}%
\providecommand \@@href[1]{\endgroup#1\@@endlink}%
\providecommand \@sanitize@url [0]{\catcode `\\12\catcode `\$12\catcode
  `\&12\catcode `\#12\catcode `\^12\catcode `\_12\catcode `\%12\relax}%
\providecommand \@@startlink[1]{}%
\providecommand \@@endlink[0]{}%
\providecommand \url  [0]{\begingroup\@sanitize@url \@url }%
\providecommand \@url [1]{\endgroup\@href {#1}{\urlprefix }}%
\providecommand \urlprefix  [0]{URL }%
\providecommand \Eprint [0]{\href }%
\providecommand \doibase [0]{https://doi.org/}%
\providecommand \selectlanguage [0]{\@gobble}%
\providecommand \bibinfo  [0]{\@secondoftwo}%
\providecommand \bibfield  [0]{\@secondoftwo}%
\providecommand \translation [1]{[#1]}%
\providecommand \BibitemOpen [0]{}%
\providecommand \bibitemStop [0]{}%
\providecommand \bibitemNoStop [0]{.\EOS\space}%
\providecommand \EOS [0]{\spacefactor3000\relax}%
\providecommand \BibitemShut  [1]{\csname bibitem#1\endcsname}%
\let\auto@bib@innerbib\@empty
\bibitem [{\citenamefont {Bergholtz}\ \emph {et~al.}(2021)\citenamefont
  {Bergholtz}, \citenamefont {Budich},\ and\ \citenamefont {Kunst}}]{emil2021}%
  \BibitemOpen
  \bibfield  {author} {\bibinfo {author} {\bibfnamefont {E.~J.}\ \bibnamefont
  {Bergholtz}}, \bibinfo {author} {\bibfnamefont {J.~C.}\ \bibnamefont
  {Budich}},\ and\ \bibinfo {author} {\bibfnamefont {F.~K.}\ \bibnamefont
  {Kunst}},\ }\bibfield  {title} {\bibinfo {title} {Exceptional topology of
  non-{H}ermitian systems},\ }\href
  {https://doi.org/10.1103/RevModPhys.93.015005} {\bibfield  {journal}
  {\bibinfo  {journal} {Rev. Mod. Phys.}\ }\textbf {\bibinfo {volume} {93}},\
  \bibinfo {pages} {015005} (\bibinfo {year} {2021})}\BibitemShut {NoStop}%
\bibitem [{\citenamefont {Gong}\ \emph {et~al.}(2018)\citenamefont {Gong},
  \citenamefont {Ashida}, \citenamefont {Kawabata}, \citenamefont {Takasan},
  \citenamefont {Higashikawa},\ and\ \citenamefont {Ueda}}]{gong2018}%
  \BibitemOpen
  \bibfield  {author} {\bibinfo {author} {\bibfnamefont {Z.}~\bibnamefont
  {Gong}}, \bibinfo {author} {\bibfnamefont {Y.}~\bibnamefont {Ashida}},
  \bibinfo {author} {\bibfnamefont {K.}~\bibnamefont {Kawabata}}, \bibinfo
  {author} {\bibfnamefont {K.}~\bibnamefont {Takasan}}, \bibinfo {author}
  {\bibfnamefont {S.}~\bibnamefont {Higashikawa}},\ and\ \bibinfo {author}
  {\bibfnamefont {M.}~\bibnamefont {Ueda}},\ }\bibfield  {title} {\bibinfo
  {title} {Topological phases of non-{H}ermitian systems},\ }\href
  {https://doi.org/10.1103/PhysRevX.8.031079} {\bibfield  {journal} {\bibinfo
  {journal} {Phys. Rev. X}\ }\textbf {\bibinfo {volume} {8}},\ \bibinfo {pages}
  {031079} (\bibinfo {year} {2018})}\BibitemShut {NoStop}%
\bibitem [{\citenamefont {Shen}\ \emph {et~al.}(2018)\citenamefont {Shen},
  \citenamefont {Zhen},\ and\ \citenamefont {Fu}}]{fu2018}%
  \BibitemOpen
  \bibfield  {author} {\bibinfo {author} {\bibfnamefont {H.}~\bibnamefont
  {Shen}}, \bibinfo {author} {\bibfnamefont {B.}~\bibnamefont {Zhen}},\ and\
  \bibinfo {author} {\bibfnamefont {L.}~\bibnamefont {Fu}},\ }\bibfield
  {title} {\bibinfo {title} {Topological band theory for non-{H}ermitian
  {H}amiltonians},\ }\href {https://doi.org/10.1103/PhysRevLett.120.146402}
  {\bibfield  {journal} {\bibinfo  {journal} {Phys. Rev. Lett.}\ }\textbf
  {\bibinfo {volume} {120}},\ \bibinfo {pages} {146402} (\bibinfo {year}
  {2018})}\BibitemShut {NoStop}%
\bibitem [{\citenamefont {Hasan}\ and\ \citenamefont {Kane}(2010)}]{HasanKane}%
  \BibitemOpen
  \bibfield  {author} {\bibinfo {author} {\bibfnamefont {M.~Z.}\ \bibnamefont
  {Hasan}}\ and\ \bibinfo {author} {\bibfnamefont {C.~L.}\ \bibnamefont
  {Kane}},\ }\bibfield  {title} {\bibinfo {title} {Colloquium: {T}opological
  insulators},\ }\href {https://doi.org/10.1103/RevModPhys.82.3045} {\bibfield
  {journal} {\bibinfo  {journal} {Rev. Mod. Phys.}\ }\textbf {\bibinfo {volume}
  {82}},\ \bibinfo {pages} {3045} (\bibinfo {year} {2010})}\BibitemShut
  {NoStop}%
\bibitem [{\citenamefont {Qi}\ and\ \citenamefont {Zhang}(2011)}]{QiZhang}%
  \BibitemOpen
  \bibfield  {author} {\bibinfo {author} {\bibfnamefont {X.-L.}\ \bibnamefont
  {Qi}}\ and\ \bibinfo {author} {\bibfnamefont {S.-C.}\ \bibnamefont {Zhang}},\
  }\bibfield  {title} {\bibinfo {title} {Topological insulators and
  superconductors},\ }\href {https://doi.org/10.1103/RevModPhys.83.1057}
  {\bibfield  {journal} {\bibinfo  {journal} {Rev. Mod. Phys.}\ }\textbf
  {\bibinfo {volume} {83}},\ \bibinfo {pages} {1057} (\bibinfo {year}
  {2011})}\BibitemShut {NoStop}%
\bibitem [{\citenamefont {Armitage}\ \emph {et~al.}(2018)\citenamefont
  {Armitage}, \citenamefont {Mele},\ and\ \citenamefont
  {Vishwanath}}]{Armitage}%
  \BibitemOpen
  \bibfield  {author} {\bibinfo {author} {\bibfnamefont {N.~P.}\ \bibnamefont
  {Armitage}}, \bibinfo {author} {\bibfnamefont {E.~J.}\ \bibnamefont {Mele}},\
  and\ \bibinfo {author} {\bibfnamefont {A.}~\bibnamefont {Vishwanath}},\
  }\bibfield  {title} {\bibinfo {title} {Weyl and {D}irac semimetals in
  three-dimensional solids},\ }\href
  {https://doi.org/10.1103/RevModPhys.90.015001} {\bibfield  {journal}
  {\bibinfo  {journal} {Rev. Mod. Phys.}\ }\textbf {\bibinfo {volume} {90}},\
  \bibinfo {pages} {015001} (\bibinfo {year} {2018})}\BibitemShut {NoStop}%
\bibitem [{\citenamefont {Kawabata}\ \emph
  {et~al.}(2019{\natexlab{a}})\citenamefont {Kawabata}, \citenamefont
  {Shiozaki}, \citenamefont {Ueda},\ and\ \citenamefont {Sato}}]{kawabataprx}%
  \BibitemOpen
  \bibfield  {author} {\bibinfo {author} {\bibfnamefont {K.}~\bibnamefont
  {Kawabata}}, \bibinfo {author} {\bibfnamefont {K.}~\bibnamefont {Shiozaki}},
  \bibinfo {author} {\bibfnamefont {M.}~\bibnamefont {Ueda}},\ and\ \bibinfo
  {author} {\bibfnamefont {M.}~\bibnamefont {Sato}},\ }\bibfield  {title}
  {\bibinfo {title} {Symmetry and topology in non-{H}ermitian physics},\ }\href
  {https://doi.org/10.1103/PhysRevX.9.041015} {\bibfield  {journal} {\bibinfo
  {journal} {Phys. Rev. X}\ }\textbf {\bibinfo {volume} {9}},\ \bibinfo {pages}
  {041015} (\bibinfo {year} {2019}{\natexlab{a}})}\BibitemShut {NoStop}%
\bibitem [{\citenamefont {Zhou}\ and\ \citenamefont {Lee}(2019)}]{symmetry2}%
  \BibitemOpen
  \bibfield  {author} {\bibinfo {author} {\bibfnamefont {H.}~\bibnamefont
  {Zhou}}\ and\ \bibinfo {author} {\bibfnamefont {J.~Y.}\ \bibnamefont {Lee}},\
  }\bibfield  {title} {\bibinfo {title} {Periodic table for topological bands
  with non-{H}ermitian symmetries},\ }\href
  {https://doi.org/10.1103/PhysRevB.99.235112} {\bibfield  {journal} {\bibinfo
  {journal} {Phys. Rev. B}\ }\textbf {\bibinfo {volume} {99}},\ \bibinfo
  {pages} {235112} (\bibinfo {year} {2019})}\BibitemShut {NoStop}%
\bibitem [{\citenamefont {Budich}\ \emph {et~al.}(2019)\citenamefont {Budich},
  \citenamefont {Carlstr\"om}, \citenamefont {Kunst},\ and\ \citenamefont
  {Bergholtz}}]{jansym}%
  \BibitemOpen
  \bibfield  {author} {\bibinfo {author} {\bibfnamefont {J.~C.}\ \bibnamefont
  {Budich}}, \bibinfo {author} {\bibfnamefont {J.}~\bibnamefont {Carlstr\"om}},
  \bibinfo {author} {\bibfnamefont {F.~K.}\ \bibnamefont {Kunst}},\ and\
  \bibinfo {author} {\bibfnamefont {E.~J.}\ \bibnamefont {Bergholtz}},\
  }\bibfield  {title} {\bibinfo {title} {Symmetry-protected nodal phases in
  non-{H}ermitian systems},\ }\href
  {https://doi.org/10.1103/PhysRevB.99.041406} {\bibfield  {journal} {\bibinfo
  {journal} {Phys. Rev. B}\ }\textbf {\bibinfo {volume} {99}},\ \bibinfo
  {pages} {041406(R)} (\bibinfo {year} {2019})}\BibitemShut {NoStop}%
\bibitem [{\citenamefont {Yoshida}\ \emph {et~al.}(2019)\citenamefont
  {Yoshida}, \citenamefont {Peters}, \citenamefont {Kawakami},\ and\
  \citenamefont {Hatsugai}}]{yoshida2019}%
  \BibitemOpen
  \bibfield  {author} {\bibinfo {author} {\bibfnamefont {T.}~\bibnamefont
  {Yoshida}}, \bibinfo {author} {\bibfnamefont {R.}~\bibnamefont {Peters}},
  \bibinfo {author} {\bibfnamefont {N.}~\bibnamefont {Kawakami}},\ and\
  \bibinfo {author} {\bibfnamefont {Y.}~\bibnamefont {Hatsugai}},\ }\bibfield
  {title} {\bibinfo {title} {Symmetry-protected exceptional rings in
  two-dimensional correlated systems with chiral symmetry},\ }\href
  {https://doi.org/10.1103/PhysRevB.99.121101} {\bibfield  {journal} {\bibinfo
  {journal} {Phys. Rev. B}\ }\textbf {\bibinfo {volume} {99}},\ \bibinfo
  {pages} {121101(R)} (\bibinfo {year} {2019})}\BibitemShut {NoStop}%
\bibitem [{\citenamefont {Kawabata}\ \emph
  {et~al.}(2019{\natexlab{b}})\citenamefont {Kawabata}, \citenamefont
  {Bessho},\ and\ \citenamefont {Sato}}]{kawabataprl}%
  \BibitemOpen
  \bibfield  {author} {\bibinfo {author} {\bibfnamefont {K.}~\bibnamefont
  {Kawabata}}, \bibinfo {author} {\bibfnamefont {T.}~\bibnamefont {Bessho}},\
  and\ \bibinfo {author} {\bibfnamefont {M.}~\bibnamefont {Sato}},\ }\bibfield
  {title} {\bibinfo {title} {Classification of exceptional points and
  non-{H}ermitian topological semimetals},\ }\href
  {https://doi.org/10.1103/PhysRevLett.123.066405} {\bibfield  {journal}
  {\bibinfo  {journal} {Phys. Rev. Lett.}\ }\textbf {\bibinfo {volume} {123}},\
  \bibinfo {pages} {066405} (\bibinfo {year} {2019}{\natexlab{b}})}\BibitemShut
  {NoStop}%
\bibitem [{\citenamefont {Delplace}\ \emph {et~al.}(2021)\citenamefont
  {Delplace}, \citenamefont {Yoshida},\ and\ \citenamefont
  {Hatsugai}}]{tsuneyaprl}%
  \BibitemOpen
  \bibfield  {author} {\bibinfo {author} {\bibfnamefont {P.}~\bibnamefont
  {Delplace}}, \bibinfo {author} {\bibfnamefont {T.}~\bibnamefont {Yoshida}},\
  and\ \bibinfo {author} {\bibfnamefont {Y.}~\bibnamefont {Hatsugai}},\
  }\bibfield  {title} {\bibinfo {title} {Symmetry-protected multifold
  exceptional points and their topological characterization},\ }\href
  {https://doi.org/10.1103/PhysRevLett.127.186602} {\bibfield  {journal}
  {\bibinfo  {journal} {Phys. Rev. Lett.}\ }\textbf {\bibinfo {volume} {127}},\
  \bibinfo {pages} {186602} (\bibinfo {year} {2021})}\BibitemShut {NoStop}%
\bibitem [{\citenamefont {Mandal}\ and\ \citenamefont
  {Bergholtz}(2021)}]{ipsitaprl}%
  \BibitemOpen
  \bibfield  {author} {\bibinfo {author} {\bibfnamefont {I.}~\bibnamefont
  {Mandal}}\ and\ \bibinfo {author} {\bibfnamefont {E.~J.}\ \bibnamefont
  {Bergholtz}},\ }\bibfield  {title} {\bibinfo {title} {Symmetry and
  higher-order exceptional points},\ }\href
  {https://doi.org/10.1103/PhysRevLett.127.186601} {\bibfield  {journal}
  {\bibinfo  {journal} {Phys. Rev. Lett.}\ }\textbf {\bibinfo {volume} {127}},\
  \bibinfo {pages} {186601} (\bibinfo {year} {2021})}\BibitemShut {NoStop}%
\bibitem [{\citenamefont {St\aa{}lhammar}\ and\ \citenamefont
  {Bergholtz}(2021)}]{marcus22}%
  \BibitemOpen
  \bibfield  {author} {\bibinfo {author} {\bibfnamefont {M.}~\bibnamefont
  {St\aa{}lhammar}}\ and\ \bibinfo {author} {\bibfnamefont {E.~J.}\
  \bibnamefont {Bergholtz}},\ }\bibfield  {title} {\bibinfo {title}
  {Classification of exceptional nodal topologies protected by $\mathcal{PT}$
  symmetry},\ }\href {https://doi.org/10.1103/PhysRevB.104.L201104} {\bibfield
  {journal} {\bibinfo  {journal} {Phys. Rev. B}\ }\textbf {\bibinfo {volume}
  {104}},\ \bibinfo {pages} {L201104} (\bibinfo {year} {2021})}\BibitemShut
  {NoStop}%
\bibitem [{\citenamefont {Lee}(2016)}]{Lee}%
  \BibitemOpen
  \bibfield  {author} {\bibinfo {author} {\bibfnamefont {T.~E.}\ \bibnamefont
  {Lee}},\ }\bibfield  {title} {\bibinfo {title} {Anomalous edge state in a
  non-{H}ermitian lattice},\ }\href
  {https://doi.org/10.1103/PhysRevLett.116.133903} {\bibfield  {journal}
  {\bibinfo  {journal} {Phys. Rev. Lett.}\ }\textbf {\bibinfo {volume} {116}},\
  \bibinfo {pages} {133903} (\bibinfo {year} {2016})}\BibitemShut {NoStop}%
\bibitem [{\citenamefont {{Martinez Alvarez}}\ \emph
  {et~al.}(2018)\citenamefont {{Martinez Alvarez}}, \citenamefont {{Barrios
  Vargas}},\ and\ \citenamefont {{Foa Torres}}}]{Alvarez}%
  \BibitemOpen
  \bibfield  {author} {\bibinfo {author} {\bibfnamefont {V.~M.}\ \bibnamefont
  {{Martinez Alvarez}}}, \bibinfo {author} {\bibfnamefont {J.~E.}\ \bibnamefont
  {{Barrios Vargas}}},\ and\ \bibinfo {author} {\bibfnamefont {L.~E.~F.}\
  \bibnamefont {{Foa Torres}}},\ }\bibfield  {title} {\bibinfo {title}
  {Non-{H}ermitian robust edge states in one dimension: {A}nomalous
  localization and eigenspace condensation at exceptional points},\ }\href
  {https://doi.org/10.1103/PhysRevB.97.121401} {\bibfield  {journal} {\bibinfo
  {journal} {Phys. Rev. B}\ }\textbf {\bibinfo {volume} {97}},\ \bibinfo
  {pages} {{121401(R)}} (\bibinfo {year} {2018})}\BibitemShut {NoStop}%
\bibitem [{\citenamefont {Xiong}(2018)}]{Xiong}%
  \BibitemOpen
  \bibfield  {author} {\bibinfo {author} {\bibfnamefont {Y.}~\bibnamefont
  {Xiong}},\ }\bibfield  {title} {\bibinfo {title} {Why does bulk boundary
  correspondence fail in some non-{H}ermitian topological models},\ }\href
  {https://doi.org/10.1088/2399-6528/aab64a} {\bibfield  {journal} {\bibinfo
  {journal} {J. Phys. Commun.}\ }\textbf {\bibinfo {volume} {2}},\ \bibinfo
  {pages} {035043} (\bibinfo {year} {2018})}\BibitemShut {NoStop}%
\bibitem [{\citenamefont {Kunst}\ \emph {et~al.}(2018)\citenamefont {Kunst},
  \citenamefont {Edvardsson}, \citenamefont {Budich},\ and\ \citenamefont
  {Bergholtz}}]{flore2018}%
  \BibitemOpen
  \bibfield  {author} {\bibinfo {author} {\bibfnamefont {F.~K.}\ \bibnamefont
  {Kunst}}, \bibinfo {author} {\bibfnamefont {E.}~\bibnamefont {Edvardsson}},
  \bibinfo {author} {\bibfnamefont {J.~C.}\ \bibnamefont {Budich}},\ and\
  \bibinfo {author} {\bibfnamefont {E.~J.}\ \bibnamefont {Bergholtz}},\
  }\bibfield  {title} {\bibinfo {title} {Biorthogonal bulk-boundary
  correspondence in non-{H}ermitian systems},\ }\href
  {https://doi.org/10.1103/PhysRevLett.121.026808} {\bibfield  {journal}
  {\bibinfo  {journal} {Phys. Rev. Lett.}\ }\textbf {\bibinfo {volume} {121}},\
  \bibinfo {pages} {026808} (\bibinfo {year} {2018})}\BibitemShut {NoStop}%
\bibitem [{\citenamefont {Yao}\ and\ \citenamefont {Wang}(2018)}]{yao2018}%
  \BibitemOpen
  \bibfield  {author} {\bibinfo {author} {\bibfnamefont {S.}~\bibnamefont
  {Yao}}\ and\ \bibinfo {author} {\bibfnamefont {Z.}~\bibnamefont {Wang}},\
  }\bibfield  {title} {\bibinfo {title} {Edge states and topological invariants
  of non-{H}ermitian systems},\ }\href
  {https://doi.org/10.1103/PhysRevLett.121.086803} {\bibfield  {journal}
  {\bibinfo  {journal} {Phys. Rev. Lett.}\ }\textbf {\bibinfo {volume} {121}},\
  \bibinfo {pages} {086803} (\bibinfo {year} {2018})}\BibitemShut {NoStop}%
\bibitem [{\citenamefont {Kunst}\ and\ \citenamefont
  {Dwivedi}(2019)}]{flore2019n}%
  \BibitemOpen
  \bibfield  {author} {\bibinfo {author} {\bibfnamefont {F.~K.}\ \bibnamefont
  {Kunst}}\ and\ \bibinfo {author} {\bibfnamefont {V.}~\bibnamefont
  {Dwivedi}},\ }\bibfield  {title} {\bibinfo {title} {Non-{H}ermitian systems
  and topology: {A} transfer-matrix perspective},\ }\href
  {https://doi.org/10.1103/PhysRevB.99.245116} {\bibfield  {journal} {\bibinfo
  {journal} {Phys. Rev. B}\ }\textbf {\bibinfo {volume} {99}},\ \bibinfo
  {pages} {245116} (\bibinfo {year} {2019})}\BibitemShut {NoStop}%
\bibitem [{\citenamefont {Herviou}\ \emph {et~al.}(2019)\citenamefont
  {Herviou}, \citenamefont {Bardarson},\ and\ \citenamefont
  {Regnault}}]{Regnault_2019}%
  \BibitemOpen
  \bibfield  {author} {\bibinfo {author} {\bibfnamefont {L.}~\bibnamefont
  {Herviou}}, \bibinfo {author} {\bibfnamefont {J.~H.}\ \bibnamefont
  {Bardarson}},\ and\ \bibinfo {author} {\bibfnamefont {N.}~\bibnamefont
  {Regnault}},\ }\bibfield  {title} {\bibinfo {title} {Defining a bulk-edge
  correspondence for non-{H}ermitian {H}amiltonians via singular-value
  decomposition},\ }\href {https://doi.org/10.1103/PhysRevA.99.052118}
  {\bibfield  {journal} {\bibinfo  {journal} {Phys. Rev. A}\ }\textbf {\bibinfo
  {volume} {99}},\ \bibinfo {pages} {052118} (\bibinfo {year}
  {2019})}\BibitemShut {NoStop}%
\bibitem [{\citenamefont {Okuma}\ \emph {et~al.}(2020)\citenamefont {Okuma},
  \citenamefont {Kawabata}, \citenamefont {Shiozaki},\ and\ \citenamefont
  {Sato}}]{Okuma_2020}%
  \BibitemOpen
  \bibfield  {author} {\bibinfo {author} {\bibfnamefont {N.}~\bibnamefont
  {Okuma}}, \bibinfo {author} {\bibfnamefont {K.}~\bibnamefont {Kawabata}},
  \bibinfo {author} {\bibfnamefont {K.}~\bibnamefont {Shiozaki}},\ and\
  \bibinfo {author} {\bibfnamefont {M.}~\bibnamefont {Sato}},\ }\bibfield
  {title} {\bibinfo {title} {Topological origin of non-{H}ermitian skin
  effects},\ }\href {https://doi.org/10.1103/PhysRevLett.124.086801} {\bibfield
   {journal} {\bibinfo  {journal} {Phys. Rev. Lett.}\ }\textbf {\bibinfo
  {volume} {124}},\ \bibinfo {pages} {086801} (\bibinfo {year}
  {2020})}\BibitemShut {NoStop}%
\bibitem [{\citenamefont {Edvardsson}\ \emph {et~al.}(2019)\citenamefont
  {Edvardsson}, \citenamefont {Kunst},\ and\ \citenamefont
  {Bergholtz}}]{Edvardsson_2019}%
  \BibitemOpen
  \bibfield  {author} {\bibinfo {author} {\bibfnamefont {E.}~\bibnamefont
  {Edvardsson}}, \bibinfo {author} {\bibfnamefont {F.~K.}\ \bibnamefont
  {Kunst}},\ and\ \bibinfo {author} {\bibfnamefont {E.~J.}\ \bibnamefont
  {Bergholtz}},\ }\bibfield  {title} {\bibinfo {title} {Non-{H}ermitian
  extensions of higher-order topological phases and their biorthogonal
  bulk-boundary correspondence},\ }\href
  {https://doi.org/10.1103/PhysRevB.99.081302} {\bibfield  {journal} {\bibinfo
  {journal} {Phys. Rev. B}\ }\textbf {\bibinfo {volume} {99}},\ \bibinfo
  {pages} {081302(R)} (\bibinfo {year} {2019})}\BibitemShut {NoStop}%
\bibitem [{\citenamefont {Longhi}(2019)}]{Longhi}%
  \BibitemOpen
  \bibfield  {author} {\bibinfo {author} {\bibfnamefont {S.}~\bibnamefont
  {Longhi}},\ }\bibfield  {title} {\bibinfo {title} {Probing non-{H}ermitian
  skin effect and non-{B}loch phase transitions},\ }\href
  {https://doi.org/10.1103/PhysRevResearch.1.023013} {\bibfield  {journal}
  {\bibinfo  {journal} {Phys. Rev. Research}\ }\textbf {\bibinfo {volume}
  {1}},\ \bibinfo {pages} {023013} (\bibinfo {year} {2019})}\BibitemShut
  {NoStop}%
\bibitem [{\citenamefont {Leykam}\ \emph {et~al.}(2017)\citenamefont {Leykam},
  \citenamefont {Bliokh}, \citenamefont {Huang}, \citenamefont {Chong},\ and\
  \citenamefont {Nori}}]{Leykam}%
  \BibitemOpen
  \bibfield  {author} {\bibinfo {author} {\bibfnamefont {D.}~\bibnamefont
  {Leykam}}, \bibinfo {author} {\bibfnamefont {K.~Y.}\ \bibnamefont {Bliokh}},
  \bibinfo {author} {\bibfnamefont {C.}~\bibnamefont {Huang}}, \bibinfo
  {author} {\bibfnamefont {Y.~D.}\ \bibnamefont {Chong}},\ and\ \bibinfo
  {author} {\bibfnamefont {F.}~\bibnamefont {Nori}},\ }\bibfield  {title}
  {\bibinfo {title} {Edge modes, degeneracies, and topological numbers in
  non-{H}ermitian systems},\ }\href
  {https://doi.org/10.1103/PhysRevLett.118.040401} {\bibfield  {journal}
  {\bibinfo  {journal} {Phys. Rev. Lett.}\ }\textbf {\bibinfo {volume} {118}},\
  \bibinfo {pages} {040401} (\bibinfo {year} {2017})}\BibitemShut {NoStop}%
\bibitem [{\citenamefont {Lee}\ and\ \citenamefont
  {Thomale}(2019)}]{LeeThomale}%
  \BibitemOpen
  \bibfield  {author} {\bibinfo {author} {\bibfnamefont {C.~H.}\ \bibnamefont
  {Lee}}\ and\ \bibinfo {author} {\bibfnamefont {R.}~\bibnamefont {Thomale}},\
  }\bibfield  {title} {\bibinfo {title} {Anatomy of skin modes and topology in
  non-{H}ermitian systems},\ }\href
  {https://doi.org/10.1103/PhysRevB.99.201103} {\bibfield  {journal} {\bibinfo
  {journal} {Phys. Rev. B}\ }\textbf {\bibinfo {volume} {99}},\ \bibinfo
  {pages} {201103(R)} (\bibinfo {year} {2019})}\BibitemShut {NoStop}%
\bibitem [{\citenamefont {Lee}\ \emph {et~al.}(2019)\citenamefont {Lee},
  \citenamefont {Li},\ and\ \citenamefont {Gong}}]{CHLee}%
  \BibitemOpen
  \bibfield  {author} {\bibinfo {author} {\bibfnamefont {C.~H.}\ \bibnamefont
  {Lee}}, \bibinfo {author} {\bibfnamefont {L.}~\bibnamefont {Li}},\ and\
  \bibinfo {author} {\bibfnamefont {J.}~\bibnamefont {Gong}},\ }\bibfield
  {title} {\bibinfo {title} {Hybrid higher-order skin-topological modes in
  nonreciprocal systems},\ }\href
  {https://doi.org/10.1103/PhysRevLett.123.016805} {\bibfield  {journal}
  {\bibinfo  {journal} {Phys. Rev. Lett.}\ }\textbf {\bibinfo {volume} {123}},\
  \bibinfo {pages} {016805} (\bibinfo {year} {2019})}\BibitemShut {NoStop}%
\bibitem [{\citenamefont {Liu}\ \emph {et~al.}(2019)\citenamefont {Liu},
  \citenamefont {Zhang}, \citenamefont {Ai}, \citenamefont {Gong},
  \citenamefont {Kawabata}, \citenamefont {Ueda},\ and\ \citenamefont
  {Nori}}]{secondorder}%
  \BibitemOpen
  \bibfield  {author} {\bibinfo {author} {\bibfnamefont {T.}~\bibnamefont
  {Liu}}, \bibinfo {author} {\bibfnamefont {Y.-R.}\ \bibnamefont {Zhang}},
  \bibinfo {author} {\bibfnamefont {Q.}~\bibnamefont {Ai}}, \bibinfo {author}
  {\bibfnamefont {Z.}~\bibnamefont {Gong}}, \bibinfo {author} {\bibfnamefont
  {K.}~\bibnamefont {Kawabata}}, \bibinfo {author} {\bibfnamefont
  {M.}~\bibnamefont {Ueda}},\ and\ \bibinfo {author} {\bibfnamefont
  {F.}~\bibnamefont {Nori}},\ }\bibfield  {title} {\bibinfo {title}
  {Second-order topological phases in non-{H}ermitian systems},\ }\href
  {https://doi.org/10.1103/PhysRevLett.122.076801} {\bibfield  {journal}
  {\bibinfo  {journal} {Phys. Rev. Lett.}\ }\textbf {\bibinfo {volume} {122}},\
  \bibinfo {pages} {076801} (\bibinfo {year} {2019})}\BibitemShut {NoStop}%
\bibitem [{\citenamefont {Fleckenstein}\ \emph {et~al.}()\citenamefont
  {Fleckenstein}, \citenamefont {Zorzato}, \citenamefont {Varjas},
  \citenamefont {Bergholtz}, \citenamefont {Bardarson},\ and\ \citenamefont
  {Tiwari}}]{fleckenstein2022nonhermitian}%
  \BibitemOpen
  \bibfield  {author} {\bibinfo {author} {\bibfnamefont {C.}~\bibnamefont
  {Fleckenstein}}, \bibinfo {author} {\bibfnamefont {A.}~\bibnamefont
  {Zorzato}}, \bibinfo {author} {\bibfnamefont {D.}~\bibnamefont {Varjas}},
  \bibinfo {author} {\bibfnamefont {E.~J.}\ \bibnamefont {Bergholtz}}, \bibinfo
  {author} {\bibfnamefont {J.~H.}\ \bibnamefont {Bardarson}},\ and\ \bibinfo
  {author} {\bibfnamefont {A.}~\bibnamefont {Tiwari}},\ }\bibfield  {title}
  {\bibinfo {title} {Non-{H}ermitian topology in monitored quantum circuits},\
  }\href {https://arxiv.org/abs/2201.05341} {\bibinfo  {journal}
  {arXiv:2201.05341}\ }\BibitemShut {NoStop}%
\bibitem [{\citenamefont {Zirnstein}\ \emph {et~al.}(2021)\citenamefont
  {Zirnstein}, \citenamefont {Refael},\ and\ \citenamefont
  {Rosenow}}]{Rosenow}%
  \BibitemOpen
\bibfield  {journal} {  }\bibfield  {author} {\bibinfo {author} {\bibfnamefont
  {H.~G.}\ \bibnamefont {Zirnstein}}, \bibinfo {author} {\bibfnamefont
  {G.}~\bibnamefont {Refael}},\ and\ \bibinfo {author} {\bibfnamefont
  {B.}~\bibnamefont {Rosenow}},\ }\bibfield  {title} {\bibinfo {title}
  {Bulk-boundary correspondence for non-{H}ermitian {H}amiltonians via {G}reen
  functions},\ }\href {https://doi.org/10.1103/PhysRevLett.126.216407}
  {\bibfield  {journal} {\bibinfo  {journal} {Phys. Rev. Lett.}\ }\textbf
  {\bibinfo {volume} {126}},\ \bibinfo {pages} {216407} (\bibinfo {year}
  {2021})}\BibitemShut {NoStop}%
\bibitem [{\citenamefont {Borgnia}\ \emph {et~al.}(2020)\citenamefont
  {Borgnia}, \citenamefont {Kruchkov},\ and\ \citenamefont {Slager}}]{Borgnia}%
  \BibitemOpen
  \bibfield  {author} {\bibinfo {author} {\bibfnamefont {D.~S.}\ \bibnamefont
  {Borgnia}}, \bibinfo {author} {\bibfnamefont {A.~J.}\ \bibnamefont
  {Kruchkov}},\ and\ \bibinfo {author} {\bibfnamefont {R.-J.}\ \bibnamefont
  {Slager}},\ }\bibfield  {title} {\bibinfo {title} {Non-{H}ermitian boundary
  modes and topology},\ }\href {https://doi.org/10.1103/PhysRevLett.124.056802}
  {\bibfield  {journal} {\bibinfo  {journal} {Phys. Rev. Lett.}\ }\textbf
  {\bibinfo {volume} {124}},\ \bibinfo {pages} {056802} (\bibinfo {year}
  {2020})}\BibitemShut {NoStop}%
\bibitem [{\citenamefont {Yokomizo}\ and\ \citenamefont
  {Murakami}(2019)}]{Murakami}%
  \BibitemOpen
  \bibfield  {author} {\bibinfo {author} {\bibfnamefont {K.}~\bibnamefont
  {Yokomizo}}\ and\ \bibinfo {author} {\bibfnamefont {S.}~\bibnamefont
  {Murakami}},\ }\bibfield  {title} {\bibinfo {title} {Non-{B}loch band theory
  of non-{H}ermitian systems},\ }\href
  {https://doi.org/10.1103/PhysRevLett.123.066404} {\bibfield  {journal}
  {\bibinfo  {journal} {Phys. Rev. Lett.}\ }\textbf {\bibinfo {volume} {123}},\
  \bibinfo {pages} {066404} (\bibinfo {year} {2019})}\BibitemShut {NoStop}%
\bibitem [{\citenamefont {Yang}\ \emph {et~al.}(2021)\citenamefont {Yang},
  \citenamefont {Morampudi},\ and\ \citenamefont {Bergholtz}}]{ESpinLiquids}%
  \BibitemOpen
  \bibfield  {author} {\bibinfo {author} {\bibfnamefont {K.}~\bibnamefont
  {Yang}}, \bibinfo {author} {\bibfnamefont {S.~C.}\ \bibnamefont
  {Morampudi}},\ and\ \bibinfo {author} {\bibfnamefont {E.~J.}\ \bibnamefont
  {Bergholtz}},\ }\bibfield  {title} {\bibinfo {title} {Exceptional spin
  liquids from couplings to the environment},\ }\href
  {https://doi.org/10.1103/PhysRevLett.126.077201} {\bibfield  {journal}
  {\bibinfo  {journal} {Phys. Rev. Lett.}\ }\textbf {\bibinfo {volume} {126}},\
  \bibinfo {pages} {077201} (\bibinfo {year} {2021})}\BibitemShut {NoStop}%
\bibitem [{\citenamefont {Edvardsson}\ \emph {et~al.}(2020)\citenamefont
  {Edvardsson}, \citenamefont {Kunst}, \citenamefont {Yoshida},\ and\
  \citenamefont {Bergholtz}}]{elisabet2020}%
  \BibitemOpen
  \bibfield  {author} {\bibinfo {author} {\bibfnamefont {E.}~\bibnamefont
  {Edvardsson}}, \bibinfo {author} {\bibfnamefont {F.~K.}\ \bibnamefont
  {Kunst}}, \bibinfo {author} {\bibfnamefont {T.}~\bibnamefont {Yoshida}},\
  and\ \bibinfo {author} {\bibfnamefont {E.~J.}\ \bibnamefont {Bergholtz}},\
  }\bibfield  {title} {\bibinfo {title} {Phase transitions and generalized
  biorthogonal polarization in non-{H}ermitian systems},\ }\href
  {https://doi.org/10.1103/PhysRevResearch.2.043046} {\bibfield  {journal}
  {\bibinfo  {journal} {Phys. Rev. Research}\ }\textbf {\bibinfo {volume}
  {2}},\ \bibinfo {pages} {043046} (\bibinfo {year} {2020})}\BibitemShut
  {NoStop}%
\bibitem [{\citenamefont {Koch}\ and\ \citenamefont {Budich}(2020)}]{koch2020}%
  \BibitemOpen
  \bibfield  {author} {\bibinfo {author} {\bibfnamefont {R.}~\bibnamefont
  {Koch}}\ and\ \bibinfo {author} {\bibfnamefont {J.~C.}\ \bibnamefont
  {Budich}},\ }\bibfield  {title} {\bibinfo {title} {Bulk-boundary
  correspondence in non-{H}ermitian systems: Stability analysis for generalized
  boundary conditions},\ }\href {https://doi.org/10.1140/epjd/e2020-100641-y}
  {\bibfield  {journal} {\bibinfo  {journal} {Eur. Phys. J. D}\ }\textbf
  {\bibinfo {volume} {74}},\ \bibinfo {pages} {70} (\bibinfo {year}
  {2020})}\BibitemShut {NoStop}%
\bibitem [{\citenamefont {Schomerus}(2020)}]{Schomerus2020}%
  \BibitemOpen
  \bibfield  {author} {\bibinfo {author} {\bibfnamefont {H.}~\bibnamefont
  {Schomerus}},\ }\bibfield  {title} {\bibinfo {title} {Nonreciprocal response
  theory of non-{H}ermitian mechanical metamaterials: {R}esponse phase
  transition from the skin effect of zero modes},\ }\href
  {https://doi.org/10.1103/PhysRevResearch.2.013058} {\bibfield  {journal}
  {\bibinfo  {journal} {Phys. Rev. Research}\ }\textbf {\bibinfo {volume}
  {2}},\ \bibinfo {pages} {013058} (\bibinfo {year} {2020})}\BibitemShut
  {NoStop}%
\bibitem [{\citenamefont {Brzezicki}\ and\ \citenamefont
  {Hyart}(2019)}]{Hyart}%
  \BibitemOpen
  \bibfield  {author} {\bibinfo {author} {\bibfnamefont {W.}~\bibnamefont
  {Brzezicki}}\ and\ \bibinfo {author} {\bibfnamefont {T.}~\bibnamefont
  {Hyart}},\ }\bibfield  {title} {\bibinfo {title} {Hidden {C}hern number in
  one-dimensional non-{H}ermitian chiral-symmetric systems},\ }\href
  {https://doi.org/10.1103/PhysRevB.100.161105} {\bibfield  {journal} {\bibinfo
   {journal} {Phys. Rev. B}\ }\textbf {\bibinfo {volume} {100}},\ \bibinfo
  {pages} {161105(R)} (\bibinfo {year} {2019})}\BibitemShut {NoStop}%
\bibitem [{\citenamefont {Budich}\ and\ \citenamefont
  {Bergholtz}(2020)}]{NTOS}%
  \BibitemOpen
  \bibfield  {author} {\bibinfo {author} {\bibfnamefont {J.~C.}\ \bibnamefont
  {Budich}}\ and\ \bibinfo {author} {\bibfnamefont {E.~J.}\ \bibnamefont
  {Bergholtz}},\ }\bibfield  {title} {\bibinfo {title} {Non-{H}ermitian
  topological sensors},\ }\href
  {https://doi.org/10.1103/PhysRevLett.125.180403} {\bibfield  {journal}
  {\bibinfo  {journal} {Phys. Rev. Lett.}\ }\textbf {\bibinfo {volume} {125}},\
  \bibinfo {pages} {180403} (\bibinfo {year} {2020})}\BibitemShut {NoStop}%
\bibitem [{\citenamefont {Zhang}\ \emph {et~al.}(2020)\citenamefont {Zhang},
  \citenamefont {Yang},\ and\ \citenamefont {Fang}}]{zhesen}%
  \BibitemOpen
  \bibfield  {author} {\bibinfo {author} {\bibfnamefont {K.}~\bibnamefont
  {Zhang}}, \bibinfo {author} {\bibfnamefont {Z.}~\bibnamefont {Yang}},\ and\
  \bibinfo {author} {\bibfnamefont {C.}~\bibnamefont {Fang}},\ }\bibfield
  {title} {\bibinfo {title} {Correspondence between winding numbers and skin
  modes in non-{H}ermitian systems},\ }\href
  {https://doi.org/10.1103/PhysRevLett.125.126402} {\bibfield  {journal}
  {\bibinfo  {journal} {Phys. Rev. Lett.}\ }\textbf {\bibinfo {volume} {125}},\
  \bibinfo {pages} {126402} (\bibinfo {year} {2020})}\BibitemShut {NoStop}%
\bibitem [{\citenamefont {Brandenbourger}\ \emph {et~al.}(2019)\citenamefont
  {Brandenbourger}, \citenamefont {Locsin}, \citenamefont {Lerner},\ and\
  \citenamefont {Coulais}}]{Brandenbourger_2019}%
  \BibitemOpen
  \bibfield  {author} {\bibinfo {author} {\bibfnamefont {M.}~\bibnamefont
  {Brandenbourger}}, \bibinfo {author} {\bibfnamefont {X.}~\bibnamefont
  {Locsin}}, \bibinfo {author} {\bibfnamefont {E.}~\bibnamefont {Lerner}},\
  and\ \bibinfo {author} {\bibfnamefont {C.}~\bibnamefont {Coulais}},\
  }\bibfield  {title} {\bibinfo {title} {Non-reciprocal robotic
  metamaterials},\ }\href {https://doi.org/10.1038/s41467-019-12599-3}
  {\bibfield  {journal} {\bibinfo  {journal} {Nat. Commun.}\ }\textbf {\bibinfo
  {volume} {10}},\ \bibinfo {pages} {4608} (\bibinfo {year}
  {2019})}\BibitemShut {NoStop}%
\bibitem [{\citenamefont {Ghatak}\ \emph {et~al.}(2020)\citenamefont {Ghatak},
  \citenamefont {Brandenbourger}, \citenamefont {van Wezel},\ and\
  \citenamefont {Coulais}}]{Ghatak2020}%
  \BibitemOpen
  \bibfield  {author} {\bibinfo {author} {\bibfnamefont {A.}~\bibnamefont
  {Ghatak}}, \bibinfo {author} {\bibfnamefont {M.}~\bibnamefont
  {Brandenbourger}}, \bibinfo {author} {\bibfnamefont {J.}~\bibnamefont {van
  Wezel}},\ and\ \bibinfo {author} {\bibfnamefont {C.}~\bibnamefont
  {Coulais}},\ }\bibfield  {title} {\bibinfo {title} {Observation of
  non-{H}ermitian topology and its bulk{\textendash}edge correspondence in an
  active mechanical metamaterial},\ }\href
  {https://doi.org/10.1073/pnas.2010580117} {\bibfield  {journal} {\bibinfo
  {journal} {PNAS}\ }\textbf {\bibinfo {volume} {117}},\ \bibinfo {pages}
  {29561} (\bibinfo {year} {2020})}\BibitemShut {NoStop}%
\bibitem [{\citenamefont {Helbig}\ \emph {et~al.}(2020)\citenamefont {Helbig},
  \citenamefont {Hofmann}, \citenamefont {Imhof}, \citenamefont {Abdelghany},
  \citenamefont {Kiessling}, \citenamefont {Molenkamp}, \citenamefont {Lee},
  \citenamefont {Szameit}, \citenamefont {Greiter},\ and\ \citenamefont
  {Thomale}}]{helbig2020}%
  \BibitemOpen
  \bibfield  {author} {\bibinfo {author} {\bibfnamefont {T.}~\bibnamefont
  {Helbig}}, \bibinfo {author} {\bibfnamefont {T.}~\bibnamefont {Hofmann}},
  \bibinfo {author} {\bibfnamefont {S.}~\bibnamefont {Imhof}}, \bibinfo
  {author} {\bibfnamefont {M.}~\bibnamefont {Abdelghany}}, \bibinfo {author}
  {\bibfnamefont {T.}~\bibnamefont {Kiessling}}, \bibinfo {author}
  {\bibfnamefont {L.}~\bibnamefont {Molenkamp}}, \bibinfo {author}
  {\bibfnamefont {C.}~\bibnamefont {Lee}}, \bibinfo {author} {\bibfnamefont
  {A.}~\bibnamefont {Szameit}}, \bibinfo {author} {\bibfnamefont
  {M.}~\bibnamefont {Greiter}},\ and\ \bibinfo {author} {\bibfnamefont
  {R.}~\bibnamefont {Thomale}},\ }\bibfield  {title} {\bibinfo {title}
  {Generalized bulk--boundary correspondence in non-{H}ermitian topolectrical
  circuits},\ }\href {https://doi.org/10.1038/s41567-020-0922-9} {\bibfield
  {journal} {\bibinfo  {journal} {Nat. Phys.}\ }\textbf {\bibinfo {volume}
  {16}},\ \bibinfo {pages} {747} (\bibinfo {year} {2020})}\BibitemShut
  {NoStop}%
\bibitem [{\citenamefont {Hofmann}\ \emph {et~al.}(2020)\citenamefont
  {Hofmann}, \citenamefont {Helbig}, \citenamefont {Schindler}, \citenamefont
  {Salgo}, \citenamefont {Brzezi\ifmmode~\acute{n}\else \'{n}\fi{}ska},
  \citenamefont {Greiter}, \citenamefont {Kiessling}, \citenamefont {Wolf},
  \citenamefont {Vollhardt}, \citenamefont {Kaba\ifmmode~\check{s}\else
  \v{s}\fi{}i}, \citenamefont {Lee}, \citenamefont {Bilu\ifmmode \check{s}\else
  \v{s}\fi{}i\ifmmode~\acute{c}\else \'{c}\fi{}}, \citenamefont {Thomale},\
  and\ \citenamefont {Neupert}}]{Neupert_2020}%
  \BibitemOpen
  \bibfield  {author} {\bibinfo {author} {\bibfnamefont {T.}~\bibnamefont
  {Hofmann}}, \bibinfo {author} {\bibfnamefont {T.}~\bibnamefont {Helbig}},
  \bibinfo {author} {\bibfnamefont {F.}~\bibnamefont {Schindler}}, \bibinfo
  {author} {\bibfnamefont {N.}~\bibnamefont {Salgo}}, \bibinfo {author}
  {\bibfnamefont {M.}~\bibnamefont {Brzezi\ifmmode~\acute{n}\else
  \'{n}\fi{}ska}}, \bibinfo {author} {\bibfnamefont {M.}~\bibnamefont
  {Greiter}}, \bibinfo {author} {\bibfnamefont {T.}~\bibnamefont {Kiessling}},
  \bibinfo {author} {\bibfnamefont {D.}~\bibnamefont {Wolf}}, \bibinfo {author}
  {\bibfnamefont {A.}~\bibnamefont {Vollhardt}}, \bibinfo {author}
  {\bibfnamefont {A.}~\bibnamefont {Kaba\ifmmode~\check{s}\else \v{s}\fi{}i}},
  \bibinfo {author} {\bibfnamefont {C.~H.}\ \bibnamefont {Lee}}, \bibinfo
  {author} {\bibfnamefont {A.}~\bibnamefont {Bilu\ifmmode \check{s}\else
  \v{s}\fi{}i\ifmmode~\acute{c}\else \'{c}\fi{}}}, \bibinfo {author}
  {\bibfnamefont {R.}~\bibnamefont {Thomale}},\ and\ \bibinfo {author}
  {\bibfnamefont {T.}~\bibnamefont {Neupert}},\ }\bibfield  {title} {\bibinfo
  {title} {Reciprocal skin effect and its realization in a topolectrical
  circuit},\ }\href {https://doi.org/10.1103/PhysRevResearch.2.023265}
  {\bibfield  {journal} {\bibinfo  {journal} {Phys. Rev. Research}\ }\textbf
  {\bibinfo {volume} {2}},\ \bibinfo {pages} {023265} (\bibinfo {year}
  {2020})}\BibitemShut {NoStop}%
\bibitem [{\citenamefont {Xiao}\ \emph {et~al.}(2020)\citenamefont {Xiao},
  \citenamefont {Deng}, \citenamefont {Wang}, \citenamefont {Zhu},
  \citenamefont {Wang}, \citenamefont {Yi},\ and\ \citenamefont
  {Xue}}]{photonicNHBBC}%
  \BibitemOpen
  \bibfield  {author} {\bibinfo {author} {\bibfnamefont {L.}~\bibnamefont
  {Xiao}}, \bibinfo {author} {\bibfnamefont {T.}~\bibnamefont {Deng}}, \bibinfo
  {author} {\bibfnamefont {K.}~\bibnamefont {Wang}}, \bibinfo {author}
  {\bibfnamefont {G.}~\bibnamefont {Zhu}}, \bibinfo {author} {\bibfnamefont
  {Z.}~\bibnamefont {Wang}}, \bibinfo {author} {\bibfnamefont {W.}~\bibnamefont
  {Yi}},\ and\ \bibinfo {author} {\bibfnamefont {P.}~\bibnamefont {Xue}},\
  }\bibfield  {title} {\bibinfo {title} {Non-{H}ermitian bulk--boundary
  correspondence in quantum dynamics},\ }\href
  {https://doi.org/10.1038/s41567-020-0836-6} {\bibfield  {journal} {\bibinfo
  {journal} {Nat. Phys.}\ }\textbf {\bibinfo {volume} {16}},\ \bibinfo {pages}
  {761} (\bibinfo {year} {2020})}\BibitemShut {NoStop}%
\bibitem [{\citenamefont {Weidemann}\ \emph {et~al.}(2020)\citenamefont
  {Weidemann}, \citenamefont {Kremer}, \citenamefont {Helbig}, \citenamefont
  {Hofmann}, \citenamefont {Stegmaier}, \citenamefont {Greiter}, \citenamefont
  {Thomale},\ and\ \citenamefont {Szameit}}]{weidemann2020}%
  \BibitemOpen
  \bibfield  {author} {\bibinfo {author} {\bibfnamefont {S.}~\bibnamefont
  {Weidemann}}, \bibinfo {author} {\bibfnamefont {M.}~\bibnamefont {Kremer}},
  \bibinfo {author} {\bibfnamefont {T.}~\bibnamefont {Helbig}}, \bibinfo
  {author} {\bibfnamefont {T.}~\bibnamefont {Hofmann}}, \bibinfo {author}
  {\bibfnamefont {A.}~\bibnamefont {Stegmaier}}, \bibinfo {author}
  {\bibfnamefont {M.}~\bibnamefont {Greiter}}, \bibinfo {author} {\bibfnamefont
  {R.}~\bibnamefont {Thomale}},\ and\ \bibinfo {author} {\bibfnamefont
  {A.}~\bibnamefont {Szameit}},\ }\bibfield  {title} {\bibinfo {title}
  {Topological funneling of light},\ }\href
  {https://doi.org/10.1126/science.aaz8727} {\bibfield  {journal} {\bibinfo
  {journal} {Science}\ }\textbf {\bibinfo {volume} {368}},\ \bibinfo {pages}
  {311} (\bibinfo {year} {2020})}\BibitemShut {NoStop}%
\bibitem [{\citenamefont {Koch}\ and\ \citenamefont {Budich}(2022)}]{QNTOS}%
  \BibitemOpen
  \bibfield  {author} {\bibinfo {author} {\bibfnamefont {F.}~\bibnamefont
  {Koch}}\ and\ \bibinfo {author} {\bibfnamefont {J.~C.}\ \bibnamefont
  {Budich}},\ }\bibfield  {title} {\bibinfo {title} {Quantum non-{H}ermitian
  topological sensors},\ }\href
  {https://doi.org/10.1103/PhysRevResearch.4.013113} {\bibfield  {journal}
  {\bibinfo  {journal} {Phys. Rev. Research}\ }\textbf {\bibinfo {volume}
  {4}},\ \bibinfo {pages} {013113} (\bibinfo {year} {2022})}\BibitemShut
  {NoStop}%
\bibitem [{\citenamefont {McDonald}\ and\ \citenamefont
  {Clerk}(2020)}]{clerk2020}%
  \BibitemOpen
  \bibfield  {author} {\bibinfo {author} {\bibfnamefont {A.}~\bibnamefont
  {McDonald}}\ and\ \bibinfo {author} {\bibfnamefont {A.~A.}\ \bibnamefont
  {Clerk}},\ }\bibfield  {title} {\bibinfo {title} {Exponentially-enhanced
  quantum sensing with non-{H}ermitian lattice dynamics},\ }\href
  {https://doi.org/10.1038/s41467-020-19090-4} {\bibfield  {journal} {\bibinfo
  {journal} {Nat. Commun.}\ }\textbf {\bibinfo {volume} {11}},\ \bibinfo
  {pages} {5382} (\bibinfo {year} {2020})}\BibitemShut {NoStop}%
\bibitem [{\citenamefont {Lindblad}(1976)}]{lindblad1976}%
  \BibitemOpen
  \bibfield  {author} {\bibinfo {author} {\bibfnamefont {G.}~\bibnamefont
  {Lindblad}},\ }\bibfield  {title} {\bibinfo {title} {On the generators of
  quantum dynamical semigroups},\ }\href
  {https://doi.org/https://doi.org/10.1007/BF01608499} {\bibfield  {journal}
  {\bibinfo  {journal} {Commun. Math. Phys.}\ }\textbf {\bibinfo {volume}
  {48}},\ \bibinfo {pages} {119} (\bibinfo {year} {1976})}\BibitemShut
  {NoStop}%
\bibitem [{\citenamefont {Langen}\ \emph {et~al.}(2015)\citenamefont {Langen},
  \citenamefont {Geiger},\ and\ \citenamefont {Schmiedmayer}}]{langen2015}%
  \BibitemOpen
  \bibfield  {author} {\bibinfo {author} {\bibfnamefont {T.}~\bibnamefont
  {Langen}}, \bibinfo {author} {\bibfnamefont {R.}~\bibnamefont {Geiger}},\
  and\ \bibinfo {author} {\bibfnamefont {J.}~\bibnamefont {Schmiedmayer}},\
  }\bibfield  {title} {\bibinfo {title} {Ultracold atoms out of equilibrium},\
  }\href {https://doi.org/10.1146/annurev-conmatphys-031214-014548} {\bibfield
  {journal} {\bibinfo  {journal} {Annu. Rev. Condens. Matter Phys.}\ }\textbf
  {\bibinfo {volume} {6}},\ \bibinfo {pages} {201} (\bibinfo {year}
  {2015})}\BibitemShut {NoStop}%
\bibitem [{\citenamefont {Diehl}\ \emph {et~al.}(2008)\citenamefont {Diehl},
  \citenamefont {Micheli}, \citenamefont {Kantian}, \citenamefont {Kraus},
  \citenamefont {B{\"u}chler},\ and\ \citenamefont {Zoller}}]{diehl2008}%
  \BibitemOpen
  \bibfield  {author} {\bibinfo {author} {\bibfnamefont {S.}~\bibnamefont
  {Diehl}}, \bibinfo {author} {\bibfnamefont {A.}~\bibnamefont {Micheli}},
  \bibinfo {author} {\bibfnamefont {A.}~\bibnamefont {Kantian}}, \bibinfo
  {author} {\bibfnamefont {B.}~\bibnamefont {Kraus}}, \bibinfo {author}
  {\bibfnamefont {H.}~\bibnamefont {B{\"u}chler}},\ and\ \bibinfo {author}
  {\bibfnamefont {P.}~\bibnamefont {Zoller}},\ }\bibfield  {title} {\bibinfo
  {title} {Quantum states and phases in driven open quantum systems with cold
  atoms},\ }\href {https://doi.org/10.1038/nphys1073} {\bibfield  {journal}
  {\bibinfo  {journal} {Nat. Phys.}\ }\textbf {\bibinfo {volume} {4}},\
  \bibinfo {pages} {878} (\bibinfo {year} {2008})}\BibitemShut {NoStop}%
\bibitem [{\citenamefont {Kraus}\ \emph {et~al.}(2008)\citenamefont {Kraus},
  \citenamefont {B{\"u}chler}, \citenamefont {Diehl}, \citenamefont {Kantian},
  \citenamefont {Micheli},\ and\ \citenamefont {Zoller}}]{kraus2008}%
  \BibitemOpen
  \bibfield  {author} {\bibinfo {author} {\bibfnamefont {B.}~\bibnamefont
  {Kraus}}, \bibinfo {author} {\bibfnamefont {H.~P.}\ \bibnamefont
  {B{\"u}chler}}, \bibinfo {author} {\bibfnamefont {S.}~\bibnamefont {Diehl}},
  \bibinfo {author} {\bibfnamefont {A.}~\bibnamefont {Kantian}}, \bibinfo
  {author} {\bibfnamefont {A.}~\bibnamefont {Micheli}},\ and\ \bibinfo {author}
  {\bibfnamefont {P.}~\bibnamefont {Zoller}},\ }\bibfield  {title} {\bibinfo
  {title} {Preparation of entangled states by quantum {M}arkov processes},\
  }\href {https://doi.org/10.1103/PhysRevA.78.042307} {\bibfield  {journal}
  {\bibinfo  {journal} {Phys. Rev. A}\ }\textbf {\bibinfo {volume} {78}},\
  \bibinfo {pages} {042307} (\bibinfo {year} {2008})}\BibitemShut {NoStop}%
\bibitem [{\citenamefont {Verstraete}\ \emph {et~al.}(2009)\citenamefont
  {Verstraete}, \citenamefont {Wolf},\ and\ \citenamefont
  {Cirac}}]{verstraete2009}%
  \BibitemOpen
  \bibfield  {author} {\bibinfo {author} {\bibfnamefont {F.}~\bibnamefont
  {Verstraete}}, \bibinfo {author} {\bibfnamefont {M.~M.}\ \bibnamefont
  {Wolf}},\ and\ \bibinfo {author} {\bibfnamefont {J.~I.}\ \bibnamefont
  {Cirac}},\ }\bibfield  {title} {\bibinfo {title} {Quantum computation and
  quantum-state engineering driven by dissipation},\ }\href
  {https://doi.org/10.1038/nphys1342} {\bibfield  {journal} {\bibinfo
  {journal} {Nat. Phys.}\ }\textbf {\bibinfo {volume} {5}},\ \bibinfo {pages}
  {633} (\bibinfo {year} {2009})}\BibitemShut {NoStop}%
\bibitem [{\citenamefont {Krauter}\ \emph {et~al.}(2011)\citenamefont
  {Krauter}, \citenamefont {Muschik}, \citenamefont {Jensen}, \citenamefont
  {Wasilewski}, \citenamefont {Petersen}, \citenamefont {Cirac},\ and\
  \citenamefont {Polzik}}]{krauter2011}%
  \BibitemOpen
  \bibfield  {author} {\bibinfo {author} {\bibfnamefont {H.}~\bibnamefont
  {Krauter}}, \bibinfo {author} {\bibfnamefont {C.~A.}\ \bibnamefont
  {Muschik}}, \bibinfo {author} {\bibfnamefont {K.}~\bibnamefont {Jensen}},
  \bibinfo {author} {\bibfnamefont {W.}~\bibnamefont {Wasilewski}}, \bibinfo
  {author} {\bibfnamefont {J.~M.}\ \bibnamefont {Petersen}}, \bibinfo {author}
  {\bibfnamefont {J.~I.}\ \bibnamefont {Cirac}},\ and\ \bibinfo {author}
  {\bibfnamefont {E.~S.}\ \bibnamefont {Polzik}},\ }\bibfield  {title}
  {\bibinfo {title} {Entanglement generated by dissipation and steady state
  entanglement of two macroscopic objects},\ }\href
  {https://doi.org/10.1103/PhysRevLett.107.080503} {\bibfield  {journal}
  {\bibinfo  {journal} {Phys. Rev. Lett.}\ }\textbf {\bibinfo {volume} {107}},\
  \bibinfo {pages} {080503} (\bibinfo {year} {2011})}\BibitemShut {NoStop}%
\bibitem [{\citenamefont {Breuer}\ and\ \citenamefont
  {Petruccione}(2007)}]{breuer2007}%
  \BibitemOpen
  \bibfield  {author} {\bibinfo {author} {\bibfnamefont {H.~P.}\ \bibnamefont
  {Breuer}}\ and\ \bibinfo {author} {\bibfnamefont {F.}~\bibnamefont
  {Petruccione}},\ }\href
  {https://doi.org/10.1093/acprof:oso/9780199213900.001.0001} {\emph {\bibinfo
  {title} {The theory of open quantum systems}}}\ (\bibinfo  {publisher}
  {Oxford University Press},\ \bibinfo {address} {Oxford},\ \bibinfo {year}
  {2007})\BibitemShut {NoStop}%
\bibitem [{\citenamefont {Diehl}\ \emph {et~al.}(2011)\citenamefont {Diehl},
  \citenamefont {Rico}, \citenamefont {Baranov},\ and\ \citenamefont
  {Zoller}}]{diehl2011}%
  \BibitemOpen
  \bibfield  {author} {\bibinfo {author} {\bibfnamefont {S.}~\bibnamefont
  {Diehl}}, \bibinfo {author} {\bibfnamefont {E.}~\bibnamefont {Rico}},
  \bibinfo {author} {\bibfnamefont {M.~A.}\ \bibnamefont {Baranov}},\ and\
  \bibinfo {author} {\bibfnamefont {P.}~\bibnamefont {Zoller}},\ }\bibfield
  {title} {\bibinfo {title} {Topology by dissipation in atomic quantum wires},\
  }\href {https://doi.org/10.1038/nphys2106} {\bibfield  {journal} {\bibinfo
  {journal} {Nat. Phys.}\ }\textbf {\bibinfo {volume} {7}},\ \bibinfo {pages}
  {971} (\bibinfo {year} {2011})}\BibitemShut {NoStop}%
\bibitem [{\citenamefont {Bardyn}\ \emph {et~al.}(2013)\citenamefont {Bardyn},
  \citenamefont {Baranov}, \citenamefont {Kraus}, \citenamefont {Rico},
  \citenamefont {{\.I}mamo{\u{g}}lu}, \citenamefont {Zoller},\ and\
  \citenamefont {Diehl}}]{bardyn2013}%
  \BibitemOpen
  \bibfield  {author} {\bibinfo {author} {\bibfnamefont {C.~E.}\ \bibnamefont
  {Bardyn}}, \bibinfo {author} {\bibfnamefont {M.~A.}\ \bibnamefont {Baranov}},
  \bibinfo {author} {\bibfnamefont {C.~V.}\ \bibnamefont {Kraus}}, \bibinfo
  {author} {\bibfnamefont {E.}~\bibnamefont {Rico}}, \bibinfo {author}
  {\bibfnamefont {A.}~\bibnamefont {{\.I}mamo{\u{g}}lu}}, \bibinfo {author}
  {\bibfnamefont {P.}~\bibnamefont {Zoller}},\ and\ \bibinfo {author}
  {\bibfnamefont {S.}~\bibnamefont {Diehl}},\ }\bibfield  {title} {\bibinfo
  {title} {Topology by dissipation},\ }\href
  {https://doi.org/10.1088/1367-2630/15/8/085001} {\bibfield  {journal}
  {\bibinfo  {journal} {New J. Phys.}\ }\textbf {\bibinfo {volume} {15}},\
  \bibinfo {pages} {085001} (\bibinfo {year} {2013})}\BibitemShut {NoStop}%
\bibitem [{\citenamefont {Budich}\ \emph {et~al.}(2015)\citenamefont {Budich},
  \citenamefont {Zoller},\ and\ \citenamefont {Diehl}}]{budich2015}%
  \BibitemOpen
  \bibfield  {author} {\bibinfo {author} {\bibfnamefont {J.~C.}\ \bibnamefont
  {Budich}}, \bibinfo {author} {\bibfnamefont {P.}~\bibnamefont {Zoller}},\
  and\ \bibinfo {author} {\bibfnamefont {S.}~\bibnamefont {Diehl}},\ }\bibfield
   {title} {\bibinfo {title} {Dissipative preparation of {C}hern insulators},\
  }\href {https://doi.org/10.1103/PhysRevA.91.042117} {\bibfield  {journal}
  {\bibinfo  {journal} {Phys. Rev. A}\ }\textbf {\bibinfo {volume} {91}},\
  \bibinfo {pages} {042117} (\bibinfo {year} {2015})}\BibitemShut {NoStop}%
\bibitem [{\citenamefont {Liu}\ \emph {et~al.}(2021)\citenamefont {Liu},
  \citenamefont {Bergholtz},\ and\ \citenamefont {Budich}}]{liu2021}%
  \BibitemOpen
  \bibfield  {author} {\bibinfo {author} {\bibfnamefont {Z.}~\bibnamefont
  {Liu}}, \bibinfo {author} {\bibfnamefont {E.~J.}\ \bibnamefont {Bergholtz}},\
  and\ \bibinfo {author} {\bibfnamefont {J.~C.}\ \bibnamefont {Budich}},\
  }\bibfield  {title} {\bibinfo {title} {Dissipative preparation of fractional
  {C}hern insulators},\ }\href
  {https://doi.org/10.1103/PhysRevResearch.3.043119} {\bibfield  {journal}
  {\bibinfo  {journal} {Phys. Rev. Research}\ }\textbf {\bibinfo {volume}
  {3}},\ \bibinfo {pages} {043119} (\bibinfo {year} {2021})}\BibitemShut
  {NoStop}%
\bibitem [{\citenamefont {He}\ \emph {et~al.}(2022)\citenamefont {He},
  \citenamefont {Liu}, \citenamefont {Wang},\ and\ \citenamefont
  {Zhu}}]{he2021}%
  \BibitemOpen
  \bibfield  {author} {\bibinfo {author} {\bibfnamefont {P.}~\bibnamefont
  {He}}, \bibinfo {author} {\bibfnamefont {Y.-G.}\ \bibnamefont {Liu}},
  \bibinfo {author} {\bibfnamefont {J.-T.}\ \bibnamefont {Wang}},\ and\
  \bibinfo {author} {\bibfnamefont {S.-L.}\ \bibnamefont {Zhu}},\ }\bibfield
  {title} {\bibinfo {title} {Damping transition in an open generalized
  {A}ubry-{A}ndr{\'e}-{H}arper model},\ }\href
  {https://doi.org/10.1103/PhysRevA.105.023311} {\bibfield  {journal} {\bibinfo
   {journal} {Phys. Rev. A}\ }\textbf {\bibinfo {volume} {105}},\ \bibinfo
  {pages} {023311} (\bibinfo {year} {2022})}\BibitemShut {NoStop}%
\bibitem [{\citenamefont {Song}\ \emph {et~al.}(2019)\citenamefont {Song},
  \citenamefont {Yao},\ and\ \citenamefont {Wang}}]{fei2019}%
  \BibitemOpen
  \bibfield  {author} {\bibinfo {author} {\bibfnamefont {F.}~\bibnamefont
  {Song}}, \bibinfo {author} {\bibfnamefont {S.}~\bibnamefont {Yao}},\ and\
  \bibinfo {author} {\bibfnamefont {Z.}~\bibnamefont {Wang}},\ }\bibfield
  {title} {\bibinfo {title} {Non-{H}ermitian skin effect and chiral damping in
  open quantum systems},\ }\href
  {https://doi.org/10.1103/PhysRevLett.123.170401} {\bibfield  {journal}
  {\bibinfo  {journal} {Phys. Rev. Lett.}\ }\textbf {\bibinfo {volume} {123}},\
  \bibinfo {pages} {170401} (\bibinfo {year} {2019})}\BibitemShut {NoStop}%
\bibitem [{\citenamefont {Wanjura}\ \emph {et~al.}(2020)\citenamefont
  {Wanjura}, \citenamefont {Brunelli},\ and\ \citenamefont
  {Nunnenkamp}}]{wanjura2020}%
  \BibitemOpen
  \bibfield  {author} {\bibinfo {author} {\bibfnamefont {C.~C.}\ \bibnamefont
  {Wanjura}}, \bibinfo {author} {\bibfnamefont {M.}~\bibnamefont {Brunelli}},\
  and\ \bibinfo {author} {\bibfnamefont {A.}~\bibnamefont {Nunnenkamp}},\
  }\bibfield  {title} {\bibinfo {title} {Topological framework for directional
  amplification in driven-dissipative cavity arrays},\ }\href
  {https://doi.org/10.1038/s41467-020-16863-9} {\bibfield  {journal} {\bibinfo
  {journal} {Nat. Commun.}\ }\textbf {\bibinfo {volume} {11}},\ \bibinfo
  {pages} {3149} (\bibinfo {year} {2020})}\BibitemShut {NoStop}%
\bibitem [{\citenamefont {Liu}\ \emph {et~al.}(2020)\citenamefont {Liu},
  \citenamefont {Zhang}, \citenamefont {Yang},\ and\ \citenamefont
  {Chen}}]{HelicalDamping}%
  \BibitemOpen
  \bibfield  {author} {\bibinfo {author} {\bibfnamefont {C.-H.}\ \bibnamefont
  {Liu}}, \bibinfo {author} {\bibfnamefont {K.}~\bibnamefont {Zhang}}, \bibinfo
  {author} {\bibfnamefont {Z.}~\bibnamefont {Yang}},\ and\ \bibinfo {author}
  {\bibfnamefont {S.}~\bibnamefont {Chen}},\ }\bibfield  {title} {\bibinfo
  {title} {Helical damping and dynamical critical skin effect in open quantum
  systems},\ }\href {https://doi.org/10.1103/PhysRevResearch.2.043167}
  {\bibfield  {journal} {\bibinfo  {journal} {Phys. Rev. Research}\ }\textbf
  {\bibinfo {volume} {2}},\ \bibinfo {pages} {043167} (\bibinfo {year}
  {2020})}\BibitemShut {NoStop}%
\bibitem [{\citenamefont {Mao}\ \emph {et~al.}(2021)\citenamefont {Mao},
  \citenamefont {Deng},\ and\ \citenamefont {Zhang}}]{mao2021}%
  \BibitemOpen
  \bibfield  {author} {\bibinfo {author} {\bibfnamefont {L.}~\bibnamefont
  {Mao}}, \bibinfo {author} {\bibfnamefont {T.}~\bibnamefont {Deng}},\ and\
  \bibinfo {author} {\bibfnamefont {P.}~\bibnamefont {Zhang}},\ }\bibfield
  {title} {\bibinfo {title} {Boundary condition independence of non-{H}ermitian
  {H}amiltonian dynamics},\ }\href
  {https://doi.org/10.1103/PhysRevB.104.125435} {\bibfield  {journal} {\bibinfo
   {journal} {Phys. Rev. B}\ }\textbf {\bibinfo {volume} {104}},\ \bibinfo
  {pages} {125435} (\bibinfo {year} {2021})}\BibitemShut {NoStop}%
\bibitem [{\citenamefont {Wanjura}\ \emph {et~al.}(2021)\citenamefont
  {Wanjura}, \citenamefont {Brunelli},\ and\ \citenamefont
  {Nunnenkamp}}]{WanjuraPRL}%
  \BibitemOpen
  \bibfield  {author} {\bibinfo {author} {\bibfnamefont {C.~C.}\ \bibnamefont
  {Wanjura}}, \bibinfo {author} {\bibfnamefont {M.}~\bibnamefont {Brunelli}},\
  and\ \bibinfo {author} {\bibfnamefont {A.}~\bibnamefont {Nunnenkamp}},\
  }\bibfield  {title} {\bibinfo {title} {Correspondence between non-{H}ermitian
  topology and directional amplification in the presence of disorder},\ }\href
  {https://doi.org/10.1103/PhysRevLett.127.213601} {\bibfield  {journal}
  {\bibinfo  {journal} {Phys. Rev. Lett.}\ }\textbf {\bibinfo {volume} {127}},\
  \bibinfo {pages} {213601} (\bibinfo {year} {2021})}\BibitemShut {NoStop}%
\bibitem [{\citenamefont {McDonald}\ \emph {et~al.}(2022)\citenamefont
  {McDonald}, \citenamefont {Hanai},\ and\ \citenamefont {Clerk}}]{NENH}%
  \BibitemOpen
  \bibfield  {author} {\bibinfo {author} {\bibfnamefont {A.}~\bibnamefont
  {McDonald}}, \bibinfo {author} {\bibfnamefont {R.}~\bibnamefont {Hanai}},\
  and\ \bibinfo {author} {\bibfnamefont {A.~A.}\ \bibnamefont {Clerk}},\
  }\bibfield  {title} {\bibinfo {title} {Nonequilibrium stationary states of
  quantum non-{H}ermitian lattice models},\ }\href
  {https://doi.org/10.1103/PhysRevB.105.064302} {\bibfield  {journal} {\bibinfo
   {journal} {Phys. Rev. B}\ }\textbf {\bibinfo {volume} {105}},\ \bibinfo
  {pages} {064302} (\bibinfo {year} {2022})}\BibitemShut {NoStop}%
\bibitem [{\citenamefont {Haga}\ \emph {et~al.}(2021)\citenamefont {Haga},
  \citenamefont {Nakagawa}, \citenamefont {Hamazaki},\ and\ \citenamefont
  {Ueda}}]{ueda2021}%
  \BibitemOpen
  \bibfield  {author} {\bibinfo {author} {\bibfnamefont {T.}~\bibnamefont
  {Haga}}, \bibinfo {author} {\bibfnamefont {M.}~\bibnamefont {Nakagawa}},
  \bibinfo {author} {\bibfnamefont {R.}~\bibnamefont {Hamazaki}},\ and\
  \bibinfo {author} {\bibfnamefont {M.}~\bibnamefont {Ueda}},\ }\bibfield
  {title} {\bibinfo {title} {Liouvillian skin effect: Slowing down of
  relaxation processes without gap closing},\ }\href
  {https://doi.org/10.1103/PhysRevLett.127.070402} {\bibfield  {journal}
  {\bibinfo  {journal} {Phys. Rev. Lett.}\ }\textbf {\bibinfo {volume} {127}},\
  \bibinfo {pages} {070402} (\bibinfo {year} {2021})}\BibitemShut {NoStop}%
\bibitem [{\citenamefont {Zhou}\ and\ \citenamefont {Yu}()}]{zhou2021}%
  \BibitemOpen
  \bibfield  {author} {\bibinfo {author} {\bibfnamefont {Z.}~\bibnamefont
  {Zhou}}\ and\ \bibinfo {author} {\bibfnamefont {Z.}~\bibnamefont {Yu}},\
  }\bibfield  {title} {\bibinfo {title} {Skin {E}ffect in {Q}uadratic
  {L}indbladian {S}ystems: An {A}djoint {F}ermion {A}pproach},\ }\href
  {https://arxiv.org/abs/2110.09874} {\bibinfo  {journal} {arXiv:2110.09874}\
  }\BibitemShut {NoStop}%
\bibitem [{\citenamefont {Prosen}(2008)}]{prosen2008}%
  \BibitemOpen
\bibfield  {journal} {  }\bibfield  {author} {\bibinfo {author} {\bibfnamefont
  {T.}~\bibnamefont {Prosen}},\ }\bibfield  {title} {\bibinfo {title} {Third
  quantization: A general method to solve master equations for quadratic open
  {F}ermi systems},\ }\href {https://doi.org/10.1088/1367-2630/10/4/043026}
  {\bibfield  {journal} {\bibinfo  {journal} {New J. Phys.}\ }\textbf {\bibinfo
  {volume} {10}},\ \bibinfo {pages} {043026} (\bibinfo {year}
  {2008})}\BibitemShut {NoStop}%
\bibitem [{\citenamefont {Prosen}\ and\ \citenamefont
  {{\v{Z}}unkovi{\v{c}}}(2010)}]{prosen2010ex}%
  \BibitemOpen
  \bibfield  {author} {\bibinfo {author} {\bibfnamefont {T.}~\bibnamefont
  {Prosen}}\ and\ \bibinfo {author} {\bibfnamefont {B.}~\bibnamefont
  {{\v{Z}}unkovi{\v{c}}}},\ }\bibfield  {title} {\bibinfo {title} {Exact
  solution of {M}arkovian master equations for quadratic {F}ermi systems:
  Thermal baths, open {XY} spin chains and non-equilibrium phase transition},\
  }\href {https://doi.org/10.1088/1367-2630/12/2/025016} {\bibfield  {journal}
  {\bibinfo  {journal} {New J. Phys.}\ }\textbf {\bibinfo {volume} {12}},\
  \bibinfo {pages} {025016} (\bibinfo {year} {2010})}\BibitemShut {NoStop}%
\bibitem [{\citenamefont {Prosen}(2010)}]{prosen2010sp}%
  \BibitemOpen
  \bibfield  {author} {\bibinfo {author} {\bibfnamefont {T.}~\bibnamefont
  {Prosen}},\ }\bibfield  {title} {\bibinfo {title} {Spectral theorem for the
  {L}indblad equation for quadratic open fermionic systems},\ }\href
  {https://doi.org/10.1088/1742-5468/2010/07/P07020} {\bibfield  {journal}
  {\bibinfo  {journal} {J. Stat. Mech.}\ }\textbf {\bibinfo {volume} {2010}},\
  \bibinfo {pages} {P07020} (\bibinfo {year} {2010})}\BibitemShut {NoStop}%
\bibitem [{\citenamefont {Prosen}\ and\ \citenamefont
  {Ilievski}(2011)}]{prosen2011}%
  \BibitemOpen
  \bibfield  {author} {\bibinfo {author} {\bibfnamefont {T.}~\bibnamefont
  {Prosen}}\ and\ \bibinfo {author} {\bibfnamefont {E.}~\bibnamefont
  {Ilievski}},\ }\bibfield  {title} {\bibinfo {title} {Nonequilibrium phase
  transition in a periodically driven {XY} spin chain},\ }\href
  {https://doi.org/10.1103/PhysRevLett.107.060403} {\bibfield  {journal}
  {\bibinfo  {journal} {Phys. Rev. Lett.}\ }\textbf {\bibinfo {volume} {107}},\
  \bibinfo {pages} {060403} (\bibinfo {year} {2011})}\BibitemShut {NoStop}%
\bibitem [{\citenamefont {van Caspel}\ \emph {et~al.}(2019)\citenamefont {van
  Caspel}, \citenamefont {Tapias~Arze},\ and\ \citenamefont
  {P{\'e}rez~Castillo}}]{caspel2019}%
  \BibitemOpen
  \bibfield  {author} {\bibinfo {author} {\bibfnamefont {M.}~\bibnamefont {van
  Caspel}}, \bibinfo {author} {\bibfnamefont {S.~E.}\ \bibnamefont
  {Tapias~Arze}},\ and\ \bibinfo {author} {\bibfnamefont {I.}~\bibnamefont
  {P{\'e}rez~Castillo}},\ }\bibfield  {title} {\bibinfo {title} {Dynamical
  signatures of topological order in the driven-dissipative {K}itaev chain},\
  }\href {https://doi.org/10.21468/SciPostPhys.6.2.026} {\bibfield  {journal}
  {\bibinfo  {journal} {SciPost Phys.}\ }\textbf {\bibinfo {volume} {6}},\
  \bibinfo {pages} {026} (\bibinfo {year} {2019})}\BibitemShut {NoStop}%
\bibitem [{\citenamefont {Lieu}\ \emph {et~al.}(2020)\citenamefont {Lieu},
  \citenamefont {McGinley},\ and\ \citenamefont {Cooper}}]{cooper2020}%
  \BibitemOpen
  \bibfield  {author} {\bibinfo {author} {\bibfnamefont {S.}~\bibnamefont
  {Lieu}}, \bibinfo {author} {\bibfnamefont {M.}~\bibnamefont {McGinley}},\
  and\ \bibinfo {author} {\bibfnamefont {N.~R.}\ \bibnamefont {Cooper}},\
  }\bibfield  {title} {\bibinfo {title} {Tenfold way for quadratic
  {L}indbladians},\ }\href {https://doi.org/10.1103/PhysRevLett.124.040401}
  {\bibfield  {journal} {\bibinfo  {journal} {Phys. Rev. Lett.}\ }\textbf
  {\bibinfo {volume} {124}},\ \bibinfo {pages} {040401} (\bibinfo {year}
  {2020})}\BibitemShut {NoStop}%
\bibitem [{\citenamefont {Minganti}\ \emph {et~al.}(2018)\citenamefont
  {Minganti}, \citenamefont {Biella}, \citenamefont {Bartolo},\ and\
  \citenamefont {Ciuti}}]{minganti2018}%
  \BibitemOpen
  \bibfield  {author} {\bibinfo {author} {\bibfnamefont {F.}~\bibnamefont
  {Minganti}}, \bibinfo {author} {\bibfnamefont {A.}~\bibnamefont {Biella}},
  \bibinfo {author} {\bibfnamefont {N.}~\bibnamefont {Bartolo}},\ and\ \bibinfo
  {author} {\bibfnamefont {C.}~\bibnamefont {Ciuti}},\ }\bibfield  {title}
  {\bibinfo {title} {Spectral theory of {L}iouvillians for dissipative phase
  transitions},\ }\href {https://doi.org/10.1103/PhysRevA.98.042118} {\bibfield
   {journal} {\bibinfo  {journal} {Phys. Rev. A}\ }\textbf {\bibinfo {volume}
  {98}},\ \bibinfo {pages} {042118} (\bibinfo {year} {2018})}\BibitemShut
  {NoStop}%
\bibitem [{\citenamefont {Brody}(2014)}]{brody2013}%
  \BibitemOpen
  \bibfield  {author} {\bibinfo {author} {\bibfnamefont {D.~C.}\ \bibnamefont
  {Brody}},\ }\bibfield  {title} {\bibinfo {title} {Biorthogonal quantum
  mechanics},\ }\href {https://doi.org/10.1088/1751-8113/47/3/035305}
  {\bibfield  {journal} {\bibinfo  {journal} {J. Phys. A}\ }\textbf {\bibinfo
  {volume} {47}},\ \bibinfo {pages} {035305} (\bibinfo {year}
  {2014})}\BibitemShut {NoStop}%
\bibitem [{\citenamefont {Hatano}\ and\ \citenamefont
  {Nelson}(1996)}]{hatano96}%
  \BibitemOpen
  \bibfield  {author} {\bibinfo {author} {\bibfnamefont {N.}~\bibnamefont
  {Hatano}}\ and\ \bibinfo {author} {\bibfnamefont {D.~R.}\ \bibnamefont
  {Nelson}},\ }\bibfield  {title} {\bibinfo {title} {Localization transitions
  in non-{H}ermitian quantum mechanics},\ }\href
  {https://doi.org/10.1103/PhysRevLett.77.570} {\bibfield  {journal} {\bibinfo
  {journal} {Phys. Rev. Lett.}\ }\textbf {\bibinfo {volume} {77}},\ \bibinfo
  {pages} {570} (\bibinfo {year} {1996})}\BibitemShut {NoStop}%
\bibitem [{\citenamefont {Hatano}\ and\ \citenamefont
  {Nelson}(1997)}]{hatano97}%
  \BibitemOpen
  \bibfield  {author} {\bibinfo {author} {\bibfnamefont {N.}~\bibnamefont
  {Hatano}}\ and\ \bibinfo {author} {\bibfnamefont {D.~R.}\ \bibnamefont
  {Nelson}},\ }\bibfield  {title} {\bibinfo {title} {Vortex pinning and
  non-{H}ermitian quantum mechanics},\ }\href
  {https://doi.org/10.1103/PhysRevB.56.8651} {\bibfield  {journal} {\bibinfo
  {journal} {Phys. Rev. B}\ }\textbf {\bibinfo {volume} {56}},\ \bibinfo
  {pages} {8651} (\bibinfo {year} {1997})}\BibitemShut {NoStop}%
\bibitem [{\citenamefont {Hatano}\ and\ \citenamefont
  {Nelson}(1998)}]{hatano98}%
  \BibitemOpen
  \bibfield  {author} {\bibinfo {author} {\bibfnamefont {N.}~\bibnamefont
  {Hatano}}\ and\ \bibinfo {author} {\bibfnamefont {D.~R.}\ \bibnamefont
  {Nelson}},\ }\bibfield  {title} {\bibinfo {title} {Non-{H}ermitian
  delocalization and eigenfunctions},\ }\href
  {https://doi.org/10.1103/PhysRevB.58.8384} {\bibfield  {journal} {\bibinfo
  {journal} {Phys. Rev. B}\ }\textbf {\bibinfo {volume} {58}},\ \bibinfo
  {pages} {8384} (\bibinfo {year} {1998})}\BibitemShut {NoStop}%
\bibitem [{\citenamefont {Altland}\ \emph {et~al.}(2021)\citenamefont
  {Altland}, \citenamefont {Fleischhauer},\ and\ \citenamefont
  {Diehl}}]{altland2021}%
  \BibitemOpen
  \bibfield  {author} {\bibinfo {author} {\bibfnamefont {A.}~\bibnamefont
  {Altland}}, \bibinfo {author} {\bibfnamefont {M.}~\bibnamefont
  {Fleischhauer}},\ and\ \bibinfo {author} {\bibfnamefont {S.}~\bibnamefont
  {Diehl}},\ }\bibfield  {title} {\bibinfo {title} {Symmetry classes of open
  fermionic quantum matter},\ }\href
  {https://doi.org/10.1103/PhysRevX.11.021037} {\bibfield  {journal} {\bibinfo
  {journal} {Phys. Rev. X}\ }\textbf {\bibinfo {volume} {11}},\ \bibinfo
  {pages} {021037} (\bibinfo {year} {2021})}\BibitemShut {NoStop}%
\bibitem [{\citenamefont {Hatano}(2019)}]{hatano2019}%
  \BibitemOpen
  \bibfield  {author} {\bibinfo {author} {\bibfnamefont {N.}~\bibnamefont
  {Hatano}},\ }\bibfield  {title} {\bibinfo {title} {Exceptional points of the
  {L}indblad operator of a two-level system},\ }\href
  {https://doi.org/10.1080/00268976.2019.1593535} {\bibfield  {journal}
  {\bibinfo  {journal} {Mol. Phys.}\ }\textbf {\bibinfo {volume} {117}},\
  \bibinfo {pages} {2121} (\bibinfo {year} {2019})}\BibitemShut {NoStop}%
\bibitem [{\citenamefont {Benatti}\ \emph {et~al.}(2021)\citenamefont
  {Benatti}, \citenamefont {Floreanini},\ and\ \citenamefont
  {Memarzadeh}}]{benatti2021}%
  \BibitemOpen
  \bibfield  {author} {\bibinfo {author} {\bibfnamefont {F.}~\bibnamefont
  {Benatti}}, \bibinfo {author} {\bibfnamefont {R.}~\bibnamefont
  {Floreanini}},\ and\ \bibinfo {author} {\bibfnamefont {L.}~\bibnamefont
  {Memarzadeh}},\ }\bibfield  {title} {\bibinfo {title} {Exact steady state of
  the open {XX}-spin chain: {E}ntanglement and transport properties},\ }\href
  {https://doi.org/10.1103/PRXQuantum.2.030344} {\bibfield  {journal} {\bibinfo
   {journal} {PRX Quantum}\ }\textbf {\bibinfo {volume} {2}},\ \bibinfo {pages}
  {030344} (\bibinfo {year} {2021})}\BibitemShut {NoStop}%
\bibitem [{\citenamefont {Sayyad}\ \emph {et~al.}(2021)\citenamefont {Sayyad},
  \citenamefont {Yu}, \citenamefont {Grushin},\ and\ \citenamefont
  {Sieberer}}]{sayyad2021}%
  \BibitemOpen
  \bibfield  {author} {\bibinfo {author} {\bibfnamefont {S.}~\bibnamefont
  {Sayyad}}, \bibinfo {author} {\bibfnamefont {J.}~\bibnamefont {Yu}}, \bibinfo
  {author} {\bibfnamefont {A.~G.}\ \bibnamefont {Grushin}},\ and\ \bibinfo
  {author} {\bibfnamefont {L.~M.}\ \bibnamefont {Sieberer}},\ }\bibfield
  {title} {\bibinfo {title} {Entanglement spectrum crossings reveal
  non-{H}ermitian dynamical topology},\ }\href
  {https://doi.org/10.1103/PhysRevResearch.3.033022} {\bibfield  {journal}
  {\bibinfo  {journal} {Phys. Rev. Research}\ }\textbf {\bibinfo {volume}
  {3}},\ \bibinfo {pages} {033022} (\bibinfo {year} {2021})}\BibitemShut
  {NoStop}%
\bibitem [{\citenamefont {Scarlatella}\ \emph {et~al.}(2021)\citenamefont
  {Scarlatella}, \citenamefont {Clerk}, \citenamefont {Fazio},\ and\
  \citenamefont {Schir{\'o}}}]{clerk2021dyn}%
  \BibitemOpen
  \bibfield  {author} {\bibinfo {author} {\bibfnamefont {O.}~\bibnamefont
  {Scarlatella}}, \bibinfo {author} {\bibfnamefont {A.~A.}\ \bibnamefont
  {Clerk}}, \bibinfo {author} {\bibfnamefont {R.}~\bibnamefont {Fazio}},\ and\
  \bibinfo {author} {\bibfnamefont {M.}~\bibnamefont {Schir{\'o}}},\ }\bibfield
   {title} {\bibinfo {title} {Dynamical mean-field theory for {M}arkovian open
  quantum many-body systems},\ }\href
  {https://doi.org/10.1103/PhysRevX.11.031018} {\bibfield  {journal} {\bibinfo
  {journal} {Phys. Rev. X}\ }\textbf {\bibinfo {volume} {11}},\ \bibinfo
  {pages} {031018} (\bibinfo {year} {2021})}\BibitemShut {NoStop}%
\bibitem [{\citenamefont {Yamamoto}\ \emph {et~al.}(2022)\citenamefont
  {Yamamoto}, \citenamefont {Nakagawa}, \citenamefont {Tezuka}, \citenamefont
  {Ueda},\ and\ \citenamefont {Kawakami}}]{yamamoto2021}%
  \BibitemOpen
  \bibfield  {author} {\bibinfo {author} {\bibfnamefont {K.}~\bibnamefont
  {Yamamoto}}, \bibinfo {author} {\bibfnamefont {M.}~\bibnamefont {Nakagawa}},
  \bibinfo {author} {\bibfnamefont {M.}~\bibnamefont {Tezuka}}, \bibinfo
  {author} {\bibfnamefont {M.}~\bibnamefont {Ueda}},\ and\ \bibinfo {author}
  {\bibfnamefont {N.}~\bibnamefont {Kawakami}},\ }\bibfield  {title} {\bibinfo
  {title} {Universal properties of dissipative {T}omonaga-{L}uttinger liquids:
  Case study of a non-{H}ermitian {XXZ} spin chain},\ }\href
  {https://doi.org/10.1103/PhysRevB.105.205125} {\bibfield  {journal} {\bibinfo
   {journal} {Phys. Rev. B}\ }\textbf {\bibinfo {volume} {105}},\ \bibinfo
  {pages} {205125} (\bibinfo {year} {2022})}\BibitemShut {NoStop}%
\bibitem [{\citenamefont {McDonald}\ and\ \citenamefont
  {Clerk}(2022)}]{clerk2022}%
  \BibitemOpen
  \bibfield  {author} {\bibinfo {author} {\bibfnamefont {A.}~\bibnamefont
  {McDonald}}\ and\ \bibinfo {author} {\bibfnamefont {A.~A.}\ \bibnamefont
  {Clerk}},\ }\bibfield  {title} {\bibinfo {title} {Exact solutions of
  interacting dissipative systems via weak symmetries},\ }\href
  {https://doi.org/10.1103/PhysRevLett.128.033602} {\bibfield  {journal}
  {\bibinfo  {journal} {Phys. Rev. Lett.}\ }\textbf {\bibinfo {volume} {128}},\
  \bibinfo {pages} {033602} (\bibinfo {year} {2022})}\BibitemShut {NoStop}%
\bibitem [{\citenamefont {S\'a}\ \emph {et~al.}()\citenamefont {S\'a},
  \citenamefont {Ribeiro},\ and\ \citenamefont {Prosen}}]{prosen2021}%
  \BibitemOpen
  \bibfield  {author} {\bibinfo {author} {\bibfnamefont {L.}~\bibnamefont
  {S\'a}}, \bibinfo {author} {\bibfnamefont {P.}~\bibnamefont {Ribeiro}},\ and\
  \bibinfo {author} {\bibfnamefont {T.}~\bibnamefont {Prosen}},\ }\bibfield
  {title} {\bibinfo {title} {Lindbladian {D}issipation of
  {S}trongly-{C}orrelated {Q}uantum {M}atter},\ }\href
  {https://arxiv.org/abs/2112.12109} {\bibinfo  {journal} {arXiv:2112.12109}\
  }\BibitemShut {NoStop}%
\bibitem [{\citenamefont {Kunst}\ \emph {et~al.}(2019)\citenamefont {Kunst},
  \citenamefont {van Miert},\ and\ \citenamefont {Bergholtz}}]{flore2019e}%
  \BibitemOpen
\bibfield  {journal} {  }\bibfield  {author} {\bibinfo {author} {\bibfnamefont
  {F.~K.}\ \bibnamefont {Kunst}}, \bibinfo {author} {\bibfnamefont
  {G.}~\bibnamefont {van Miert}},\ and\ \bibinfo {author} {\bibfnamefont
  {E.~J.}\ \bibnamefont {Bergholtz}},\ }\bibfield  {title} {\bibinfo {title}
  {Extended {B}loch theorem for topological lattice models with open
  boundaries},\ }\href {https://doi.org/10.1103/PhysRevB.99.085427} {\bibfield
  {journal} {\bibinfo  {journal} {Phys. Rev. B}\ }\textbf {\bibinfo {volume}
  {99}},\ \bibinfo {pages} {085427} (\bibinfo {year} {2019})}\BibitemShut
  {NoStop}%
\end{thebibliography}%


\end{document}